\documentclass[12pt]{article}
\usepackage{epsfig,amssymb,amsmath,psfrag,subfigure,rotate}
\pdfoutput=1

\allowdisplaybreaks
\newcommand{\insertfig}[2]{\includegraphics[width=#1cm]{#2}}

\def \be  {\begin{equation}}
\def \ee  {\end{equation}}
\def \ba  {\begin{eqnarray}}
\def \ea  {\end{eqnarray}}
\def \baa {\begin{eqnarray*}}
\def \eaa {\end{eqnarray*}}
\def \lab #1 {\label{#1}}

\newcommand\re[1]{(\ref{#1})}
\def\d{\hbox{{d}\kern-.20em\hbox{l}}}

\def \matrix #1 {\left(\begin{array}{cc} #1 \end{array}\right)}

\def \tr {\mathop{\rm tr}\nolimits}

\def\II{\hbox{{1}\kern-.25em\hbox{l}}}

\newcommand \vev [1] {\langle{#1}\rangle}
\newcommand \VEV [1] {\left\langle{#1}\right\rangle}
\newcommand \ket [1] {|{#1}\rangle}
\newcommand \bra [1] {\langle {#1}|}
\newcommand{\bit}[1]{\mbox{\boldmath$#1$}}

\newcommand{\ft}[2]{{\textstyle\frac{#1}{#2}}}
\numberwithin{equation}{section}
\begin{document}

\begin{titlepage}

\thispagestyle{empty}

\vspace*{2cm}

\centerline{\large \bf Quantum mechanics of null polygonal Wilson loops}
\vspace*{1cm}

\centerline{\sc A.V.~Belitsky$^{a}$,  S.E.~Derkachov$^b$, A.N~Manashov$^{c,d}$}

\vspace{10mm}

\centerline{\it $^a$Department of Physics, Arizona State University}
\centerline{\it Tempe, AZ 85287-1504, USA}

\vspace{3mm}

\centerline{\it $^b$St. Petersburg Department of Steklov Mathematical Institute}
\centerline{\it Fontanka 27, 191023 St. Petersburg, Russia}

\vspace{3mm}

\centerline{\it $^c$Institut f\"ur Theoretische Physik, Universit\"at Regensburg}
\centerline{\it D-93040 Regensburg, Germany}

\vspace{3mm}
\centerline{\it $^d$Department of Theoretical Physics, St.-Petersburg State University}
\centerline{\it 199034, St.-Petersburg, Russia}

\vspace{1cm}

\centerline{\bf Abstract}

\vspace{5mm}

Scattering amplitudes in maximally supersymmetric gauge theory are dual to super-Wilson loops on null polygonal
contours. The operator product expansion for the latter revealed that their dynamics is governed by the evolution of
multiparticle GKP excitations. They were shown to emerge from the spectral problem of an underlying open spin chain.
In this work we solve this model with the help of the Baxter $\mathbb{Q}$-operator and Sklyanin's  Separation of Variables
methods. We provide an explicit construction for eigenfunctions and eigenvalues of GKP excitations. We demonstrate how
the former define the so-called multiparticle hexagon transitions in super-Wison loops and prove their factorized form
suggested earlier.

\end{titlepage}

\newpage

\pagestyle{plain}
\setcounter{page} 1

{\small \tableofcontents}

\newpage

\section{Motivation}\label{sect:motivation}

Recently, it was suggested that null polygonal super-Wilson loop may serve as a generating function for all scattering amplitudes in
(regularized) planar maximally supersymmetric $SU(N)$ Yang-Mills theory. The equivalence was first established for the so-called
Maximally Helicity Violating (MHV) gluon amplitudes, that involve two particles with helicities opposite to the rest, and the holonomy of a gauge
connection circulating around a closed contour composed of straight light-like segments \cite{AldMal08,KorDruSok08,BraHesTra07}.
These segments are identified with momenta of incoming gluons involved in the scattering process and one regards the resulting loop
as living in a dual coordinates space \cite{DruHenSmiSok06,KorDruSok08}. The conventional soft and collinear divergences of massless
gluon amplitudes emerge in this picture from the ultraviolet singularities stemming from virtual corrections in the vicinity of the cusps where
the two adjacent segments meet in a point. However, not only the divergent parts were shown to agree, but also finite contributions to both
coincide as well. By uplifting the contour to superspace, the duality establishes the equivalence\footnote{A proper care has to be taken to
handle the regularization of ultraviolet divergences \cite{BelKorSok11} in order to maintain supersymmetry of the super-loop and
conformal symmetry of the so-called remainder function \cite{Bel12a,BelCarHuo12}.} between the vacuum expectation value of the
super-Wilson loop $\mathcal{W}_n$ in $\mathcal{N} = 4$ SYM and the reduced $N$-particle matrix element $\widehat{A}_n$ of the S-matrix of the
theory \cite{ManSki10,Car10} that is expected to hold nonperturbatively in 't Hooft coupling $a = g^2_{\rm\scriptscriptstyle YM} N_c/(4 \pi^2)$. The
former is determined by bosonic $\mathcal{A}_{\alpha\dot\alpha}$ and fermionic $\mathcal{F}_{\alpha A}$ connections
\begin{align}
\vev{\mathcal{W}_n (x_i, \theta_i; a)}
=
\frac{1}{N_c}
\VEV{
\tr \, P \, \exp
\left(
\frac{i g_{\rm\scriptscriptstyle YM}}{2} \int_{\mathcal{C}_n} dx^{\dot\alpha\alpha} \mathcal{A}_{\alpha\dot\alpha}
+
i g_{\rm\scriptscriptstyle YM} \int_{\mathcal{C}_n} d \theta^{\alpha A} \mathcal{F}_{\alpha A}
\right)
}
\, ,
\end{align}
that live on a piece-wise contour in superspace $\mathcal{C}_n = [X_1, X_2]  \cup [X_2, X_3] \cup \dots \cup [X_n, X_1]$
with cusps located at superpoints $X_i = (x_i, \theta_i)$. Both superfields $\mathcal{A}_{\alpha\dot\alpha}$ and $\mathcal{F}_{\alpha A}$
receive a terminating series in the Grassmann variable $\theta$, with their expansion coefficients being determined by the field content of
the $\mathcal{N} = 4$ SYM (see, e.g., Ref.\ \cite{BelKorSok11} for details). The dynamical content of the duality is summarized by the following
equation
\begin{align}
\label{SWLSAduality}
\vev{\mathcal{W}_n (X_i; a)}
=
\widehat{A}_n (\mathcal{Z}_i; a)
\, ,
\end{align}
where the right-hand side corresponds to the reduced $n$-particle S-matrix element upon extraction of the conservation laws for the
supermomentum
$(p_{\alpha\dot\alpha}, q^A_\alpha)$
\begin{align}
S_n = i (2 \pi)^4 \frac{\delta^{(4)} (p_{\alpha\dot\alpha}) \delta^{(8)} (q^A_\alpha)}{\vev{12} \vev{23} \dots \vev{n-1n}}
\widehat{A}_n (\mathcal{Z}_i; a)
\, .
\end{align}
The kinematical variables entering the super-Wilson loop are related to the momentum supertwistors $\mathcal{Z}^{\widehat{A}}_i
= (\lambda^\alpha_i, \mu_{i \dot\alpha} ,\chi^{A}_i)$ defining the superscattering amplitude via the relation\footnote{A brief summary
of our conventions is summarized in Appendix \ref{MomentumTwistorApp}.}
\begin{align}
X_i = \mathcal{Z}_{i - i} \wedge \mathcal{Z}_i
\, .
\end{align}
In addition to the conservation-law delta functions, we traditionally factored out the denominator of the Parke-Taylor tree amplitude \cite{ParTay86}, where
the angle brackets between bosonic twistors are conventionally defined as $\vev{ij} = \lambda_i^\alpha \lambda_{j\alpha}$.  A brute force calculation of the
super-Wilson loop represents a challenge on its own even at lowest orders in coupling $a$, thus exposing inefficiency of usual techniques based of Feynman
diagrams. It was noticed however that the left-hand side of the above relation \re{SWLSAduality} is advantageous in computing the S-matrix
since the Wilson loop formalism allows one to break free from limitations of perturbation theory, being more amenable to nonperturbative techniques
\cite{AldMal08}. Such a techniques was put forward a little while ago and is based on the operator product expansion (OPE) \cite{AldGaiMalSevVie11}.
Recently, a complementary view on this framework was offered by exhibiting its connection to the expansion in terms of light-cone operators with
boundary fields set by light-like Wilson lines, or the $\Pi$-shaped Wilson loop with field insertions \cite{Bel11,SevVieWan11}. The renormalization group
evolution of the latter was shown to be governed by a noncompact open spin chain \cite{Bel11}, whose eigenenergies provide corrections to leading
discontinuities of the Wilson loop, while the corresponding eigenfunctions define the coupling of the Gubser-Klebanov-Polyakov (GKP) excitations
\cite{BelGorKor03,Bas10} to the Wilson loop contour as well as their multiparticle transition amplitudes. In this paper we explore the emerging open spin
chain. We will focus on the component formalism leaving a fully supersymmetric superspace formulation for future publication. Notice that quite recently a
framework that combines the OPE and integrability was proposed in Ref.\ \cite{BasSevVie13}. It is based on a few conjectured equations obeyed by building
blocks of the Wilson loops that are matched to explicit calculations of amplitudes and bootstrapped to all orders in 't Hooft coupling. The next section follows
a similar construction.

\subsection{Soft-collinear expansion of polygons}
\label{OPEpolygons}

To start with, let us recall the emergence of non-local light-cone operators in the OPE of the super-Wilson loop. We will discuss it in a simplified set-up of
restricted kinematics \cite{GaiMalSevVie10}, i.e., when the bosonic part of the Wilson loop contour is embedded in a two-dimensional plane\footnote{To
preserve superconformal symmetry this implies that out of four components of the Grassmann $\chi^A$ two of them have to vanish as well. However, this
subtlety is not relevant for our present discussion and  we will assume that all components of $\chi_i^A$ are nonzero at even and odd sites.}. The first
nontrivial Wilson loop in this kinematics is the octagon. Let us focus on NMHV amplitudes and choose a specific Grassmann component that admits a
clear OPE interpretation, say $\chi_2 \chi_3 \chi_6 \chi_7$. It was calculated diagrammatically in Ref.\ \cite{Bel12b} and also making use of the so-called
$\bar{Q}$-equation in Ref.\ \cite{CarHe11}. The subtracted superloop, with ultraviolet divergences eliminated in a superconformally invariant fashion, reads
to two loop accuracy
\begin{align}
\vev{\mathcal{W}_8^R}
=
\frac{\chi_2 \chi_3 \chi_5 \chi_6}{\vev{2367}}
\bigg[
a
+ a^2
\left(
\ln u^+(1+u^+) \ln u^-(1+u^-) -2 \ln (1 + u^+) \ln (1 + u^-)
\right)
+
O(a^3)
\bigg]
\, ,
\end{align}
where the conformal cross ratios are
\begin{align}
u^+ = \frac{x_{34}^+ x_{78}^+}{x_{37}^+ x_{48}^+}
\, , \qquad
u^- = \frac{x_{23}^- x_{67}^-}{x_{26}^- x_{37}^-}
\, .
\end{align}
The leading contribution in the OPE of this channel stems from the scalar exchange between the cusps at points $x_3$ and $x_4$, see Fig.\ \ref{OPEoctagon} (a).
The soft-collinear expansion corresponds to the limit $x_{32}^+ \to 0$ and $x_{67}^+ \to 0$, which results in flattening of the octagon to a square, see Appendix
\ref{RefSquare}. The subleading perturbative corrections come from a number of Feynman graphs encoding the interaction of the scalar field with the contour,
Fig.\ \ref{OPEoctagon} (b), vertex corrections and independent propagation of a gauge field along with the original scalar, Fig.\ \ref{OPEoctagon} (c).  A natural
set of the cusp coordinates, or equivalently twistors, to discuss the OPE is introduced in Appendix \ref{DecagonParametrization}. In this frame, the remainder
function admits the form
\begin{align}
\label{W8ConfFrame}
\vev{\mathcal{W}_8^R}
=
\frac{\chi_2 \chi_3 \chi_5 \chi_6}{4 \cosh\tau \cosh\sigma}
\left[
a + a^2
\left(
\tau \ln (1 + {\rm e}^{2 \sigma}) (1 + {\rm e}^{-2 \sigma})
+
\ln (1 + {\rm e}^{- 2 \tau}) (1 + {\rm e}^{2 \sigma})
\right)
+
O(a^3)
\right]
\, . \nonumber
\end{align}
The afore-introduced soft-collinear asymptotics translates into the $\tau \to \infty$ limit, with the remainder function corresponding to a single scalar GKP
excitation (also known as a hole) propagating in the OPE channel\footnote{Notice that our definition of the rapidity variable differs by a sign from the
conventions of Ref.\ \cite{BasSevVie13}.}
\begin{align}
\vev{\mathcal{W}_8^R}
\stackrel{\tau\to\infty}{=}
\chi_2 \chi_3 \chi_5 \chi_6
\int_{-\infty}^{\infty} d u \, \mu_{\rm h} (u; a)
\exp \left( - \tau E_{\rm h} (u; a) - i \sigma \, p_{\rm h} (u; a) \right) + O ({\rm e}^{- 3 \tau})
\, .
\end{align}
Here the energy and momentum of the hole excitation read to one-loop accuracy \cite{BelGorKor03,Bas10}
\begin{align}
\label{HoleEandP}
E_{\rm h} (u; a)
&= 1 - a \left[ 2 \psi(1) - \psi ( \ft12 + i u) - \psi (\ft12 - i u) \right] + O (a^2)
\, , \\
p_{\rm h} (u; a)
&= 2u - 2 \pi a \tanh (\pi u) + O (a^2) ,
\end{align}
while the measure encoding the coupling of the excitation to the Wilson loop contour is
\begin{align}
\label{HoleMeasure}
\mu_{\rm h} (u; a) = \mu_{\rm h} (u)
\left[
a + a^2 \pi^2 \left( 1 - \frac{2}{\cosh^2 (\pi u)} \right)
+
O (a^3)
\right]
\, , \qquad
\mu_{\rm h} (u)
=
\frac{\pi}{\cosh (\pi u)}
\, .
\end{align}
In this equation, the $O(a^2)$ term is merely needed to compensate the $O(a^2)$ correction to the momentum \re{HoleEandP} to match the OPE
expansion into the exact two-loop data \re{W8ConfFrame}. By expanding the right-hand side to next-to-leading order in $a$, we immediately reproduce
the $\tau$-enhanced contribution in Eq.\ \re{W8ConfFrame}.

A similar consideration can be performed for the decagon. Using the conformal frames discussed in Appendix \ref{DecagonParametrization}, we can write
the remainder function of the $\chi_2 \chi_3 \chi_8 \chi_9$ NMHV component \cite{CarHe11} of the super-Wilson loop in the form
\begin{align}
\label{W10ConfFrame}
\vev{\mathcal{W}_{10}^R}
&=
\frac{\chi_2 \chi_3 \chi_8 \chi_9}{4 \cosh(\tau_1 - \tau_2)}
\frac{1}{2 \cosh (\sigma_1 - \sigma_2) + {\rm e}^{- \sigma_1 - \sigma_2}}
\\
&\times
\bigg[
a + a^2
\bigg(
2 \tau_1 \ln \frac{(2 \cosh (\sigma_1 - \sigma_2) + {\rm e}^{- \sigma_1 - \sigma_2})^2}{1 + {\rm e}^{- 2 \sigma_2}}
+
2 \tau_2 \ln \frac{(2 \cosh (\sigma_1  - \sigma_2) + {\rm e}^{- \sigma_1 - \sigma_2})^2}{1 + {\rm e}^{- 2 \sigma_1}}
\nonumber\\
&\qquad\
+
\ln (1 + {\rm e}^{- 2 \tau_1}) \ln \frac{1 + {\rm e}^{- 2 \sigma_1}}{1 + {\rm e}^{- 2 \sigma_2}}
-
\ln (1 + {\rm e}^{- 2 \tau_2}) \ln (1 + {\rm e}^{- 2 \sigma_1})(1 + {\rm e}^{- 2 \sigma_2})
\nonumber\\
&\qquad\
+
2
\ln (1 + {\rm e}^{- 2 \tau_1 - 2 \tau_2}) \ln (1 + {\rm e}^{- 2 \sigma_2} + {\rm e}^{2 \sigma_1 - 2 \sigma_2})
\bigg)
\bigg]
\, . \nonumber
\end{align}
As above, the leading asymptotic is governed by a single hole excitation
\begin{align}
\vev{\mathcal{W}_{10}^R}
\stackrel{\tau_{1,2} \to \infty}{=}&
\chi_2 \chi_3 \chi_8 \chi_9
\int_{- \infty}^\infty du \, dv \, \mu_{\rm h} (u; a) H_{\rm h}(u | v; a) \mu_{\rm h} (v; a)
\nonumber\\
\times&
\exp \left( - \tau_1 E_{\rm h} (u; a) - \tau_2 E_{\rm h} (v; a) - i \sigma_1 \, p_{\rm h} (u; a) - i \sigma_2 \, p_{\rm h} (v; a) \right)
+ O ({\rm e}^{-3 \tau_{1,2}})
\, ,
\end{align}
where the hexagon transition kernel reads to this order in coupling
\begin{align}
H_{\rm h} (u | v; a) = H_{\rm h} (u | v)
\left[
a^{-1}
+
2 \psi^\prime (\ft12 + i u) + 2 \psi^\prime (\ft12 - i v) - \pi^2
+
O (a)
\right]
\, ,
\end{align}
with the leading term being
\begin{align}\label{hex-hole-half}
H_{\rm h} (u | v)
=
\frac{\Gamma (i v - i u)}{\Gamma (\ft12 + i v) \Gamma (\ft12 - i u)}
\, .
\end{align}
As we recall in the next section, these leading contributions in the asymptotic expansion, i.e., $O ({\rm e}^{- \tau_{1,2}})$, come from the renormalization
of a nonlocal operator with a $\Pi$-shaped contour, formed by two Wilson lines building up the contour in a given OPE channel and an elementary scalar
field insertion, while the subleading effects correspond to more than one GKP excitations sandwiched between the boundary Wilson lines.

\subsection{Light-cone operators}
\label{LCoperatorsHamiltonian}

\begin{figure}[t]
\begin{center}
\mbox{
\begin{picture}(0,170)(250,0)
\put(0,0){\insertfig{17.5}{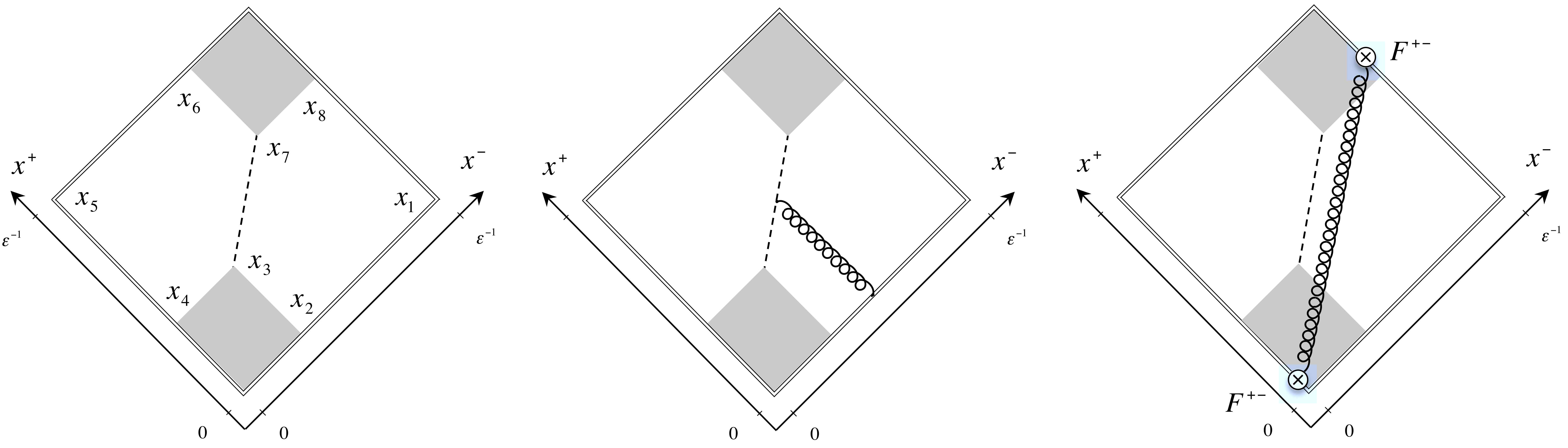}}
\put(3,0){$(a)$}
\put(173,0){$(b)$}
\put(343,0){$(c)$}
\end{picture}
}
\end{center}
\caption{ \label{OPEoctagon} Single (a,b) and two-particle (c) contributions to OPE of the octagon. The one-loop graph in (a) given
by the scalar propagator exchanged between the cusps produces a components of the tree NHMV amplitude. The graph in panel (b)
displays one of the perturbative corrections due to the Hamiltonian acting on the light-cone operator. In (c) we show the two-particle contribution
that produces subleading effects in the OPE.}
\end{figure}
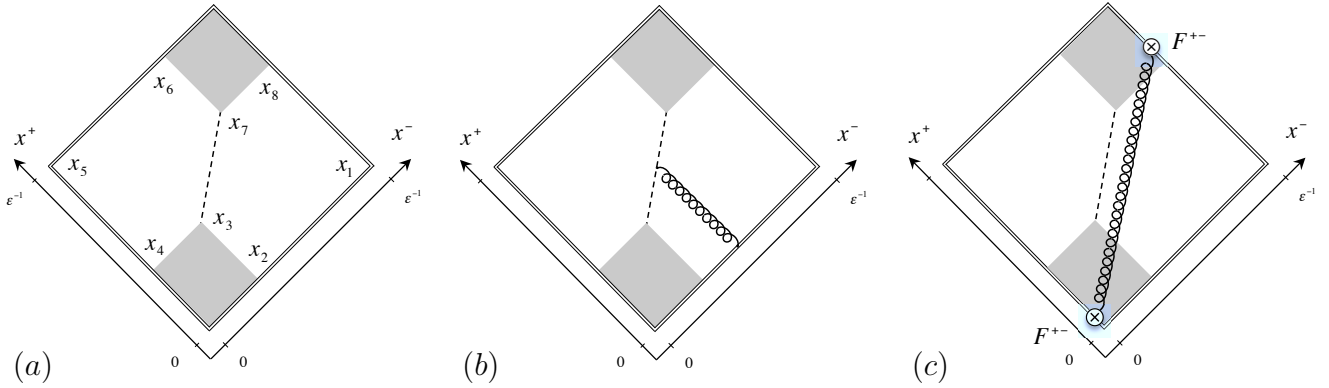

Let us cast the above heuristic discussion into an operator language. We will do it for the octagon at two-loop order and then generalize
it to an arbitrary number of GKP excitations exchanged in a given OPE channel. As one takes the limit $\tau \to \infty$, the top and bottom portions of
the contour of the octagon Wilson loop get flatten out and one ends up with a scalar field $Z$ inserted into the top and bottom straight-line segments.
Each of these correspond to a $\Pi$-shaped light-cone operator of the form
\begin{align}
\label{LCoperator}
O_{\rm h} (x) = W^\dagger (0) [0, x^+]_- Z (x^+) [x^+, \infty]_-  W (\infty)
\, ,
\end{align}
where the boundary fields $W (x^+)$ stand for the Wilson lines stretched along the light-like direction $x^-$ starting at
$0$ and going to the null infinity $1/\varepsilon \to \infty$, localized at the position $x^+$ on the tangent light-cone,
\begin{align}
W (x^+) = P \exp \left( i g_{\rm\scriptscriptstyle YM} \int_0^\infty dx^- A^+ (x^+, x^-, \bit{0}_\perp) \right)
\, .
\end{align}
By a suitable gauge choice, i.e., the light-cone gauge $A^- = 0$, one can even eliminate the gauge links between the fields in the composite operator,
i.e., $[\dots]_- \to 1$. The contribution of \re{LCoperator} to the  the leading ${\rm e}^{- \tau}$ term in the OPE reads
\begin{align}
\vev{W_8^R}
&= \frac{a \chi_2 \chi_3 \chi_6 \chi_7}{\vev{23}\vev{67}}
\vev{ O^\dagger_{\rm h} (x_7) (1 + a \tau \mathcal{H}_1) O_{\rm h} (x_3)}
\, .
\end{align}
For $a = 0$, this can be easily identified as a propagation of a free scalar from the bottom to the top of the loop, while the $O(a)$ effect induces the
$\tau$-enhanced contribution to the two-loop remainder function and arises from the interaction of the scalar field with the Wilson loop contour by
means of the light-cone Hamiltonian
\begin{align}
\mathcal{H}_1 = H_{01} + H_{1\infty}
\, ,
\end{align}
with individual Hamiltonians for the interaction with the left and right boundaries being \cite{Bel11}
\begin{align}
\label{BoundaryHamiltonians}
H_{01} O (x_1)
&= \int_0^1 \frac{d \alpha}{1 - \alpha} \left[ \alpha^{2s - 1} O (\alpha x_1)- O (x_1)  \right]
\, , \\
H_{1\infty} O (x_1)
&= \int_1^\infty \frac{d \alpha}{\alpha - 1} \left[  O(\alpha x_1) - \alpha ^{-1} O (x_1) \right]
\, ,
\end{align}
respectively. Here we presented them in a generic form suitable for studies of elementary fields of conformal spin $s$. While for the case at hand, $s = 1/2$.
For a one-particle GKP excitation, the diagonalization of the above Hamiltonian can be easily performed with a plane-wave eigenfunction
\begin{align}
\label{1Peigenfunction}
\Psi_{u_1} (x_1) = x_1^{- i u_1 - s}
\, ,
\end{align}
that carries the momentum $p(u_1) = 2 u_1$ and the energy
\begin{align}
\mathcal{H}_1 \Psi_{u_1} (x_1) = \left[ 2 \psi(1) - \psi (s + i u_1) - \psi (s - i u_1) \right] \Psi_u (x_1)
\, ,
\end{align}
cf.\ Eq.\ \re{HoleEandP}. Analogously, we can perform the soft-collinear expansion for the decagon in terms of the light-cone operators \re{LCoperator}.
The formalism can be extended to account for multiparticle effects as well. As one collapses the top and bottom of the loop in the soft limit it gives rise to
curvature corrections, actually an infinite series of them in increasing powers of the gluon field strength $F^{+-}$. The first correction to the asymptotic
behavior stems from the diagram shown in Fig.\ \ref{OPEoctagon} (c) for the octagon. At even higher orders, the OPE involves multiparticle light-cone
operators
\begin{align}
\label{MultiPartLCoperator}
O_N (x_1, \dots, x_N)
=
W^\dagger (0) [0, x_1] X_1 (x_1)  [x_1, x_2] X_2 (x_2) \dots X_N (x_N) [x_N, \infty]  W (\infty)
\, ,
\end{align}
where we stripped the $+$ superscripts off the $x$-coordinates for brevity since it is the only component that will appear in the analysis that follows.
Under renormalization group evolution, these operators mix and one has to solve the resulting eigensystem problem. In this work we will consider a
simplified setup when all GKP excitations are of the same type $X_n = X$ and possess a generic conformal spin $s$. The Hamiltonian
\begin{align}
\label{MultiParticleH}
\mathcal{H}_N = H_{01} + \sum_{n=1}^{N-1} H_{n,n+1} + H_{N\infty}
\, ,
\end{align}
built-up from nearest-neighbor interactions only, in planar limit, is the one-loop dilatation operator acting on the operator \re{MultiPartLCoperator}
\begin{align}
\frac{d}{d \ln \mu} O_N  (x_1, \dots, x_N) = - a \mathcal{H}_N O_N  (x_1, \dots, x_N)
\, .
\end{align}
Here the pair-wise Hamiltonians (including the boundary ones) can be encoded in a single formula
\begin{align}
\label{LCpairwiseHamiltonian}
&
H_{n,n+1} O_N (\dots, x_n, x_{n+1}, \dots)
\\
&=
\int_{x_n/x_{n+1}}^1 \ \frac{d \alpha }{1 - \alpha}
\left[
\left( \frac{\alpha x_{n+1} - x_n}{x_{n+1} - x_n} \right)^{2 s - 1} O_N (\dots, x_n, \alpha x_{n+1}, \dots) - O_N (\dots, x_n, x_{n+1}, \dots)
\right]
\nonumber\\
&+
\int_1^{x_{n+1}/x_n} \frac{d \alpha }{\alpha - 1}
\left[
\left( \frac{x_{n+1} - \alpha x_n}{x_{n+1} - x_n} \right)^{2 s - 1} O_N (\dots, \alpha x_n, x_{n+1}, \dots) - \frac{1}{\alpha} O_N (\dots, x_n, x_{n+1}, \dots)
\right]
\, . \nonumber
\end{align}
Their complementary representation in terms of differential operators reads
\begin{align}
H_{01}
&
= \psi (1) - \psi (x_1 \partial_{1} + 2 s)
\, , \nonumber\\
H_{n,n+1}
&
= 2 \psi (1) - \psi (x_{n,n+1} \partial_{n} + 2 s)  - \psi (x_{n+1,n} \partial_{n+1}  + 2 s) - \ln x_n/x_{n+1}
\, , \\
H_{N\infty}
&
= \psi (1) - \psi (- x_N \partial_{N})
\, , \nonumber
\end{align}
where $\psi (x) = d \ln \Gamma (x)/dx$ is the Euler digamma function. It is important to notice that this is not the form naturally coming from the
computation of Feynman graphs contributing to the renormalization of \re{MultiPartLCoperator}. Namely, the one-loop light-cone Hamiltonians for
interaction of GKP excitations among themselves are not sensitive to the boundary Wilson lines and thus have to enjoy full $SL(2, \mathbb{R})$
invariance \cite{BraKorMul03,BelRad04}. However, it is not the case in the above form due to the presence of the logarithms. The latter emerge
from $1/\alpha$-factor in the second integral in Eq.\ \re{LCpairwiseHamiltonian} that was introduced in order to match it to the boundary Hamiltonians
\re{BoundaryHamiltonians}, with all intermediate logarithms canceling telescopically with one another. Therefore, in order to restore the conformal
symmetry property, one can shift the logarithms into the boundary terms, such that the reshuffled Hamiltonians will become
\begin{align}
\label{BoundaryAndIntermediateH}
h_{01}
&
= \psi (1) - \psi (x_1 \partial_{1} + 2 s) - \ln x_1 = - \ln (x_1^2 \partial_1 + 2 s x_1)
\, , \nonumber\\
h_{n,n+1}
&
= 2 \psi (1) - \psi (x_{n,n+1} \partial_{n} + 2 s)  - \psi (x_{n+1,n} \partial_{n+1}  + 2 s)
\, , \\
h_{N\infty}
&
= \psi (1) - \psi (- x_N \partial_{N}) + \ln x_N = - \ln \partial_N
\, , \nonumber
\end{align}
with $\mathcal{H}_N = h_{01} + \sum_{n = 1}^{N-1} h_{n,n+1} + h_{N\infty}$.
Here we used identity~(\ref{magic}) from Appendix~\ref{FeynmanAppendix} to recast
the boundary Hamiltonians in terms of logarithms.

At this point we would like to point out that similar Hamiltonians emerged in the discussion of the
Regge limit of scattering amplitudes in Ref.\ \cite{Lip09}.

Thus, the goal of this paper is to solve the spectral problem for the multiparticle Hamiltonian \re{MultiParticleH}
\begin{align}
\label{SpectralProblem}
\mathcal{H}_N \Psi_{\bit{\scriptstyle u}}  (x_1, \dots, x_N) = E_{\bit{\scriptstyle u}} \Psi_{\bit{\scriptstyle u}} (x_1, \dots, x_N)
\, .
\end{align}
The problem turns out to be integrable since the Hamiltonian $\mathcal{H}_N$ possesses enough integrals of motion with eigenvalues
$\bit{u} = (u_1, \dots, u_N)$ to be exactly solvable. Thus we have the powerful machinery of integrable spin chain at our disposal for constructing
the explicit eigenfunctions and finding the corresponding eigenvalues. For the case at hand we are dealing with an open spin chain that does
not possess $SL(2,\mathbb{R})$ symmetry due to boundary interactions. In addition, as we can conclude from the one-particle eigenfunction
\re{1Peigenfunction}, the system is not endowed with a vacuum states. As a consequence of this last fact, the traditional methods based on the
Algebraic Bethe Ansatz \cite{TakFad79,Fad96} are not applicable and instead one has to rely on the method of the Baxter $\mathbb{Q}$-operator
\cite{Bax82} and the Separation of Variables (SoV) \cite{Skl85}.

Our subsequent consideration is organized as follows. After introducing a natural Hilbert space for the Hamiltonian and a scalar product that
makes it self-adjoint in the next section, we find a complete set of integrals motion commuting with $\mathcal{H}_N$. As a next step on the way to
solve the spectral problem, we recall the factorization of $R$-matrices  \cite{TakFad79,Fad96} into intertwiners that play a crucial role in our construction.
In Section \ref{HamiltonianSection}, we use them to construct the Hamiltonians and demonstrate their equivalence with the ones arising in the problem of
renormalization of the light-cone operators \re{MultiPartLCoperator}. Then in Section \ref{BaxterSection}, we construct the Baxter $\mathbb{Q}$-operator
as well as its conjugate making use of available techniques developed for $SL(2)$ invariant magnets. Both operators are then used to generate commuting
Hamiltonians. In Section \ref{EigenFunctionSection}, we devise an iterative procedure to construct the eigenfunctions of the Hamiltonian. We then cast the
latter in a form of a contour integral representation as well as an integral representation in the upper half-plane. The latter is instrumental in demonstration of
their orthogonality under the scalar product introduced earlier. To adopt our construction for multiparticle contributions to the null polygonal Wilson loops, we
define a new scalar product on the real half-line in Section \ref{ScalarLine} and establish its relation to the one for the analytically continued functions in the
upper half-plane. Then we explicitly compute the square and hexagon transitions in Section \ref{SquareAndHexagon}, where we also prove the factorized
ansatz for the latter suggested in Ref.\ \cite{BasSevVie13}. Finally, we conclude by pointing applications of the current construction to the operator product
expansion of polygonal Wilson loops and outline the generalization to include GKP excitations of different spins. Several appendices contain conventions
and computational details that we found inappropriate to include in the main body of the paper as well as a complementary construction the wave function
in the representation of Separated Variables and its relation to the eigenvalues of the Baxter $\mathbb{Q}$-operator.

\section{Open spin chain}

The Hamiltonian \re{MultiParticleH} defines a non-periodic one-dimensional lattice model of interacting spins $\bit{S}_n = (S^0_n, S^+_n, S^-_n)$ acting at
each site defined by the coordinate of the GKP excitation. They form an infinite-dimensional representation of the $sl(2,\mathbb{R})$ algebra,
\begin{align}
[S_n^+, S_n^-] = 2 S^0_n \, , \qquad [S_n^0, S_n^\pm] = \pm S_n^\pm
\, .
\end{align}
Our choice of the representation $S_n^{\pm,0}$ for the spin generators $\mathbb{S}^{\pm,0}$ on the fields $X$ is driven by the form of the pairwise
Hamiltonians \re{BoundaryAndIntermediateH} acting on the elementary fields $X (x)$, i.e. $\,[\mathbb{S}^{\pm, 0}, X (x_n)] = i S^{\pm, 0}_n X(x_n)$
\cite{BraKorMul03,BelRad04}. They are taken in the form of differential operators acting on lattice sites
\begin{align}
\label{SpinEGenerators}
S^+_n = x_n^2 \partial_n + 2 s x_n
\, , \qquad
S^-_n = - \partial_n
\, , \qquad
S^0_n = x_n \partial_n + s
\, ,
\end{align}
where all labels of the representation $s$ are the same for any $n$. The latter condition defines a homogeneous open spin chain. The spin variable $s$ is
assumed to be real and bounded from below by $1/2$.

\subsection{Scalar product}

It is convenient to define a scalar product on the space of functions $\Psi_{\bit{\scriptstyle u}} (x_1, \dots,
x_N)$. Its choice is a matter of convenience but it has to be adapted to the physical problem under consideration.
As was explained in Section \ref{sect:motivation}, we are interested in the $\tau$-dependence of correlation
functions of the light-cone operators $O_N$ which read schematically
\begin{align}
\vev{0|O^\prime_N(x^\prime_1, \dots, x^\prime_N) {\rm e}^{ a \tau \mathcal{H}_N } O_N(x_1, \dots, x_N)|0}
=
\sum_{{\bit{\scriptstyle u}}} {\rm e}^{a \tau E_{\bit{\scriptscriptstyle u}}}
\left( \Psi^{\prime}_{\bit{\scriptstyle u}} (x^\prime_1, \dots, x^\prime_N) \right)^\dagger \Psi_{\bit{\scriptstyle u}} (x_1, \dots, x_N)\, .
\end{align}
Here the functions $\Psi_{\bit{\scriptstyle u}}$ are determined by the operator matrix elements between the  eigenstate $\ket{E_{\bit{\scriptstyle u}}}$
of Hamiltonian and the vacuum\footnote{In QCD terminology $\Psi_{\bit{\scriptstyle u}} (x_1, \dots, x_N)$ is known as the distribution amplitude.},
$\Psi_{\bit{\scriptstyle u}} (x_1, \dots, x_N)= \vev{E_{\bit{\scriptstyle u}}| O_N(x_1, \dots, x_N )|0}$, and the sum runs over all eigenstates.
Although the variables  $x_n$ are the light-cone coordinates of the fields $X$ entering the Lagrangian of the theory and are, therefore, real,
nevertheless distribution amplitudes $\Psi_{\bit{\scriptstyle u}} (z_1, \dots, z_N)$ can be regarded as functions of $z_n$ in complex plane. It is known
that they are analytic functions in upper half-plane and vanish at infinity. As a consequence of this change, the $sl (2, \mathbb{R})$ generators in
Eq.\ \re{SpinEGenerators} depend and act on the $z_n$ variables accordingly.

In what follows we will consider the spectral problem~(\ref{SpectralProblem}) on the space of functions of $N$
complex variables $z_n$ analytic in the upper half-plane and endowed with a standard $SL(2,\mathbb{R})$ invariant
scalar product~\cite{GelGraVil66}. Such  a formulation appeared to be very useful for constructing the SoV
representation for the closed and open $SL(2,\mathbb{R})$ spin chains~\cite{DerKorMan02,DerKorMan03}. The scalar
product on the space of functions holomorphic in the upper  half-plane\footnote{The Hilbert spaces of holomorphic
functions  are well studied mathematical topic, for a review, see Ref.~\cite{BHall}.} is defined  as follows
\cite{GelGraVil66}
\begin{align}\label{scalar-product}
\vev{\Phi | \Psi} = \int D z_n \, \left({\Phi} (z_n)\right)^*\, \Psi (z_n)
\,,
\end{align}
where $z_n = x_n + i y_n$. The integration measure reads
\begin{align}
\label{MeasureSL2}
D z_n = \frac{2s - 1}{\pi}\, {dx_n d {y}_n}\, ({2y_n)^{2 s-2}} \,\theta (y_n)
\,
\end{align}
and the integration runs over the upper half-plane due to the presence of the step-function $\theta (y_n)$.
Under this scalar product the generators
\re{SpinEGenerators} are anti-self-adjoint
\begin{align}
\label{AntiHermitS}
\left( S_n^{0, \pm} \right)^\dagger = - S_n^{0, \pm}
\, ,
\end{align}
with hermitian conjugation defined conventionally as
\begin{align}
\vev{ \Phi | G \Psi}  = \vev{ G^\dagger \Phi | \Psi}
\, .
\end{align}

The  Hilbert space of the model  is given by the  tensor product of Hilbert spaces at each site $\otimes^N_{n = 1} \mathbb{V}_n$, such that the
generalization of the scalar product \re{scalar-product} to multivariable functions is straightforward,
\begin{align}
\label{NScalarProduct}
\VEV{\Phi|\Psi}=\int \prod_{k=1}^N D z_k\, (\Phi(z_1,\ldots,z_N))^*\, \Psi(z_1,\ldots,z_N)\,.
\end{align}
One can immediately verify that the Hamiltonian $\mathcal{H}_N$ is a hermitian operator
\begin{align}
\mathcal{H}_N^\dagger = \mathcal{H}_N
\end{align}
by virtue of Eq.\ \re{AntiHermitS}. Notice that this is obvious for the boundary Hamiltonians \re{BoundaryAndIntermediateH}
that are merely functions
of the spin $S^\pm$ generators acting on the right/left-most sites. These Hamiltonians are not $SL (2, \mathbb{R})$ invariant.
The intermediate
Hamiltonians can be cast in an $SL(2, \mathbb{R})$ symmetric form \cite{KulResSkl81} and read
\begin{align}
h_{n,n+1} = 2 \psi (1) - 2 \psi (J_{n,n+1})
\, .
\end{align}
Here $J_{n,n+1}$ is an operator related to the two-particle Casimir via the formula
\begin{align}
J_{n,n+1}(J_{n,n+1}  -1) = (\bit{S}_n + \bit{S}_{n+1})^2
\, .
\end{align}
These are explicitly self-adjoint. In the following section, we show that there exists a complete set of integrals of motion (commuting with the Hamiltonian)
that are hermitian with respect to this scalar product and thus their eigenstates form an orthogonal set.

Closing this section we remark that to any operator $A$ acting on the Hilbert space \re{NScalarProduct} we can associate a function $\mathcal{A}$
of $N$ holomorphic and $N$ anti-holomorphic variables, i.e., the so-called integral kernel, in a unique way via the relation
\begin{align}\label{integral-kernel}
[A \Psi](z_1,\ldots,z_N)=\int\prod_{k=1}^N D w_k\, \mathcal{A} (z_1,\ldots, z_N|\bar w_1,\ldots,\bar w_N) \Psi(w_1,\ldots,w_N)\,.
\end{align}
It turns out that integral kernels of operators which we construct in the following sections take a form of
two-dimensional Feynman diagrams. All operator identities can be cast into the identities between Feynman diagrams
which can be verified with the help of a standard diagram technique that we remind in Appendix
\ref{FeynmanAppendix}.

\subsection{Integrals of motion}

We start the construction of the integrals of motion following the conventional $R$-matrix approach \cite{TakFad79,Fad96}.
The Lax operator acting
on the direct product $\mathbb{C}^2 \otimes \mathbb{V}_n$ of an auxiliary two-dimensional space $\mathbb{C}^2$ and
the quantum space at the
$n$-th site $\mathbb{V}_n$ is defined as
\begin{align}
\label{Lax}
\mathbb{L}_n (u, s)
=
u + i (\bit{\sigma} \cdot \bit{S}_n)
=
\left(
\begin{array}{cc}
u + i S^0_n & i S^-_n \\
i S^+_n        & u - i S_n^0
\end{array}
\right)
\, ,
\end{align}
with a complex spectral parameter $u$. We also displayed its dependence on the spin parameter $s$. The product of
$N$ copies of this operator in the auxiliary space determines the monodromy matrix $\mathbb{T}^{(s)} (u)$,
\begin{align}
\label{Monodromy}
\mathbb{T}^{(s)}_N (u)
=
\mathbb{L}_1 (u, s) \dots \mathbb{L}_N (u, s)
=
\left(
\begin{array}{cc}
A_N(u) & B_N(u) \\
C_N(u) & D_N(u)
\end{array}
\right)
\end{align}
with its elements acting on the quantum space of the chain $\otimes_{n = 1}^N \mathbb{V}_n$. As can be easily
established from the Yang-Baxter equation with a rational $R$-matrix \cite{TakFad79,Fad96}, each element of the
monodromy matrix commutes with itself for arbitrary spectral parameters, i.e., $[D_N (u), D_N (v)] = 0$ etc. As
we will demonstrate below, the operator $D_N (u)$ commutes also with the Hamiltonian \re{MultiParticleH}
\begin{align}
[ D_N (u), \mathcal{H}_N] = 0
\, .
\end{align}
It can be easily seen from the form of the Lax operator that the entries of the monodromy matrix are polynomials of
degree $N$ in the spectral parameter $u$ with operator valued coefficients. In particular,
\begin{align}
\label{DNelement}
D_N (u) = u^N + \widehat{d}_1 u^{N - 1} + \dots + \widehat{d}_N\,.
\end{align}
The expansion coefficients (integrals of motions)  $\widehat{d}_k$, $k=1,\ldots,N$ commute with each other and with
the Hamiltonian, $[\widehat{d}_k,\widehat{d}_n]=[\widehat{d}_k,\mathcal{H}_N]=0$.
Their eigenvalues define a complete set of quantum numbers of the eigenstate.
Since the eigenvalues of the operator $D_N(u)$ are polynomials in $u$, its
eigenstates can be labelled by the zeroes of the corresponding polynomial,
$\bit{u} = (u_1, \dots, u_N)$
\begin{align}
\label{AuxProblem}
D_N (u)  \Psi_{\bit{\scriptstyle u}}  (z_1, \dots, z_N) = \prod_{n=1}^N (u - u_n)  \Psi_{\bit{\scriptstyle u}}  (z_1, \dots, z_N)
=\prod_{n=1}^N (u - \widehat u_n)  \Psi_{\bit{\scriptstyle u}}  (z_1, \dots, z_N)
\, .
\end{align}
Here the operator $\widehat{u}_n$ are the so-called  Sklyanin's operator zeroes  \cite{Skl85}, defined by the relation
$\widehat{u}_k\Psi_{\bit{\scriptstyle u}}=u_k\Psi_{\bit{\scriptstyle u}}$.

As follows from Eq.\ \re{Lax}, the integrals of motion $\widehat{d}_n$ are polynomials in the spin generators $\bit{S}_n$ ($n = 1 ,\dots, N$),
for instance,
\begin{align}
\widehat{d}_1 = - i (S_1^0 + \dots + S_N^0)
\, ,
\end{align}
with the rest $\widehat{d}_n$ being given by $n$-th order differential operators. Thus, Eq.\ \re{AuxProblem} defines a system of $N$ differential
equations that once solved yields the eigenfunctions of the open spin chain. By virtue of the property \re{AntiHermitS}, the operator $D_N$ is
self-conjugate (for real $u$) and thus $\widehat{d}_n$ (and as a consequence $\widehat{u}_n$) are hermitian as well. This implies that all
eigenvalues $u_n$ are real.

\section{Factorized $R$-matrices and Hamiltonians}
\label{HamiltonianSection}

Let us now demonstrate that local Hamiltonians, arising from the $R$-matrices acting on the direct product of two copies of the quantum
space $\mathbb{V} \otimes \mathbb{V}$, indeed coincide with the one coming from diagrammatic calculations alluded to in Section
\ref{LCoperatorsHamiltonian}.

\subsection{Intertwiners}

To start with let us recall relevant facts about the aforementioned quantum $R$-matrices.
Let us represent the  $R-$matrix in the form ${\mathcal{R}}_{12} =\Pi_{12} \check{\mathcal{R}}_{12}$,
where $\Pi_{12}$ is a permutation operator on the product of two spaces, $\Pi_{12} \Psi (z_1, z_2) = \Psi (z_2,
z_1)$. For the operator $\check{\mathcal{R}}_{12}$ the conventional $RLL-$relation \cite{TakFad79,Fad96} takes the following form
\begin{align}\label{Rcheck}
\check{\mathcal{R}}_{12} (u - v)\, \mathbb{L}_1 (u,s_1) \mathbb{L}_2 (v,s_2)
=
\mathbb{L}_1 (v,s_2) \mathbb{L}_2 (u,s_1)\, \check{\mathcal{R}}_{12} (u-v)
\,.
\end{align}
Here $\mathbb{L}_{1(2)}$ are the differential operators in $z_1(z_2)$, according to their expression (\ref{Lax}) continued in the
upper half-plane. In the above equation, we have clearly displayed all parameters which these operators depend on (notice that we
will not assume in this section that $s_1=s_2$). Equation~(\ref{Rcheck}) shows that the operator $\check{\mathcal{R}}_{12}$ interchanges the
parameters of the Lax operators acting on the first and the second spaces, $(u,s_1)\leftrightarrow(v,s_2)$.

Making use of the explicit form of the generators it is possible to demonstrate that the Lax operator can be factorized
into a product of triangular matrices
\begin{equation}\label{LaxFactorization}
\mathbb{L}_n (u,s_n)
=
\left(
\begin{array}{cc}
1 & 0 \\
z_n & 1 \\
\end{array}
\right)
\left(
\begin{array}{cc}
u+is_n-i & -i\partial_n \\
0 & u-is_n \\
\end{array}
\right)
\left(
\begin{array}{cc}
1 & 0 \\
-z_n & 1 \\
\end{array}
\right)\, ,
\end{equation}
As it is obvious from this representation,
it is convenient to introduce the following combinations of the spectral parameters $u$, $v$ and the spins $s_1$, $s_2$
\begin{align}
\label{uplusminus}
u_\pm = u \pm i s_1 \ \ ;\ \ v_\pm = v \pm i s_2
\, ,
\end{align}
such that $\mathbb{L}_1 (u,s_1) \equiv \mathbb{L}_1 (u_+, u_-)$ and $\mathbb{L}_2 (v,s_2) \equiv \mathbb{L}_2 (v_+, v_-)$. As Eq.\ \re{Rcheck}
suggests, the operator $\check{\mathcal{R}}$ interchanges simultaneously $u^\pm$ with $v^\pm$.

A fruitful approach to solving Eq.~(\ref{Rcheck}) was pioneered in Ref.\ \cite{Der05}.
It was suggested there to look for the solution of  Eq.~(\ref{Rcheck}) in the form of a product of two operators
\begin{align}
\check{\mathcal{R}}_{12} (u - v) = \mathcal{R}^{+}_{12} (u_+ |v_+, u_-) \mathcal{R}^{-}_{12} (u_+, u_-| v_-)
\, ,
\end{align}
that exchange corresponding spectral parameters \re{uplusminus} independently, i.e.,
\begin{subequations}
\begin{align}
\label{RLLplus}
&
\mathcal{R}^{+}_{12} (u_+ |v_+, v_-)\,
\mathbb{L}_1 (u_+, u_-) \mathbb{L}_2 (v_+, v_-)
=
\mathbb{L}_1 (v_+, u_-) \mathbb{L}_2 (u_+, v_-)\,
\mathcal{R}^{+}_{12} (u_+ |v_+, v_-)
\, , \\[2mm]
\label{RLLminus}
&
\mathcal{R}^{-}_{12} (u_+, u_- | v_-)\,
\mathbb{L}_1 (u_+, u_-) \mathbb{L}_2 (v_+, v_-)
=
\mathbb{L}_1 (u_+, v_-) \mathbb{L}_2 (v_+, u_-)\,
\mathcal{R}^{-}_{12} (u_+ , u_- | v_-)
\, .
\end{align}
\end{subequations}
These operators change the spin of the $sl(2, \mathbb{R})$ representations and map the Hilbert spaces of the spin chain sites as follows
\begin{subequations}
\begin{align}
&
\mathcal{R}^{+}_{12} (u_+ |v_+, v_-) :
\mathbb{V}_{s_1} \otimes \mathbb{V}_{s_2} \to \mathbb{V}_{s_1 - i(v_+ - u_+)/2} \otimes \mathbb{V}_{s_2 + i (v_+ - u_+)/2}
\,, \\[2mm]
&
\mathcal{R}^{-}_{12} (u_+ , u_- | v_-) :
\mathbb{V}_{s_1} \otimes \mathbb{V}_{s_2} \to
\mathbb{V}_{s_1 + i (v_- - u_-)/2} \otimes
\mathbb{V}_{s_2 - i (v_- - u_-)/2}
\,,
\end{align}
\end{subequations}
where we explicitly displayed the parameters labelling the representation. The solution for intertwiners was found in Ref.\ \cite{Der05} and they
read
\begin{subequations}
\begin{align}
\label{Rplus}
&
\mathcal{R}^{+}_{12} (u_+ |v_+, v_-)
=
\frac{\Gamma (z_{21} \partial_2  + i v_- - i u_+)}{\Gamma (z_{21} \partial_2 + i v_- - i v_+)}
\frac{\Gamma(i v_- - i v_+)}{\Gamma( i v_- - i u_+)}
\, , \\[2mm]
\label{Rminus}
&
\mathcal{R}^{-}_{12} (u_+ , u_- | v_-)
=
\frac{\Gamma (z_{12} \partial_1 + i v_- - i u_+ )}{\Gamma (z_{12} \partial_1 + i  u_- - i u_+)}
\frac{\Gamma(i u_- - i u_+)}{\Gamma( i v_- - i u_+)}
\, ,
\end{align}
\end{subequations}
where $z_{12} = z_1 - z_2$.
These can also be easily cast in the integral form making use of the definition of the Euler Beta function.
Note also that the operators $\mathcal{R}_{12}^\pm$ depend only on the difference of spectral parameters, i.e.
\begin{align}\label{shift-invariance}
\mathcal{R}^{+}_{12} (u_+ |v_+, v_-)=\mathcal{R}^{+}_{12} (0 |v_+-u_+, v_--u_+)\,,\notag\\
\mathcal{R}^{-}_{12} (u_+ , u_- | v_-)=\mathcal{R}^{-}_{12} (u_+ -v_-, u_--v_- | 0)\,.
\end{align}

\subsection{Hamiltonians}

The local $SL(2, \mathbb{R})$ invariant Hamiltonians  arise as coefficients of the leading power in the expansion of  the $R$-matrix in Taylor series
in the vicinity of $u - v = \varepsilon \to 0$ \cite{KulResSkl81}. The representation in terms of the Euler Gamma functions \re{Rplus} and \re{Rminus}
then yields,
\begin{align}
&
\mathcal{R}^{+}_{12} (u_+ |v_+, v_-)  = 1 - i \varepsilon h_{12}^+ + O (\varepsilon^2)
\, , \\
&
\mathcal{R}^{-}_{12} (u_+ , u_- | v_-)  = 1 - i \varepsilon h_{12}^- + O (\varepsilon^2)
\, ,
\end{align}
where
\begin{align}
h_{12}^+ = \psi(2s)- \psi (z_{21} \partial_2 + 2s)
\, , \qquad
h_{12}^- = \psi(2s)- \psi (z_{12} \partial_1 + 2s)
\, .
\end{align}
Comparing their sum $h_{12} = h_{12}^+ + h_{12}^- + 2 (\psi (1)-\psi(2s))$ with the intermediate Hamiltonian \re{BoundaryAndIntermediateH}, we immediately find
that they coincide up to an additive constant. The commutation relation of these Hamiltonians with the monodromy matrix \re{Monodromy}
can easily be established from the Yang-Baxter equations \re{RLLplus} and \re{RLLminus} for the intertwining operators by expanding both
sides around the point $v = u - \varepsilon$ as $\varepsilon \to 0$. We get for $O(\varepsilon)$ terms
\begin{align}
\label{LocalCommhLL}
&
[h^\pm_{12},  \mathbb{L}_1 (u, s) \mathbb{L}_2 (u, s)] =  i \mathbb{M}^\pm_1 \mathbb{L}_2 (u, s) - i \mathbb{L}_1 (u, s) \mathbb{M}^\pm_2
\, ,
\end{align}
where
\begin{align}\label{Mpm}
\mathbb{M}^+_n
=
\left(
\begin{array}{cc}
1      & 0 \\
z_n & 0
\end{array}
\right)
\, , \qquad
\mathbb{M}^-_n =
\left(
\begin{array}{cc}
0       & 0 \\
- z_n & 1
\end{array}
\right)
\, .
\end{align}
For $N$-site Hamiltonians, the commutation relations with the monodromy matrix can be established from the above equations \re{LocalCommhLL} to be
\begin{align}
\left[ \sum\nolimits_{n = 1}^{N - 1} h^\pm_{n,n+1}, \mathbb{T}^{(s)}_N (u) \right]
=
i \mathbb{M}^\pm_1 \mathbb{L}_2 (u, s) \dots \mathbb{L}_{N} (u, s)
-
i \mathbb{L}_1 (u, s) \dots \mathbb{L}_{N-1} (u, s) \mathbb{M}^\pm_n
\, .
\end{align}
There are leftover boundary terms in the above equations which have to be cancelled against some boundary Hamiltonians $h_{01}$ and
$h_{N \infty}$ to enforce the commutativity. We are going to find their form next. To project out the $D_N (u)$ entry from the monodromy matrix, one
sandwiches the above commutation relations between the two-dimensional vector $\ket\downarrow = (0, 1)^T$ and its transposed $\bra\downarrow = (0, 1)$.
These equations in turn can be reduced to a set of local equations for the boundary Hamiltonians, namely,
\begin{align}
\bra\downarrow\, [h^\pm_{01}, \mathbb{L}_1 (u, s)] = - i
\bra\downarrow\, \mathbb{M}^\pm_1
\, , \qquad
[h^\pm_{N\infty}, \mathbb{L}_N (u, s)]
\ket\downarrow = i \mathbb{M}^\pm_N\,\ket\downarrow
\, .
\end{align}
They receive the following solutions
\begin{align}
h^+_{01} = - \psi (z_1 \partial_1 + 2 s)
\, , \qquad
h^-_{01} =  - \ln z_1
\, , \qquad
h^+_{N \infty} = 0
\, , \qquad
h^-_{N \infty} = - \ln \partial_N
\, .
\end{align}
Summing these expressions up, we recognize the boundary Hamiltonians \re{BoundaryAndIntermediateH} that emerged from the gauge theory
analysis (again up to an additive constant). Thus, finding the eigenfunctions of the Hamiltonian is equivalent to the diagonalization of the operator
$D_N (u)$, which we address in Section \ref{EigenFunctionSection} below.

\section{Baxter operators and Baxter equations}
\label{BaxterSection}

Having in mind a goal of the present study to construct the eigenfunctions for the spin chain in SoV \cite{Skl85}, we find in this section the
Baxter operators $\mathbb{Q}^\pm$. The latter form a conjugate pair with respect to the scalar product \re{NScalarProduct}. We start our consideration
by establishing their explicit expression as operators commuting with the element $D_N$ of the monodromy matrix.  Then we devise finite difference
equations that they obey and, finally, we find the local Hamiltonians stemming from them.

\subsection{Baxter operator $\mathbb{Q}^+$}

To start with, let us derive a defining relation that will play a distinguished role in the analyses which follow.
Recalling that the intertwiner
$\mathcal{R}^+$ exchanges $u_+$ and $v_+$ spectral parameters,
we can immediately find that for a product of Lax operators. Namely,  the string of the
$\mathcal{R}^+$-operators~\footnote{Here we take into account that the intertwiners depend only on the difference of
the spectral parameters, see Eq.~(\ref{shift-invariance})}
\begin{align}
\label{Splus}
\mathcal{S}_N^+ (u_+-v_+ , u_--v_+)
\equiv
\mathcal{R}^{+}_{N-1, N} (v_+| u_+, u_-)
\mathcal{R}^{+}_{N-2, N-1} (v_+| u_+, u_-)
\dots
\mathcal{R}^{+}_{12} (v_+| u_+, u_-)
\,
\end{align}
moves the spectral parameter $v_+$ from left to right in the product of the Lax operators
\begin{align}
&
\mathcal{S}_N^+ (u_+-v_+, u_--v_+)\,
\mathbb{L}_1 (v_+, u_-)
\mathbb{L}_2 (u_+, u_-)
\dots
\mathbb{L}_N (u_+, u_-)
\nonumber\\
&
=
\mathbb{L}_1 (u_+, u_-)
\mathbb{L}_2 (u_+, u_-)
\dots
\mathbb{L}_N (v_+, u_-)
\,\mathcal{S}_N^+ (u_+-v_+, u_--v_+)
\, .
\end{align}
This can easily be proved by using the defining relation \re{RLLplus} $N$ times. Notice that we cannot immediately identify this product of the Lax operators
with the monodromy matrix $\mathbb{T}_N$ since the first and last Lax operator on the left- and right-hand side, respectively, depend on an alien spectral
parameter $v_+$ rather than $u_+$. However, the correct dependence can be recovered owing to the (right) factorization property of $\mathbb{L}$
\begin{align}
\mathbb{L}_N (v_+, u_-) = \mathbb{L}_N (u_+, u_-) \mathbb{F}_N (u_+|v_+)
\, , \qquad
\mathbb{F}_N (u_+|v_+)
\equiv
 \left(
\begin{array}{cc}
\frac{v_+ - i}{u_+ - i} & 0 \\
z_N \frac{v_+ - u_+}{u_+ - i}  & 1
\end{array}
\right)
\, .
\end{align}
Then we find
\begin{align}
\label{DefRel}
\mathcal{S}_N^+ (u_+-v_+, u_--v_+)\,
\mathbb{L}_1 (v_+, u_-) \mathbb{T}_{N-1} (u)
=
\mathbb{T}_N (u) \mathbb{F}_N (u_+|v_+)\, \mathcal{S}_N^+ (u_+-v_+, u_--v_+)
\, ,
\end{align}
where the $N-1$-site monodromy is $\mathbb{T}_{N-1} (u) =
\mathbb{L}_2 (u_+, u_-) \dots \mathbb{L}_N (u_+, u_-)$.
Next we notice that the intertwiner
\begin{align}
\mathcal{R}^+_1 (v_+| u_+, u_-) \equiv
\mathcal{R}^+_{01}  (v_+| u_+, u_-) |_{z_0 = 0}
=
\frac{\Gamma (z_1 \partial_1 + i u_- - i v_+)}
{\Gamma (z_1 \partial_1 + i u_- - i u_+)}\frac{\Gamma(i u_- - i u_+)}{\Gamma( i u_- - i v_+)}
\end{align}
restores the correct spectral-parameter dependence in the $21$- and $22$-elements of the Lax operator on the left-hand side,
\begin{align}
\mathbb{L}_1 (v_+, u_-)
=
\mathcal{R}^+_1 (v_+| u_+, u_-)
\mathbb{N}_1^+ (u_+, u_-; v_+)\,
[\mathcal{R}^+_1 (v_+| u_+, u_-) ]^{-1}
\, ,
\end{align}
where the second-row entries of the matrix  operator $\mathbb{N}_1^+$ are the same as in the Lax operator $\mathbb{L}_1 (u_+, u_-)$, i.e.,
$\bra\downarrow\mathbb{N}_1^+ = \bra\downarrow\mathbb{L}_1$,
\begin{align}
\mathbb{N}_1^+ (u_+, u_-; v_+)
=
\left(
\begin{array}{cc}
v_+ + i z_1 \partial_1
&
- i \frac{i z_1 \partial_1 + v_+ - u_-}
{i z_1 \partial_1 + u_+ - u_-} \partial_1
\\
i z_1^2 \partial_1 + (u_+ - u_-) z_1
&
u_- - i z_1 \partial_1
\end{array}
\right)
\, .
\end{align}
Substituting this relation into the left-hand side of Eq.\ \re{DefRel} and freely dragging $(\mathcal{R}^+_1)^{-1}$ to the right of $\mathbb{T}_{N-1}$,
since the latter does not depend on the first lattice site, we project on the 22 entry of the resulting matrix equation with the two-vector
$ \bra\downarrow$, to find after multiplication by $\mathcal{R}^+_1$ from the right
\begin{align}
&
\mathcal{S}_N^+ (u_+-v_+, u_--v_+) \mathcal{R}^+_1 (v_+| u_+, u_-) D_N (u)
\nonumber\\
&
=
D_N (u) \mathcal{S}_N^+ (u_+-v_+, u_--v_+) \mathcal{R}^+_1 (v_+| u_+, u_-)
\, .
\end{align}

\begin{figure}[t]
\centerline{\includegraphics[width=0.9\linewidth]{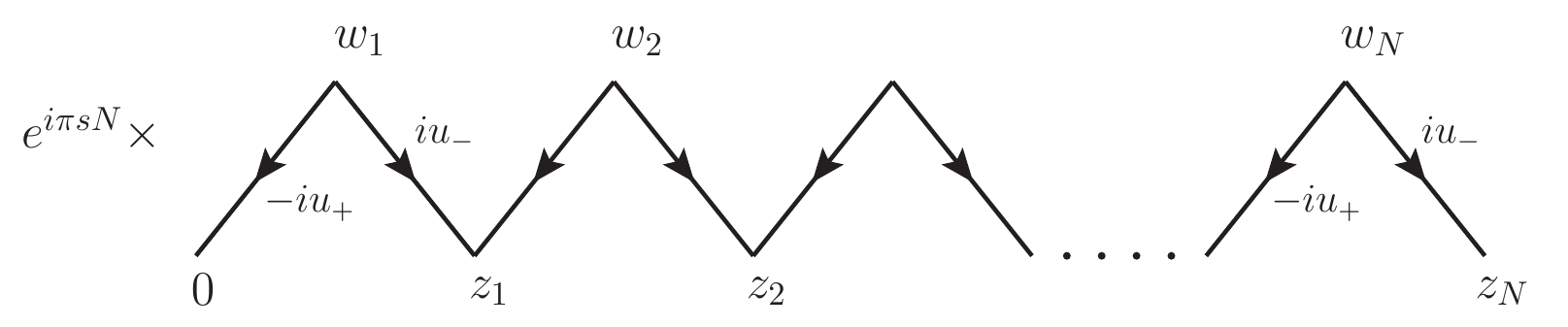}}
\caption{\label{fig:Qplus}The diagrammatic  representation of the kernel of the Baxter operator $\mathbb{Q}^+$, Eq.~(\ref{Q+kernel}).}
\end{figure}

Having established the commutativity of $\mathbb{S}_N^+ \mathcal{R}^+_1$ with $D_N (u)$, we demonstrate in Section \ref{BaxterEquation} that the former taken
at the point $v_+ = 0$ obeys a finite-difference Baxter equation \re{QplusBaxter} thus allowing us to identify it with the Baxter operator
\begin{align}
\mathbb{Q}^+ (u) \equiv
\mathcal{S}_N^+ (u_+, u_-) \mathcal{R}^+_1 (0| u_+, u_-)
\, .
\end{align}
The integral kernel (see Eq.~(\ref{integral-kernel})) of the Baxter operator $\mathbb{Q}^+ (u)$ takes a rather simple form
\begin{align}\label{Q+kernel}
\mathbb{Q}_u^+ (z_1,\ldots,z_N|\bar w_1,\ldots,\bar w_N)=e^{i\pi s N}
\prod_{n=1}^N (z_{n-1}-\bar w_n)^{-iu_-}\,  (z_n-\bar w_n)^{iu_+}\,,
\end{align}
where $z_{0}=0$. The diagrammatic representation for the kernel~(\ref{Q+kernel}) is shown in Fig.~\ref{fig:Qplus}.
Using the Feynman trick to combine the propagators attached to the same integration vertices, $w_n$,   and taking into
Eq.~(\ref{RepKernel}) for the reproducing kernel one gets the following representation for the Baxter operator
\begin{align}
\label{QplusIntegral}
&\mathbb{Q}^+ (u) \Psi (z_1, \dots, z_N)
=
\left( \frac{\Gamma (2 s)}{\Gamma (i u_-) \Gamma (- i u_+)} \right)^N
\\
&\qquad\times \int_0^1 \prod_{n = 1}^N d \alpha_n \, \alpha_n^{i u_- - 1} \bar\alpha_n^{- i u_+ - 1}
\,
\Psi (\alpha_1 z_1, \bar\alpha_2 z_1 + \alpha_2 z_2, \dots, \bar\alpha_N z_{N-1} + \alpha_N z_N)
\, , \nonumber
\end{align}
which is particularly useful in deriving the light-cone Hamiltonian \re{MultiParticleH}.


\subsection{Baxter operator $\mathbb{Q}^-$}
\begin{figure}[t]
\centerline{\includegraphics[width=0.9\linewidth]{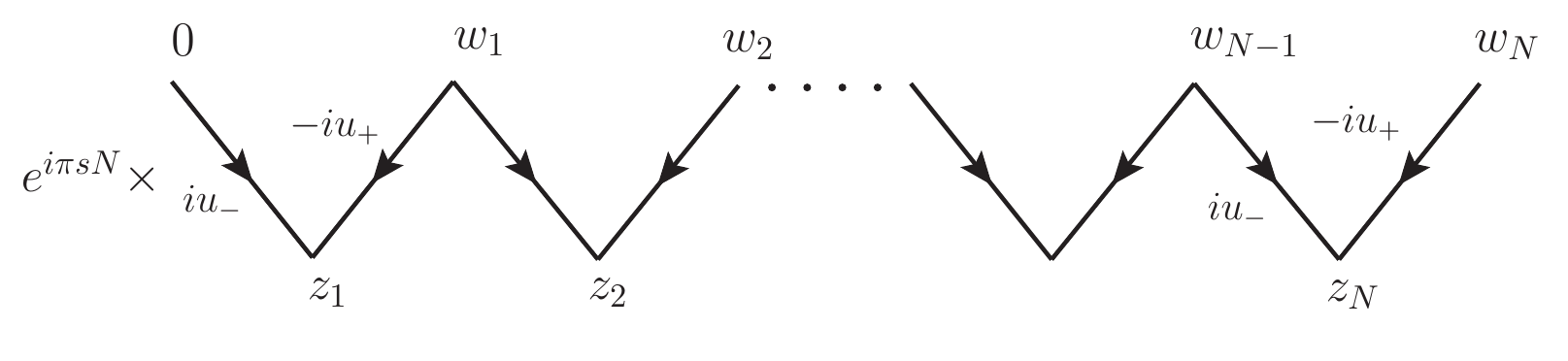}}
\caption{\label{fig:Qminus}The diagrammatic  representation of the kernel of the Baxter operator $\mathbb{Q}^-$.}
\end{figure}
Making use of the scalar product \re{NScalarProduct}, we can find the conjugate Baxter operator $\mathbb{Q}^- (u)$,
\begin{align}
\label{Qconjugation}
\mathbb{Q}^- (u)
=
\left( \mathbb{Q}^+ (u^\ast) \right)^\dagger
\, .
\end{align}
A kernel  of this operator can be
easily obtained from~(\ref{Q+kernel})
\begin{align}\label{Q-kernel}
\mathbb{Q}_u^- (z_1,\ldots,z_N|\bar w_1,\ldots,\bar w_N)=e^{i\pi s N}
\prod_{k=1}^N (z_{k}-\bar w_{k-1})^{-iu_-}\,  (z_{k}-\bar w_k)^{iu_+}\,,
\end{align}
where $w_{0}=0$. In the diagrammatic form it  is shown in Fig.~\ref{fig:Qminus}.

At the same time, the operator $\mathbb{Q}^-$ can be constructed explicitly by means of the defining relation
\re{RLLminus} for the intertwiner $\mathcal{R}^-$. The main steps in its construction mimic the derivation of the
operator
$\mathbb{S}_N^+ \mathcal{R}^+_1$ commuting with the matrix element $D_N$ of the monodromy operator $\mathbb{T}_N$ that we deduced in the preceding
section. By analogy with \re{Splus}, we define the operator
\begin{align}
\label{Sminus}
\mathcal{S}_N^- (u_+-v_-, u_--v_- )
\equiv
\mathcal{R}^{-}_{12} (u_+, u_-| v_-)
\mathcal{R}^{-}_{23} (u_+, u_-| v_-)
\dots
\mathcal{R}^{-}_{N-1, N} (u_+, u_-| v_-)
\, ,
\end{align}
that moves the spectral parameter $v_-$ from the right- to the left-most argument in a string of Lax operators
\begin{align}
&
\label{SLLminus}
\mathcal{S}_N^- (u_+-v_-, u_--v_-)\,
\mathbb{L}_1 (u_+, u_-)
\mathbb{L}_2 (u_+, u_-)
\dots
\mathbb{L}_N (u_+, v_-)
\nonumber\\
&
=
\mathbb{L}_1 (u_+, v_-)
\mathbb{L}_2 (u_+, u_-)
\dots
\mathbb{L}_N (u_+, u_-)\,
\mathcal{S}_N^- (u_+-v_-, u_--v_-)
\, .
\end{align}
In the right-hand side of this equation, we can recover the $N$-site monodromy matrix by using the left factorization of the Lax operator
such that
\begin{align}
\label{LeftFactUpper}
\mathbb{L}_1 (u_+, v_-) = z_1^{i u_- - i v_-}
\mathbb{G}_1 (v_-|u_-) \mathbb{L}_1 (u_+, u_-)  z_1^{i v_- - i u_-}
\, , \qquad
\mathbb{G}_1 (v_-|u_-)
\equiv
 \left(
\begin{array}{cc}
\frac{v_-}{u_-} & \frac{1}{z_1} \frac{u_- - v_-}{u_-} \\
0 & 1
\end{array}
\right)
\, .
\end{align}
To restore the correct spectral parameter in the right-most Lax operator in the left-hand side of Eq.\ \re{SLLminus}, we notice that
this can be achieved for the (relevant) second column of $\mathbb{L}_N$ via the equation
\begin{align}
\label{LtoN}
\mathbb{L}_N (u_+, v_-)
=
\partial_N^{i v_- - i u_-} \mathbb{N}^-_N  (u_+, u_-; v_-)
\partial_N^{i u_- - i v_-}
\, ,
\end{align}
where
\begin{align}
&
\mathbb{N}^-_N  (u_+, u_-; v_-)
\nonumber\\
&\quad=
\left(
\begin{array}{cc}
u_+ - u_- + v_- + i z_N \partial_N
&
- i \partial_N
\\
i z_N^2 \partial_N + (v_- + u_+ -2 u_-) z_N
+ i (u_- -v_-) (u_+ - u_- - i) \partial_N^{-1}
&
u_- - i z_N \partial_N
\end{array}
\right)
\, .
\end{align}
Substituting Eqs.\ \re{LeftFactUpper} and \re{LtoN} back into the permutation relation \re{SLLminus}, multiplying the result from the left and right by
$z_1^{i v_- - i u_-}$ and $\partial_N^{i v_- - i u_-}$, respectively, we project out the $22$-entry of the matrix equation to find the commutativity
of $D_N (u)$ with the operator $z_1^{i v_- - i u_-} \mathcal{S}_N^-\, \partial_N^{i v_- - i u_-}$. Comparing it with the conjugate to $\mathbb{Q}^+$,
we can verify that they coincide for the choice of the spectral parameter $v_- = 0$
\begin{align}
\label{QminusDef}
\mathbb{Q}^- (u) \equiv (\Gamma (2 s)/\Gamma (- i u_+))
(- z_1)^{- i u_-} \mathcal{S}^-_N (u_+, u_-) \partial_N^{- i u_-}
\, ,
\end{align}
with the diagrammatic representation for its integral kernel shown in Fig.\ \ref{fig:Qminus}. In the integral form, it reads
\begin{align}
\label{QminusIntegral}
&
\mathbb{Q}^- (u) \Psi (z_1, \dots, z_N) =
\left( \frac{\Gamma (2s)}{\Gamma (i u_-) \Gamma (- i u_+)} \right)^N
\left(
\frac{z_1}{z_N}
\right)^{- i u_-}
\nonumber\\
&\qquad
\times\int_0^1 \prod_{n=1}^{N} d \alpha_n \, \alpha_n^{- i u_+ - 1} \bar\alpha_n^{i u_- - 1}\alpha_N^{-2s}\,
\Psi (\alpha_1 z_1 + \bar\alpha_1 z_2, \dots, \alpha_{N - 1} z_{N - 1} + \bar\alpha_{N - 1} z_N , z_N/\alpha_N)
\, ,
\end{align}
where we used the results of Appendix \ref{FeynmanAppendix}.

\subsection{Baxter equations}
\label{BaxterEquation}

In this section we will show that the operator $\mathbb{Q}^\pm$ satisfy the first order difference equations in the spectral parameter $u$.
We perform the analysis for the Baxter operator $\mathbb{Q}^-$ and merely state the result for its hermitian conjugate. In our
consideration\footnote{We can also derive the Baxter equation for the operator $\mathbb{Q}^-$ making use of the invariance of the
monodromy matrix under gauge rotations of the Lax operators,
$$
\mathbb{L}_n
\to\begin{pmatrix}
1 & 0 \\
-\bar w_{n - 1} &1
\end{pmatrix}
\mathbb{L}_n
\begin{pmatrix}
1 & 0 \\
\bar w_n &1
\end{pmatrix}
\, .
$$
For detail, see Ref.\ \cite{Der99,DerKorMan02}.} we will follow the procedure developed in Ref.\ \cite{Der05}.
The stating point is the
relation \re{RLLminus} and the factorized form of the Lax operator \cite{PasGau92,Der99} analogous to \re{LaxFactorization}
\begin{align}
\mathbb{L}_n (u_+, u_-)
=
i z_n^{i u_+ - i u_-}  \mathbb{Z}_n \mathbb{W}_n (u_+, u_-) \mathbb{Z}_n^{-1} z_n^{i u_- - i u_+}
\, .
\end{align}
where
\begin{align}
 \mathbb{Z}_n
 =
\left(
\begin{array}{cc}
1  & 1/z_n \\
0  & 1
\end{array}
\right)
\, , \qquad
\mathbb{W}_n (u_+, u_-)
=
\left(
\begin{array}{cc}
- i u_-         & 0 \\
z_n^2 \partial_n  & - i u_+ - 1
\end{array}
\right)
\, .
\end{align}
Multiplying \re{RLLplus} from the left by $ \mathbb{Z}_1^{-1}$ and from the right sequentially first by $\mathbb{Z}_2 z_2^{i v_+ - i v_-}$ and
then by $\mathbb{W}_2^{-1}$ followed by $z_2^{i v_- - i v_+}$, we find that the defining equation takes the form
\begin{align}
\label{RecRel1step}
&
\mathbb{Z}_1^{-1} \mathcal{R}^{-}_{12} (u_+ ,u_-| v_-)
\mathbb{L}_1 (u_+, u_-) \mathbb{Z}_2=
\nonumber\\
&
\qquad=
\left(
\begin{array}{cc}
\ast
&
v_- (1/z_2 - 1/z_1) \mathcal{R}^{-}_{12}  (u_+ ,u_-| v_-)
\\
\ast
&
v_- \mathcal{R}^-_{12} (u_+ ,u_-| v_-)  + i (z_1/z_2)
\mathcal{R}^-_{12} (u_+ + i, u_- + i| v_-)
\end{array}
\right)\, .
\end{align}
As can easily be
seen from the explicit form of the matrix in the right-hand side of Eq.\ \re{RecRel1step}, it can be brought into a
triangular form by setting the spectral parameter $v_-$ to zero. This defines the main building block of the recursion,
\begin{align}
\mathbb{Z}_n^{-1} \mathcal{R}^-_{n, n+1} (u_+, u_-| 0) \mathbb{L}_n(u_+, u_-) \mathbb{Z}_{n+1}
=
\left(
\begin{array}{cc}
\ast
&
0\\
\ast
&
i (z_n/z_{n+1}) \mathcal{R}^-_{n,n+1} (u_+ + i, u_- + i | 0)
\end{array}
\right)
\, .
\end{align}
Taking the product of $N-1$ of these local relations with increasing value of indices enumerating the quantum spaces in the chain from left to right, we
can identify the result with the Baxter operator up to boundary multiplicative factors. To complete the right boundary, we use the following result
\begin{align}
\mathbb{M}_N^{-1} \partial_N^{- i u_-} \mathbb{L}_N (u)
=
\left(
\begin{array}{cc}
\ast
&
0\\
\ast
&
- i z_N^{\phantom{1}} \partial_N^{- i (u_- + i)}
\end{array}
\right)
\, ,
\end{align}
such that sandwiching the emerging matrix equation between the states $\bra\downarrow$ and $\ket\downarrow$, we immediately find
\begin{align}
\mathcal{S}_N^- (u_+, u_-)\,
\partial_N^{- i u_-} D_N (u)
=
- i^N z_1\,
\mathcal{S}_N^- (u_+ + i, u_- + i)\,
\partial_N^{- i (u_- + i)}
\, .
\end{align}
Introducing the Baxter operator via Eq.\ \re{QminusDef}, we recover a finite-difference equation that it obeys, i.e.,
\begin{align}
u_+^N \mathbb{Q}^- (u + i) =D_N (u) \mathbb{Q}^- (u)
\, ,
\end{align}
where we used the commutativity of the two operators in the right-hand side that was established in the previous section. Taking the hermitian conjugate
of this equation, it yields the Baxter equation for operator $\mathbb{Q}^+ (u)$,
\begin{align}
\label{QplusBaxter}
u_-^N \mathbb{Q}^+ (u - i) = D_N (u) \mathbb{Q}^+ (u)
\, .
\end{align}

Being first order finite-difference equations, both of them can be easily solved in the operator form with the result
\begin{align}
\label{QpmOperators}
\mathbb{Q}^\pm (u)
=
\prod_{n=1}^N \frac{\Gamma (2s) \Gamma (\pm i u \mp i \widehat{u}_n)}{\Gamma (s\pm i u) \Gamma (s \mp i \widehat{u}_n)}
\, .
\end{align}
Here we chose a normalization that matches the integral representations \re{QplusIntegral} and \re{QminusIntegral} for $\mathbb{Q}^\pm$.
Herefrom it is obvious that $\mathbb{Q}^\pm$ are conjugate to one another \re{Qconjugation} for real spectral parameter $u$ by virtue of the
fact that the Sklyanin's zeroes $\widehat{u}_n$ are hermitian, $\widehat{u}_n^\dagger = \widehat{u}_n$.

\subsection{Hamiltonians from Baxter operators}

Finally, having constructed the Baxter operators, we can find Hamiltonians from their expansion in the vicinity of the points $u = \pm i s$.
A simple calculation yields the result of interest
\begin{align}
\mathbb{Q}^\pm (\mp i s \pm \varepsilon) = 1 - i \varepsilon \left( \mathcal{H}^\pm_N + N \psi (2s) - N \psi(1) \right)+ O (\varepsilon^2)
\, ,
\end{align}
where
\begin{align}
\mathcal{H}^+_N
&=
N \psi (1)
- \psi (z_1 \partial_1 + 2 s) - \psi (z_{21} \partial_2 + 2 s) - \ldots - \psi (z_{N,N-1} \partial_N + 2 s)
\, , \\
\mathcal{H}^-_N
&=
N \psi (1)
- \ln z_1 - \psi (z_{12} \partial_1 + 2 s) - \ldots - \psi (z_{N-1,N} \partial_{N-1} + 2 s)  - \ln \partial_N
\, .
\end{align}
An immediate inspection demonstrates that their sum is indeed equivalent to \re{MultiParticleH}, $\mathcal{H}_N = \mathcal{H}^+_N
+ \mathcal{H}^-_N$. An alternative representation that makes the diagonalization of the Hamiltonians particularly straightforward
can be read off from Eq.\ \re{QpmOperators},
\begin{align}
\mathcal{H}_N
=
i \left(
\ln \frac{\mathbb{Q}^+ (- i s)}{\mathbb{Q}^- (+ i s)}
\right)^\prime
-
2N \left( \psi (2s) - \psi (1) \right)
=
\sum_{n = 1}^N \left[ 2 \psi(1) - \psi (s + i \widehat{u}_n) - \psi (s - i \widehat{u}_n) \right]
\, ,
\end{align}
with its eigenvalue coinciding with the well-known one-loop energy of $N$-particle spin-$s$ GKP excitation \cite{BelGorKor03,Bas10}
\begin{align}
\label{GKPenergies}
E_{\bit{\scriptstyle{u}}} = \sum_{n = 1}^N \left[  2 \psi (1) - \psi (s + i u_n) - \psi (s - i u_n) \right]
\, ,
\end{align}
when computed on the eigenfunction $\Psi_{\bit{\scriptstyle u}}$ \re{AuxProblem} of the operator $D_N$ that solves the spectral problem \re{SpectralProblem}
in question. We are turning to the construction of the wave functions $\Psi_{\bit{\scriptstyle u}}$ next.

\section{Eigenfunctions}
\label{EigenFunctionSection}

There is a brute force method to find the eigenfunction of the Hamiltonian $\mathcal{H}_N$ by solving differential equations stemming from $D_N$, see Eq.\
\re{AuxProblem}. However, this endeavor is hopeless for generic values of $N$ and thus we take a route of devising an algebraic construction of the former.

\subsection{Recurrence relation}

The formalism is based on a recurrence relation originating from the defining relation \re{DefRel}. Namely, projecting both sides of
the latter on the $22$-component of the matrix equation, we immediately find
\begin{align}
&
\mathcal{S}_N^+ (u_+-v_+, u_--v_+)
\bigg[
\left( i z_1^2 \partial_1 + (v_+ - u_-) z_1 \right) B_{N-1} (u)
+ (u_- - i z_1 \partial_1) D_{N-1} (u)
\bigg]
\nonumber\\
&
=
D_N (u)\,\mathcal{S}_N^+ (u_+-v_+, u_--v_+)
\, .
\end{align}
As it stands, this relation does not define a recursion for $D_N$ since it gets a contamination from the element $B_{N-1}$. We notice however, that we
can eliminate the unwanted first term in the left-hand side of this relation by an educated choice of a function $\Psi_N(z_1, \dots, z _N)$ that both sides
of this equation act on. Let us take it in the form $\Psi_N(z_1, \dots, z _N)=z_1^{i v_+ - i u_-} \Psi_{N-1}(z_2,\ldots, z_N)$ then
\begin{align}
&
v_+ \mathcal{S}_N^+ (u_+-v_+, u_--v_+)  z_1^{i v_+ - i u_-}
D_{N-1} (u) \Psi (z_2, \dots, z _N)
\nonumber\\
=
&
D_N (u) \mathcal{S}_N^+ (u_+-v_+, u_--v_+)  z_1^{i v_+ - i u_-}
\Psi (z_2, \dots, z _N)
\, .
\end{align}
Choosing the spectral parameter as $v_+ = u - u_1$ and introducing the notation
\begin{align}
\Lambda_N(u_1)=\mathcal{S}_N^+ (u_1+is, u_1-is)\, z_1^{- i u_1 - s}
\end{align}
one gets
\begin{align}
D_N (u)\,\Lambda_N(u_1)\Psi_{N-1}(z_2, \dots, z _N)=(u-u_1)\Lambda_N(u_1)D_{N-1} (u)\Psi_{N-1}(z_2, \dots, z _N)
\end{align}
The solution to this one-term recursion relation is simple and we find for the eigenfunction
\begin{align}
\label{LambdaNeigen}
&
\Psi_{\bit{\scriptstyle u}} (z_1, \dots, z_N)
=
{\rm e}^{-i \pi s N (N-1)/2}
\Lambda_N(u_1)\Lambda_{N - 1}(u_2)\dots
\Lambda_{2}(u_{N-1})\, z_N^{- i u_N - s}
\, , \nonumber
\end{align}
where
\begin{align}
\Lambda_{N-n}(u_{n+1})=\mathcal{S}_n^+ (u_{n+1}+is, u_{n+1}-is)\,
z_{n+1}^{- i u_{n+1} - s}
\end{align}
and the operators $\mathcal{S}^+_n$ are obtained from $\mathcal{S}_N^+$ by removing $n$ of the rightmost $\mathcal{R}^+$-factors, i.e.,
\begin{align}
\mathcal{S}_n^+ = \mathcal{R}_{N-1, N}^+ \dots \mathcal{R}_{N - n + 1, N - n + 2}^+
\, .
\end{align}
We introduced an overall normalization constant in Eq.\ \re{LambdaNeigen} for later convenience. The above eigenfunction diagonalizes
$D_N$
\begin{align}
D_N (u) \Psi_{\bit{\scriptstyle u}} (z_1, \dots, z_N)
=
\prod_{n=1}^N (u - u_n )  \Psi_{\bit{\scriptstyle u}} (z_1, \dots, z_N)
\, .
\end{align}
and is labelled by the zeroes $u_n$.

\subsection{Integral representation of eigenfunctions}

Having found a compact representations for the eigenfunctions \re{LambdaNeigen}, we are now in a position to cast them in a
more explicit integral form that will be extremely useful in proving their orthogonality. This discussion essentially follows an analogous one
devised for $SL(2)$ invariant spin chains in Ref.\ \cite{DerKorMan02}.

\begin{figure}[t]
\begin{center}
\mbox{
\begin{picture}(0,280)(240,0)
\put(0,0){\insertfig{17}{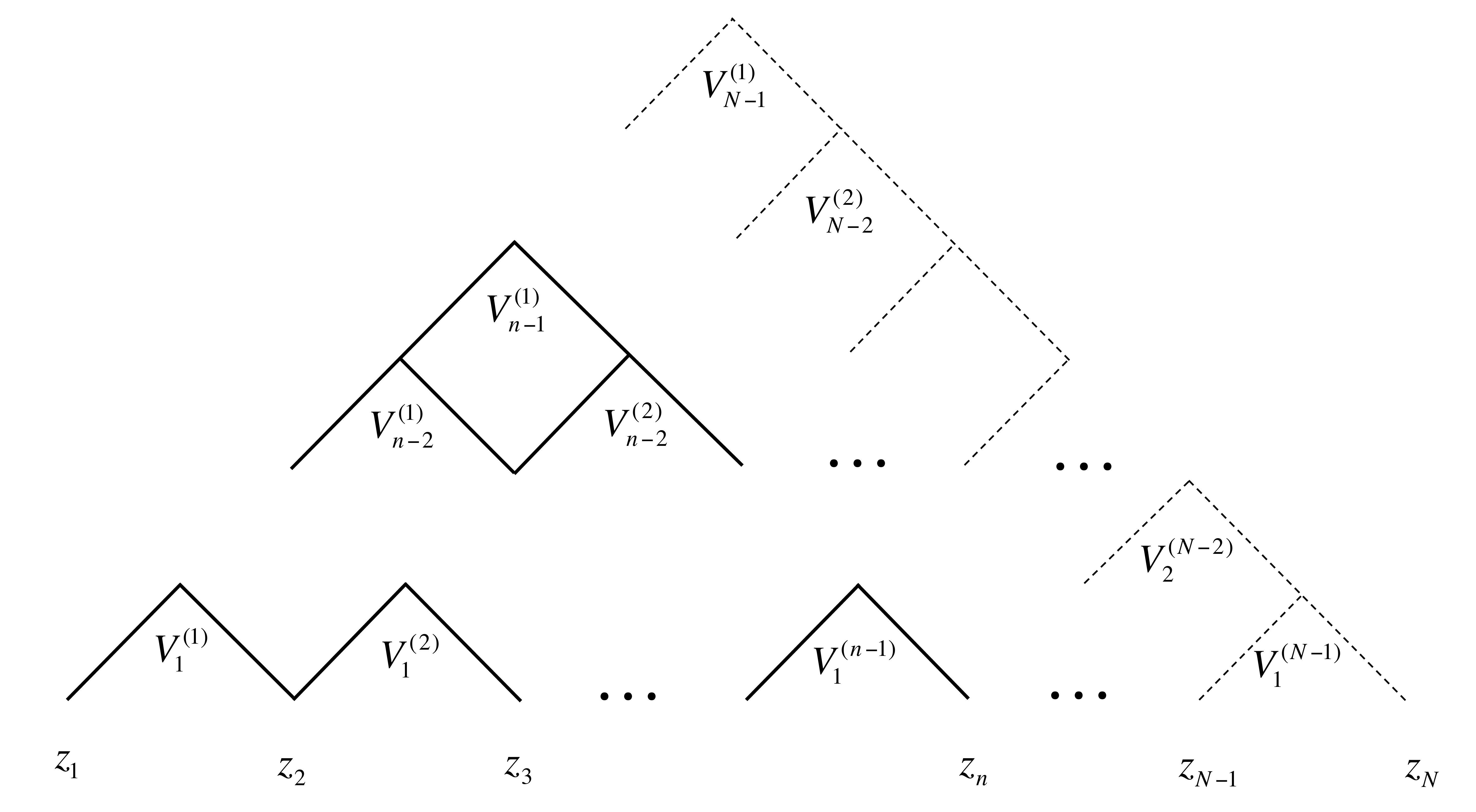}}
\end{picture}
}
\end{center}
\caption{ \label{LinePyramid} A subtree (shown in solid lines) in $N$-site pyramid (solid and dashed lines) originating from the vertex $V_{n-1}^{(1)}$
ending on sites from $z_1$ to $z_n$.}
\end{figure}

To start with, let us demonstrate the key manipulations using the two-site eigenfunction as an example,
\begin{align}
\Psi_{(u_1, u_2)} (z_1, z_2)
=
{\rm e}^{- i \pi s}
z_1^{- i u_1 - s} \frac{\Gamma (z_{21} \partial_2 + s + i u_1) \Gamma (2s)}{\Gamma (z_{21} \partial_2 + 2 s) \Gamma (s + i u_1)} z_2^{- i u_2 - s}
\, .
\end{align}
Recalling the integral representation of the Euler Beta function, we immediately find
\begin{align}
\label{2ptLineInt}
\Psi_{(u_1, u_2)} (z_1, z_2) = z_1^{- i u_1 - s} \frac{{\rm e}^{- i \pi s} \Gamma (2s)}{\Gamma (s - i u_1)\Gamma (s + i u_1)}
\int_0^1 d \tau \tau^{s - i u_1 - 1} \bar\tau^{s + i u_1 - 1} (\tau z_1 + \bar\tau z_2)^{- i u_2 - s}
\, .
\end{align}
This integral is proportional to the hypergeometric function $z_2^{- i u_2 - s} {_2 F_1} (s - i u_1, s + i u_2, 2s | 1 - z_1/z_2)$.
Applying the well-known connection formulas for hypergeometric functions, we can rewrite $\Psi_{(u_1, u_2)}$ as follows,
agreeing with finding of Ref.\  \cite{BasSevVie13},
\begin{align}
\Psi_{(u_1, u_2)} (z_1, z_2)
=
\frac{{\rm e}^{- i \pi s} \Gamma (i u_1 - i u_2)}{\Gamma (s + i u_1)\Gamma (s - i u_2)}
\bigg\{
 &
 z_1^{- i u_1 - s} z_2^{- i u_2 - s}
{_2 F_1}
\left.\left(
{s - i u_1, s + i u_2 \atop 1 - i u_1 + i u_2}
\right|
\frac{z_1}{z_2}
\right)
\\
+
S (u_2 | u_1)
& z_1^{- i u_2 - s} z_2^{- i u_1 - s}
{_2 F_1}
\left.\left(
{s + i u_1, s - i u_2 \atop 1 - i u_2 + i u_1}
\right|
\frac{z_1}{z_2}
\right)
\bigg\}
\, . \nonumber
\end{align}
The wavefunction admits the form of two plane waves $z_n^{- i u_n}$ with momenta $p(u_n) = 2 u_n$ scattering on each other by exchanging their
momenta with the two-particle S-matrix of GKP excitations\footnote{We would like to point out that due to our choice of momenta in the definition of the
plane waves, the $S$-matrix enters with exchanged rapidities since $S (u_2|u_1) = S (-u_1 |-u_2)$ compared to Ref.\ \cite{BasSevVie13}.}
\cite{BelGorKor03,Bas10,BasBel11}
\begin{align}
\label{GKPSmatrix}
S (u_1|u_2)
=
\frac{\Gamma (s + i u_2) \Gamma (s -  i u_1) \Gamma (i u_1 -  i u_2)}{\Gamma (s - i u_2)
\Gamma (s +  i u_1) \Gamma (i u_2 -  i u_1)}\,.
\end{align}

Another integral representation for $\Psi_{(u_1, u_2)} (z_1, z_2)$, regarding it as a holomorphic function in the upper half-plane, can be deduced
by means of the formula \re{FromZtoLine} derived in Appendix \ref{FeynmanAppendix} such that
\begin{align}
\Psi_{(u_1, u_2)} (z_1, z_2)
=
\int Dw \, z_1^{- i u_1 - s} (z_1 - \bar{w})^{i u_1 - s} (z_2 - \bar{w})^{- i u_1 - s}  w^{i u_2 - s}
\, .
\end{align}

\begin{figure}[t]
\begin{center}
\mbox{
\begin{picture}(0,270)(220,0)
\put(0,0){\insertfig{16}{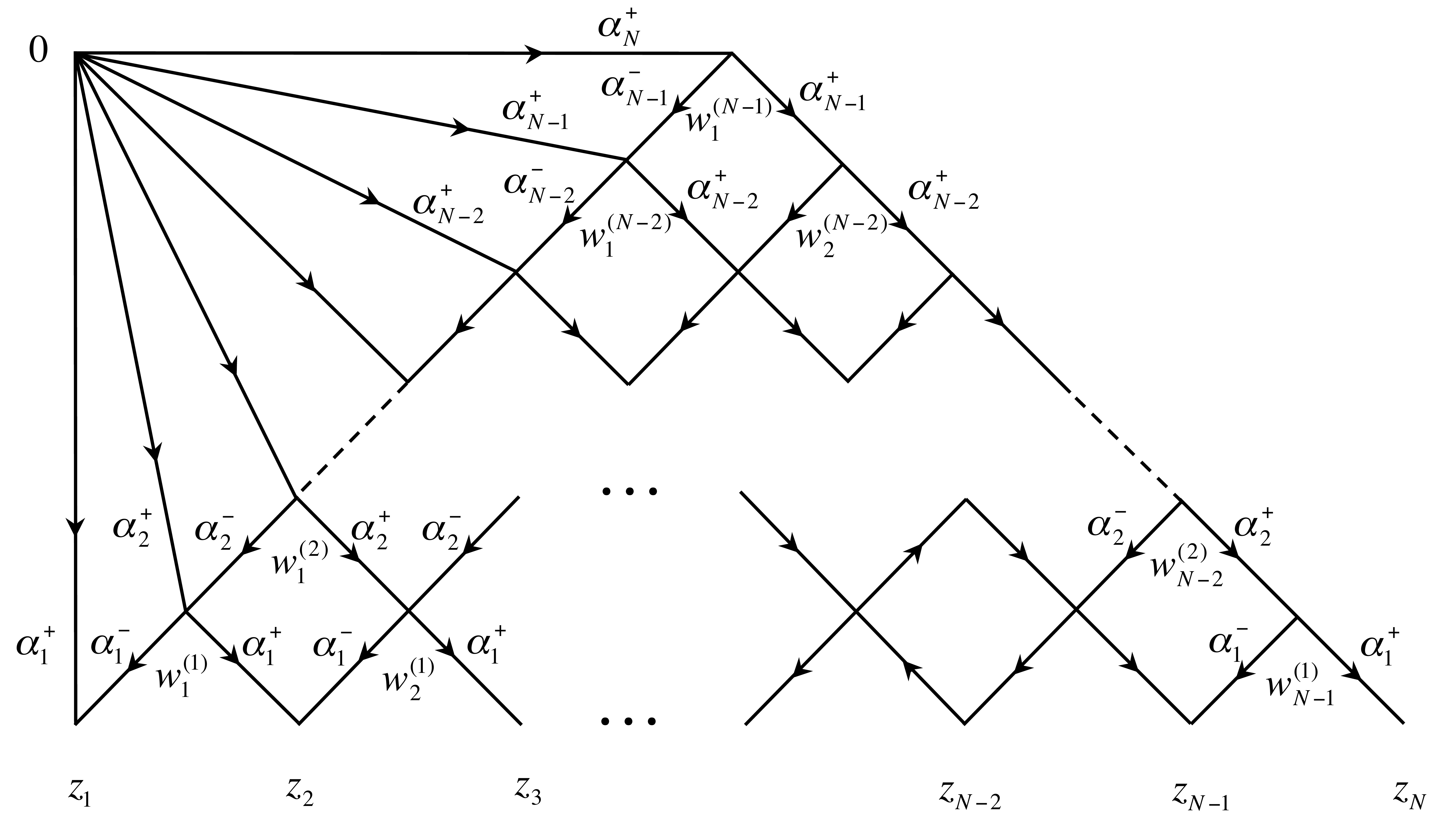}}
\end{picture}
}
\end{center}
\caption{ \label{NparticleWF} Diagrammatic representation for $N$-particle eigenstate \re{Deigenfunction}. Here $\alpha_n^\pm \equiv s \pm i u_n$
stands for the power of the propagator (see Appendix \ref{FeynmanAppendix} for conventions). Each vertex corresponds to a coordinate
in the pyramid that is integrated with the measure \re{MeasureSL2}. The top leftmost vertex is located at $w = 0$.}
\end{figure}

This construction can easily be extended to $N$-site eigenstates. The latter admit a convenient diagrammatic form shown in Fig.\
\ref{NparticleWF} for the line-integral representation generalizing Eq.\ \re{2ptLineInt}. The explicit eigenfunction in terms of multiple integrals reads
\begin{align}
\label{ExactWaveFunctLine}
\Psi_{\bit{\scriptstyle u}}  (z_1, z_2, \dots, z_N)
&=
{\rm e}^{- i \pi s N (N-1)/2}
\prod_{n=1}^{N-1} \left( \frac{\Gamma (s+i u_n) \Gamma (s -  i u_n)}{\Gamma (2s)} \right)^{n - N}
\\
&\times\int_0^1\prod_{n=1}^{N-1} \prod_{k=1}^{N-n} d \tau_n^{\scriptscriptstyle(k)}
\left( \tau_n^{\scriptscriptstyle(k)} \right)^{s - i u_n - 1}
\left( 1 - \tau_n^{\scriptscriptstyle(k)} \right)^{s + i u_n - 1}
\prod_{j=0}^{N-1} \left( V_j^{(1)} \right)^{- i u_{j + 1} - s}
\nonumber
\end{align}
where the integrand contains a product of vertices along the left edge of the pyramid (see Fig.~\ref{LinePyramid}) which branch down in trees
ending on $z_n$'s sites.  A few low-order trees read
\begin{align}
\notag
V_0^{(1)}
&= z_1
\, , \\
\notag
V_1^{(1)}
&=
\tau_1^{\scriptscriptstyle(1)} z_1 + (1 - \tau_1^{\scriptscriptstyle(1)}) z_2
\, , \\
\notag
V_2^{(1)}
&=
\tau_2^{\scriptscriptstyle(1)}
\left(
\tau_1^{\scriptscriptstyle(1)} z_1 + (1-\tau_1^{\scriptscriptstyle(1)}) z_2
\right)
+
(1-\tau_2^{\scriptscriptstyle(1)})
\left(
\tau_1^{\scriptscriptstyle(2)} z_2 + (1-\tau_1^{\scriptscriptstyle(2)}) z_3
\right)
\, , \\
&\dots
\nonumber
\end{align}
The rest can be easily constructed from the above Feynman graph using the fact that any internal vertex of the tree is parametrized by $\tau$-parameters
as $V_{n+1}^{(k)} = \tau_n^{\scriptscriptstyle(k)} V_n^{(k)} + (1 - \tau_n^{\scriptscriptstyle(k)}) V_n^{(k+1)}$.

Making use of the representation~(\ref{ExactWaveFunctLine})  it is easy to verify that as a function of real
variables the eigenfunction
$\Psi_{\bit{\scriptstyle u}}$ satisfies the following relation
\begin{align}\label{XmX}
\Psi_{\bit{\scriptstyle u}}(-x_1,\ldots, -x_N)=e^{-i\pi\sum_{n=1}^N (s+iu_n)} \,\Psi_{\bit{\scriptstyle u}}(x_1,\ldots, x_N)\,,
\end{align}
where all $x_k>0$.

Finally, the representation for the eigenfunction, determined by the pyramid in Fig.\ \ref{NparticleWF}, in terms of multiple integrals over the upper
half-plane yields
\begin{align}
\Psi_{\bit{\scriptstyle u}}  (z_1, z_2, \dots, z_N)
&
= z_1^{- i u_1 - s}
\int D w_{1}^{\scriptscriptstyle (1)} \dots D w_{N-1}^{\scriptscriptstyle (1)} \, (w_1^{\scriptscriptstyle (1)})^{- i u_2 - s} \, Y_{u_1} \left( z_1, z_2 | \bar{w}_1^{\scriptscriptstyle (1)} \right)
\dots
Y_{u_1} \left( z_{N-1}, z_N | \bar{w}_{N-1}^{\scriptscriptstyle (1)} \right)
\nonumber\\
&\times \int D w_{1}^{\scriptscriptstyle (2)} \dots D w_{N-2}^{\scriptscriptstyle (2)} \,
\left( w_{1}^{\scriptscriptstyle (2)} \right)^{- i u_3 - s}
\,
Y_{u_2} \left( w_1^{\scriptscriptstyle (1)}, w_2^{\scriptscriptstyle (1)} | \bar{w}_1^{\scriptscriptstyle (2)} \right)
\dots
Y_{u_2} \left( w_{N-2}^{\scriptscriptstyle (1)}, w_{N-1}^{\scriptscriptstyle (1)} | \bar{w}_{N-2}^{\scriptscriptstyle (2)} \right)
\nonumber\\
&\qquad\quad\vdots
\nonumber\\
&\times \int D w_1^{\scriptscriptstyle (N-1)}
\left( w_{1}^{\scriptscriptstyle (N-1)} \right)^{- i u_N - s}
\,
Y_{u_{N-1}} \left( w_1^{\scriptscriptstyle (N-2)}, w_2^{\scriptscriptstyle (N-2)} | \bar{w}_1^{\scriptscriptstyle (N-1)} \right)
\, ,
\label{Deigenfunction}
\end{align}
where we introduced the function
\begin{align}
\label{Yfunction}
Y_u (z_n, z_{n+1}|\bar{w}_n) \equiv (z_n - \bar{w}_n)^{i u - s} (z_{n+1} -\bar{w}_n)^{- i u - s}
\, .
\end{align}
Since these are eigenfunction of a hermitian operator $D_N$, they are orthogonal to each other by default. However, let us demonstrate this
explicitly.

\subsection{Proof of orthogonality}
\label{BNorthogonality}

Let us start with proving the orthogonality of one-site eigenfunction \re{1Peigenfunction}, $\Psi_{u_1} (z_1) = z_1^{- i u_1 - s}$.
A simple calculation making use of integral identities from Appendix \ref{FeynmanAppendix} yields
\begin{align}
\vev{\Psi_{v_1} | \Psi_{u_1}}
=
\frac{2 \pi {\rm e}^{\pi u_1}  \Gamma (2s)}{\Gamma (s + i u_1) \Gamma (s - i u_1)} \delta (u_1 - v_1)
\, .
\end{align}
It can be regarded as a chain rule \re{ChainRule} for end-points at $w = 0$ and $w^\prime = 0$, i.e.,
\begin{align}
\label{ChainToNorm}
\vev{\Psi_{v_1} | \Psi_{u_1}}
=
{\rm e}^{i \pi \beta^-_1} \int D z \, (z - \bar{w})^{- \alpha^+_1} (w^\prime - \bar{z})^{- \beta^-_1} |_{w = w^\prime = 0}
\, ,
\end{align}
where $\beta^\pm=s\pm i v$.
This will be an instrumental building block of the construction that follows.

The orthogonality can be shown inductively, so it is sufficient to demonstrate it for reduction of the scalar product from $N$ to $N-1$
sites. First, we have to define the complex conjugate pyramid. It is represented by the same graph as in Fig.\ \ref{NparticleWF} where,
however, one reverses the direction of arrows on all propagators. As a consequence of this, the integral is accompanied by an additional phase
\begin{align}
{\rm e}^{i \pi s N (N-1)} \prod_{n=1}^N {\rm e}^{i \pi (s - i u_n)}
\, ,
\end{align}
where the first factor in it comes from the complex conjugation of each of $N (N-1)/2$ $Y$-functions
\begin{align}
\notag
Y^\ast_u (z_n, z_{n+1}|\bar{w}_n)
\equiv
{\rm e}^{2 \pi i s} \,
\overline{Y}_u (w_n| \bar{z}_n, \bar{z}_{n+1})
=
{\rm e}^{2 \pi i s}
(w_n - \bar{z}_n)^{- i u - s} (w_n - \bar{z}_{n+1} )^{i u - s}
\, ,
\end{align}
while the second one originates from the conjugation of propagators connecting vertices with the point $w = 0$, namely,
\begin{align}
\notag
(z_n^{- i u_n - s})^\ast = {\rm e}^{i \pi (- i u_n + s)} (- \bar{z}_n)^{i u_n - s}
\, .
\end{align}

\begin{figure}[h]
\begin{center}
\mbox{
\begin{picture}(0,150)(225,0)
\put(0,0){\insertfig{16}{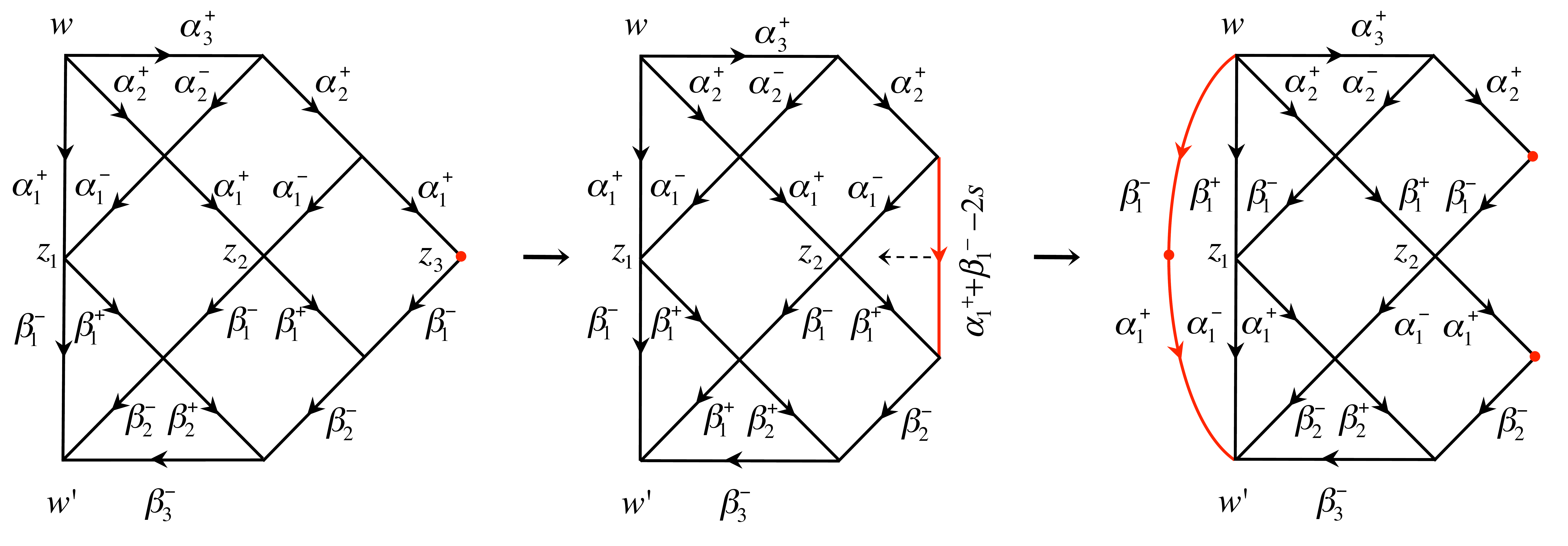}}
\end{picture}
}
\end{center}
\caption{ \label{NormLevel1} The scalar product of the eigenfunctions \re{Deigenfunction} for $N=3$. The points $w$ and $w^\prime$ are set
to zero. This chain of diagrams exhibits the first level of reduction that yields the delta function $\delta (u_1 - v_1)$. The parts of the graph shown
in red are the once that are involved in application of chain rule (for vertices) or permutation identity (for bonds).}
\end{figure}

\begin{figure}[h]
\begin{center}
\mbox{
\begin{picture}(0,150)(245,0)
\put(0,0){\insertfig{17}{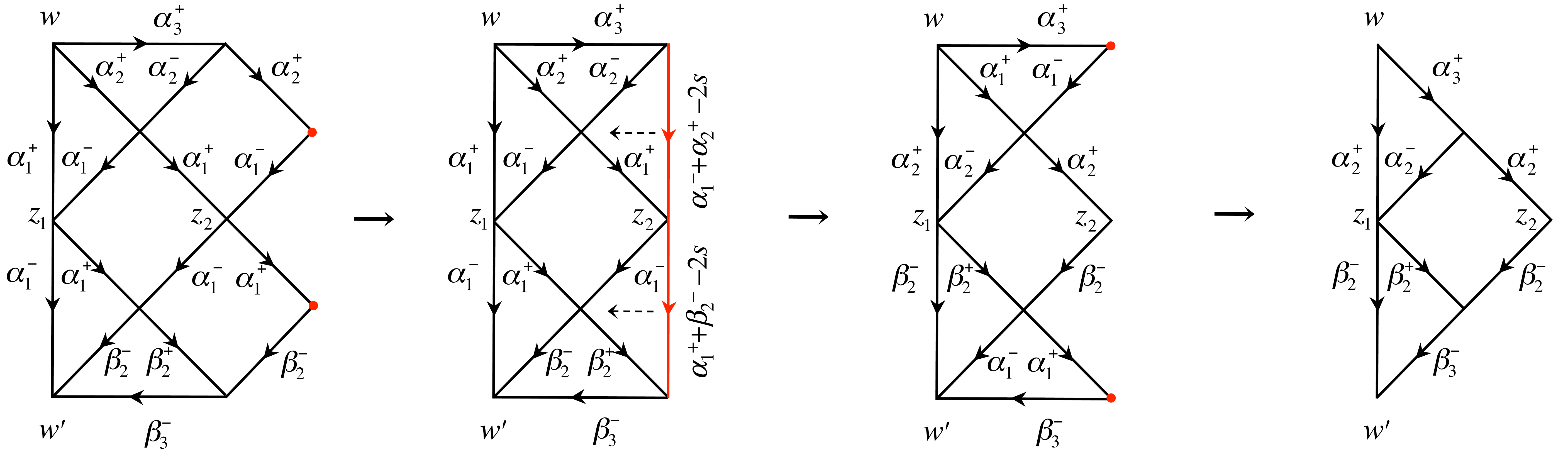}}
\end{picture}
}
\end{center}
\caption{ \label{NormLevel2} The second level of reduction of $N=3$ scalar product yielding the scalar product product of two-site eigenfunctions.
Here $w = w^\prime = 0$.}
\end{figure}

Without loss of generality let us perform the reduction from $N=3$ to $N=2$ since it can be easily visualized diagrammatically. We will split the
induction in two step, i.e., the first one that yields the delta-function and then second one that induces an overall normalization factor. They
are shown in Fig.\ \ref{NormLevel1} and \ref{NormLevel2}, respectively. We start with the right-most vertex $z_3$ in Fig.\ \ref{NormLevel1}
and integrate it out making use of the chain rule \re{ChainRule} from Appendix \ref{FeynmanAppendix} such that after this step the middle graph
gets multiplied by the factor ${\rm e}^{- i \pi s} a (\alpha_1^+, \beta_1^-)$. Next, we use the permutation identity \re{PermutationIdentity} twice and move
the (red) bond produced in the previous step all the way to the left such that it connects the points $w$ and $w^\prime$, both set at zero. If we represent this
bond using the chain rule backwards as in Eq.\ \re{ChainToNorm}, we recognize it as the scalar product of one-particle eigenfunction that factorizes off
from the graph. Therefore, the starting graph in Fig.\ \ref{NormLevel1} is equal to the left-most graph in Fig.\ \ref{NormLevel2} multiplied by
\begin{align}
{\rm e}^{- i \pi (s - i u_1)} \vev{\Psi_{v_1} | \Psi_{u_1}}
\, .
\end{align}
Notice that the delta function in the inner product $\vev{\Psi_{v_1} | \Psi_{u_1}}$ forces us to set $v_1 = u_1$ from now on in the starting graph
of the second level of reduction examined in Fig.\ \ref{NormLevel2}.

In the following step, we start by integrating out the two right-most vertices of the left graph in Fig.\ \ref{NormLevel2} located at $w_1^{\scriptscriptstyle (2)}$
and $w_1^{\prime \scriptscriptstyle (2)}$ and produce the graph to its right times the factor
\begin{align}
{\rm e}^{- 2 i \pi s} a (\alpha_2^+, \alpha_1^-) a (\alpha_1^+, \beta_2^-)
\, .
\end{align}
Next we move the vertical rightmost propagators (shown in red in Fig.\ \ref{NormLevel2}) from right to left. This step exchanges the momenta
$u_1 \leftrightarrow u_2$ and $v_2 \leftrightarrow u_1$, respectively. Finally, integrating the rightmost top and bottom vertices at $w_2^{\scriptscriptstyle (1)}$
and $w_2^{\prime \scriptscriptstyle (1)}$, we acquire extra factors emerging from the chain rules
\begin{align}
{\rm e}^{- 2 i \pi s} a (\alpha_3^+, \alpha_1^-) a (\alpha_1^+, \beta_3^-)
\, .
\end{align}
Putting everything together we find that the leftmost graph in Fig.~\ref{NormLevel1} is equal to the rightmost  one in Fig.~\ref{NormLevel2}
up to an overall factor
\begin{align}
{\rm e}^{- i \pi (s - i u_1)} {\rm e}^{- 4 i \pi s}
a (\alpha_2^+, \alpha_1^-) a (\alpha_1^+, \beta_2^-)
a (\alpha_3^+, \alpha_1^-) a (\alpha_1^+, \beta_3^-)
\vev{\Psi_{v_1} | \Psi_{u_1}}
\, .
\end{align}
Considering the fact that for a generic $N$, these two steps yield a phase ${\rm e}^{- i \pi (s - i u_1)} {\rm e}^{- 2 i \pi s (N-1)}$,
when combined with the overall one stemming from the complex conjugation, it reduces to the overall phase to the one for $N-1$-site
inner product, namely,
\begin{align}
{\rm e}^{- i \pi (s - i u_1)} {\rm e}^{- 2 i \pi s (N-1)}
{\rm e}^{i \pi s N (N-1)} \prod_{n=1}^N {\rm e}^{i \pi (s - i u_n)}
=
{\rm e}^{i \pi s (N-1) (N-2)} \prod_{n=2}^N {\rm e}^{i \pi (s - i u_n)}
\, .
\end{align}
The above construction demonstrates the induction
\begin{align}
\vev{\Psi_{(v_1, v_2, v_3)} | \Psi_{(u_1, u_2, u_3)}}
=
a (\alpha_2^+, \alpha_1^-) a (\alpha_1^+, \beta_2^-)
a (\alpha_3^+, \alpha_1^-) a (\alpha_1^+, \beta_3^-)
\vev{\Psi_{v_1} | \Psi_{u_1}}
\vev{\Psi_{(v_2, v_3)} | \Psi_{(u_2, u_3)}}
\, ,
\end{align}
that can be used to prove the orthogonality. For instance, to finish-up with the three-site case, repeating the same steps for $\vev{\Psi_{(v_2, v_3)} |
\Psi_{(u_2, u_3)}}$ inner product, one finds
\begin{align}
\vev{\Psi_{(v_2, v_3)} | \Psi_{(u_2, u_3)}}
=
a (\alpha_3^+, \beta_2^-) a (\alpha_2^+, \beta_3^-)
\vev{\Psi_{v_2} | \Psi_{u_2}}
\vev{\Psi_{v_3} | \Psi_{u_3}}
\, .
\end{align}
Without further ado, we just quote the orthogonality relation for arbitrary $N$
\begin{align}
\label{NparticleOrthogonality}
\vev{
\Psi_{\bit{\scriptstyle v}}|\Psi_{\bit{\scriptstyle u}}}
=
\sum_P
\prod_{n=1}^N\frac{ 2\pi e^{\pi u_n} \Gamma (2s)}{\Gamma(s - iu_n)\Gamma(s + iu_n)}
\delta \left(u_n - v_{P(n)}\right)
\prod_{k\neq j}\frac{\Gamma(i u_j- i u_k)}{\Gamma(s-iu_k)\Gamma(s+iu_j)}
\, ,
\end{align}
where we symmetrized with respect to the set of quantum numbers $\bit{v}$ to account for situations with $u_n = v_k$ for $n \neq k$.

In spite of the fact we found an explicit form of the eigenfunctions $\Psi_{\bit{\scriptstyle u}} (z_1, \dots, z_N)$ in Eq.\ \re{LambdaNeigen}, we find it
instructive to provide yet another representation which is based on the SoV formalism \cite{Skl85} that is discussed in Appendix \ref{SoVappendix}.

\section{Scalar product on real line}
\label{ScalarLine}

Having determined the eigenfunction of the renormalization group problem for the light-cone operators $O_N$ in Eq.\ \re{MultiPartLCoperator}, we can
now use them in the calculation of the multiparticle transitions in the OPE of the null polygonal Wilson loops discussed in Section \ref{OPEpolygons}.
However, one realizes that while we dealt in the solution of the open spin chain with integrals defined over the upper half-plane, the transition amplitudes
entering the Wilson loops are determined by the overlap integrals on the real positive half-line. The wave functions on the latter space are boundary
values of the eigenfunction holomorphic in the upper half-plane. Thus, we have to define a new scalar product. Its choice is driven by the existence of the
intertwining factor
\begin{align}\label{WN}
W_N (x_1, \dots, x_N) = \left( x_1 x_{21} \dots x_{N, N-1} \right)^{2s-1}
\, ,
\end{align}
that changes the spin label of the representation of the operator $D_N$
\begin{align}
\label{WNintertwiner}
D_N^{(1-s)} W_N = W_N D_N^{(s)}
\, ,
\end{align}
where we explicitly displayed them as superscripts. A proof of this intertwining relation is presented in Appendix \ref{Intappendix}.

This allows us to propose the following definition of the inner product
on the real positive half-line
\begin{align}
\label{RealLineScalarProduct}
( \Phi | \Psi ) = \int_0^\infty \mathcal{D}^N x \, \Phi^\ast (x_1, \dots, x_N) W_N (x_1, \dots, x_N) \Psi (x_1, \dots, x_N)
\, ,
\end{align}
where the measure is
\begin{align}
\label{RealMeasure}
\mathcal{D}^N x = dx_1\dots dx_N \theta (x_N > x_{N-1} > \dots > x_1)
\, .
\end{align}
Notice that the integrand is positive-definite in the integration domain.

In order to prove that the eigenfunctions $\Psi_{\bit{\scriptstyle u}}$ of the previous section are orthogonal with respect to the inner product
\re{RealLineScalarProduct} we have to demonstrate that the operator $D_N$ is self-adjoint,
\begin{align}
(\Phi| D_N (u) \Psi) = (D_N (u^\ast) \Phi | \Psi)
\, ,
\end{align}
for any $s \geq 1/2$. This can be verified by assuming that the functions entering \re{RealLineScalarProduct} vanish sufficiently fast as $x_n$ goes
to infinity and confirming that all boundary terms are absent for $x_n = x_{n+1}$.

\subsection{Relation between scalar products}

Let us establish a relation between  the scalar products  (\ref{NScalarProduct}) and (\ref{RealLineScalarProduct}). Taking into account  the
defining property of the reproducing kernel \re{RepKernel}, we can represent a function $\Psi$ with arguments belonging to the real half-line
in terms of its analytic continuation in the upper half-plane as follows
\begin{align}
\Psi(x_1,\ldots, x_N)
=
\int \prod_{n=1}^N \left( Dz_n\,\mathcal{K}(x_n,\bar z_n) \right) \Psi(z_1,\ldots, z_N)
\, ,
\end{align}
where the reproducing kernel reads (see Appendix \ref{FeynmanAppendix} for details)
\begin{align}
\mathcal{K}(x_n,\bar z_n)  =  {\rm e}^{i \pi s} (x_n - \bar{z}_n)^{- 2s}
\, .
\end{align}
Analogous relation holds for $\Phi$. Using this representation for the functions $\Psi$ and $\Phi$ and substituting then into the scalar product
\re{RealLineScalarProduct}, we get
\begin{align}
( \Phi | \Psi )= \bra{\Phi} X \ket{\Psi} \, ,
\end{align}
where $X$ is an operator with the integral representation
\begin{align}
[X\Psi](w_1, \dots, w_N)=\int \prod_{n=1}^N Dz_n \, \mathcal{X} (w_1, \dots, w_N; \bar{z}_1, \dots \bar{z}_N) \Psi(z_1,\ldots,z_N)
\end{align}
and the kernel
\begin{align}
\mathcal{X} (w_1, \dots, w_N; \bar{z}_1, \dots \bar{z}_N)
=
( \mathcal{K}_{w_1, \dots, w_N} | \mathcal{K}_{\bar{z}_1, \dots, \bar{z}_N})
\, ,
\end{align}
written in the form of the scalar product~(\ref{RealLineScalarProduct}) of the products of reproducing kernels
\begin{align}
\mathcal{K}_{w_1, \dots, w_N} (x_1, \dots, x_N) = \prod_{n=1}^N\,\mathcal{K}(x_n, w_n)\,.
\end{align}
Taking into account the action of $sl(2, \mathbb{R})$ generators on the reproducing kernel is symmetric with respect to the interchange of its
arguments $S^{\pm,0}_w\,\mathcal{K}(w,x)=S^{\pm, 0}_x\,\mathcal{K}(w,x)$, we immediately establish the commutativity of the operator
$D_N(u)$ with $X$
\begin{align}
D_N(u)\, X= X\, D_N(u)\,.
\end{align}
Therefore, it shares the same eigenfunctions $\Psi_{\boldsymbol{u}}$ with the operator $D_N$ that was diagonalized in previous sections. One can
easily calculate its eigenvalues  $X(\boldsymbol{u})$ from the equation
\begin{align}\label{XU}
[X\,\Psi_{\boldsymbol{u}}](w)=X(\boldsymbol{u})\,\Psi_{\boldsymbol{u}}(w)\,.
\end{align}
A direct iterative proof of this relation is given in Appendix \ref{Intappendix}. However, a much easier way to accomplish the same goal is is by comparing
the left- and right-hand sides of Eq.\ \re{XU} in the region $0\ll z_1\ll z_2\ll\ldots \ll z_N$. The eigenfunction in this asymptotic domain merely takes the form
of a product of free plane waves $\Psi_{\boldsymbol{u}}(z)\sim z_1^{- iu_1-s}\ldots z_N^{- iu_N - s} + \ldots$. The leading term in the left-hand side of Eq.\
\re{XU}, that reads
\begin{align}\label{Xuu}
[X\,\Psi_{\boldsymbol{u}}](w)
=
\int \mathcal{D}^N x \, \mathcal{K}_{w_1, \dots, w_N} (x_1, \dots, x_N)  W_N(x_1, \dots, x_N) \Psi_{\boldsymbol{u}}(x_1,\ldots,x_N)
\, ,
\end{align}
comes from the integration in the vicinity $x_k\sim w_k$. It implies that we can replace the eigenfunction by the leading term in the asymptotic
expansion.  Obviously, since $x_{n-1} \ll x_n$, we can replace accordingly the intertwining factor by its leading order form
$$
W_N(x_1, \dots, x_N) \to \prod_{n=1}^N x_n^{2s-1}
$$
such that the integrals in Eq.~(\ref{Xuu}) decouple and can be easily calculated individually yielding the result that we sought for
\begin{align}
X(\boldsymbol{u})=\prod_{n=1}^N {\rm e}^{-\pi u_n} \frac{\Gamma(s-iu_n)\Gamma(s+iu_n)}{\Gamma(2s)}\,.
\end{align}
Thus for the scalar product of an arbitrary function $\Phi$ with the eigenfunction $\Psi_{\boldsymbol{u}}$, we get
\begin{align}\label{s-product-relation}
(\Phi|\Psi_{\boldsymbol{u}})=X(\boldsymbol{u})\,\vev{\Phi|\Psi_{\boldsymbol{u}}}\, .
\end{align}
This equation provides a relation of the scalar products on the line and half plane and it will be instrumental in calculating various integrals which
are hopeless to be computed analytically otherwise.

Let us note an important difference between the two scalar products. The scalar product  on the upper half-plane is an $SL(2,\mathbb{R})$ invariant
scalar product. Namely, the group transformations $\mathbb{T}^s(g)=(\otimes T^{s}(g))^N$ of the wave function is defined as
\begin{align}
\label{Tg-definition}
[\mathbb{T}^s(g) \Psi](z_1,\ldots, z_N)
=
\prod_{n=1}^N (cz_n+d)^{- 2s}
\Psi\left(
\frac{a z_1 + b}{c z_1 + d},\ldots, \frac{a z_N + b}{c z_N + d}
\right)\,,
\end{align}
with elements of the group $g^{-1}= \left( { a \ b\atop c \ d} \right)$ being real unimodular matrices, $ad-bc=1$. So that $\mathbb{T}^s(g)$ are unitary
operators with respect to this scalar product,
\begin{align}
\vev{\mathbb{T}^s(g)\Phi|\mathbb{T}^s(g)\Psi}=\vev{\Phi|\Psi}
\, .
\end{align}
On the other hand, contrary to this, the scalar product on the real positive half-line~(\ref{RealLineScalarProduct}) is invariant only under scale
transformations $g_\lambda$,
\begin{align}
(\mathbb{T}^s (g_\lambda)\Phi| \mathbb{T}^s (g_\lambda)\Psi)=(\Phi|\Psi)\,.
\end{align}
with the group elements $g_\lambda^{-1}= \left( {\lambda \ \ \ 0 \atop 0 \ \lambda^{-1} } \right)$.

\section{Multiparticle transitions}
\label{SquareAndHexagon}

While in the above discussion we analyzed the eigenfunction of the matrix elements
$\bra{E_{\bit{\scriptstyle u}}} O_N \ket{0}$
of the light-cone operators \re{MultiPartLCoperator}, to apply our findings to the analysis of $N$-particle
contributions in the Wilson loops, we define an $N$-particle state in the following fashion
\begin{align}
\label{Neigenstate}
\ket{E_{\bit{\scriptstyle u}}}
&= \int_0^\infty \mathcal{D}^N x \, \psi^\ast_{\bit{\scriptstyle u}} (x_1, \dots, x_N)\, O_N (x_1, \dots, x_N) \ket{0}
\, ,
\end{align}
with $\psi_{\bit{\scriptstyle u}} (x_1, \dots, x_N)$ being the
wave function of the GKP excitations in the position space \cite{BasSevVie13}. These are related to the
eigenfunctions \re{ExactWaveFunctLine} found in Section~\ref{EigenFunctionSection}
via the relation
\begin{align}
\label{psiLambdaRelation}
\psi_{\bit{\scriptstyle u}} (x_1, \dots, x_N)
=
\mathcal{N}_{\bit{\scriptstyle u}}
\,W_N(x_1, \dots, x_N)\,\Psi_{\bit{\scriptstyle u}} (x_1, \dots, x_N)
\, ,
\end{align}
where $\mathcal{N}_{\bit{\scriptstyle u}}$ is a normalization coefficient. Indeed, the state
$\ket{E_{\bit{\scriptstyle u}}}$ takes the form of the scalar product~(\ref{RealLineScalarProduct}) of the operator
$O_N$ with the eigenfunction $\Psi_{\bit{\scriptstyle u}}$
\begin{align}
\ket{E_{\bit{\scriptstyle u}}}=\mathcal{N}_{\bit{\scriptstyle u}}\, (\Psi_{\bit{\scriptstyle u}}|O_N)\,\ket{0}.
\end{align}
Obviously it is the eigenstate of the operator $D_N(u)$ and, hence, the Hamiltonian.
In Ref.~\cite{BasSevVie13} the function $\psi_{\bit{\scriptstyle u}}$ was normalized
in a manner consistent with the Coordinate Bethe Ansatz, such that for large separation between the GKP excitations,
$z_{k}/z_{k+1}\ll 1$, the initial state consists of noninteracting plane waves with unit amplitude
\begin{align}
\label{CBAform}
\left. \psi_{\bit{\scriptstyle u}} (z_1, \dots, z_N) \left( \prod\nolimits_{n = 1}^N z_n^{1 - s}\right) \right|_{z_N \gg \dots \gg z_1}
=
z_1^{-i u_1} \dots z_N^{-i u_N} + \ldots
\, .
\end{align}
The ellipses stand for the terms $z_1{\!}^{-i u_{k_1}} \dots z_N{\!}^{-i u_{k_N}}$ accompanied by the GKP $S$-matrices
\re{GKPSmatrix} for each permutation $P$ as well as power-suppressed $O(z_k/z_{k+1})$ contributions. The coefficient
$\mathcal{N}_{\bit{\scriptstyle u}}$ that corresponds to the normalization~(\ref{CBAform}) takes the form
\begin{align}\label{N_u}
\mathcal{N}_{\bit{\scriptstyle u}}={\rm e}^{i \pi s N (N-1)/2}
\prod_{n > k}^N \frac{\Gamma (s - i u_n) \Gamma (s + i u_k)}{\Gamma (2s) \Gamma (i u_k - i u_n)}\,.
\end{align}
It is evident that for physical quantities all dependence on the normalization coefficient  drops out. From a
technical point of view it is more convenient and natural to accept normalization which preserves the symmetry
function under permutation of the separated variables $u_1,\ldots,u_N$. However, to make comparison with the results
of Ref.\ \cite{BasSevVie13} easier, we present our calculation of the square and hexagon transitions in the
normalization~(\ref{N_u}).

\subsection{Square transitions}

To start with, let us address the square transitions. Within the context of the operator product expansion for the octagon Wilson loop
discussed in Section \ref{OPEpolygons}, they define its leading asymptotic behavior. For the $\chi_2 \chi_3 \chi_5 \chi_6$
component, it reads
\begin{align}
\vev{\mathcal{W}_8^R}
\stackrel{\tau\to\infty}{=}
\chi_2 \chi_3 \chi_5 \chi_6 \frac{a}{2 \pi}
\int_{-\infty}^{\infty} d u \int_{-\infty}^{\infty} d v \, B_{\rm h} (u|v)
\exp \left( - \tau E_{\rm h} (u; a) - i \sigma \, p_{\rm h} (u; a) \right)
\, .
\end{align}
When generalized to multiparticle spin-$s$ GKP excitations, the square transition is determined by the overlap integral of the $D_N$
eigenfunction and its complex conjugate both evaluated in the same conformal frame
\begin{align}
\label{SquareDef}
B_s (\bit{u} | \bit{v})
=
\int_0^\infty \mathcal{D}^N x^\prime \int_0^\infty \mathcal{D}^N x \,
\psi^\ast_{\bit{\scriptstyle v}} \left( x_1^\prime, \dots, x_N^\prime \right)
\prod_{k=1}^N G_s \left( x_k^\prime , x_k \right)
\psi_{\bit{\scriptstyle u}} (x_1, \dots, x_N)
\, ,
\end{align}
connected point-by-point by the propagators for the spin-$s$ excitation which read
\begin{align}
\label{PropagatorGs}
G_s (x_k^\prime, x_k) = (x_k^\prime + x_k)^{- 2s}
\, .
\end{align}
It is a generalization of the one obtained from the OPE of the Wilson loop for the hole excitation, which is a
Fourier transform of the measure $\mu_{\rm h} (u)$ in Eq.\ \re{HoleMeasure}.  The integration over the bottom wave function with
attached propagators can be performed explicitly. Indeed, taking into account Eqs.\ (\ref{Xuu}), (\ref{XU}) and the relation (\ref{XmX}), we find
\begin{align}\label{first-x-int}
\int_0^\infty \mathcal{D}^N x \, \prod_{k=1}^N G \left( x_k^\prime , x_k \right) \psi_{\bit{\scriptstyle u}} (x_1, \dots, x_N)
&=
{\rm e}^{i\pi s N} \mathcal{N}_{\bit{\scriptstyle{u}}}\, [X \Psi_{\bit{\scriptstyle u}}](-x^\prime_1,\ldots, -x^\prime_N)
\nonumber\\
&=
\mathcal{N}_{\bit{\scriptstyle{u}}}
\prod_{n=1}^N \mu_s (u_n) \Psi_{\bit{\scriptstyle u}} (x_1^\prime, \dots, x_N^\prime)
\, ,
\end{align}
where the right-hand side is accompanied by the product of the one-particle measures for spin-$s$ GKP excitations
\begin{align}\label{musn}
\mu_s (u) = \frac{\Gamma (s + i u) \Gamma (s - i u)}{\Gamma (2s)}
\, .
\end{align}
It reduces to Eq.\ \re{HoleMeasure} for $s = 1/2$ corresponding to hole excitation. The above equation allows us to relate the square transitions to the
inner product on the real positive half-line
\begin{align}
B_s (\bit{u} | \bit{v})
=
\mathcal{N}_{\bit{\scriptstyle{u}}}\,\mathcal{N}^\ast_{\bit{\scriptstyle{v}}}\, \prod_{n=1}^N \mu_s (u_n)  \,
(\Psi_{\bit{\scriptstyle{v}}} | \Psi_{\bit{\scriptstyle{u}}})
\, .
\end{align}
This in turn can be rewritten via the scalar product in the upper half-plane, see Eq.~(\ref{NparticleOrthogonality}), and immediately yields the
result
\begin{align}\label{Box}
B_s (\bit{u} | \bit{v})
=
(2\pi)^N \prod_{n=1}^N \mu_s (u_n)
\Big( 1 + S (u_2| u_1) P_{u_1u_2} + \dots \Big) \delta^N (\bit{v} - \bit{u})
\,,
\end{align}
where the sum goes over all permutations accompanied by scattering matrices of GKP excitations. For $N=1,2$ our expressions for the square
transitions (\ref{Box}) reproduces those in Ref.~\cite{BasSevVie13}.

\subsection{Hexagon transitions}

The $N$-particle hexagon\footnote{To this order in perturbation theory, these are identical to the pentagon transitions discussed in Ref.\ \cite{BasSevVie13}.}
transitions are defined as follows
\begin{align}
\label{HexagonDef}
H_s (\bit{u} | \bit{v})
=
\prod_{n=1}^N \mu^{-1}_s (u_n) \mu^{-1}_s (v_n)
\int_0^1 \mathcal{D}^N x^\prime \int_0^\infty \mathcal{D}^N x \,
\psi^{\prime \ast}_{\bit{\scriptstyle v}} \left( x_1^\prime, \dots, x_N^\prime \right)
\prod_{k=1}^N G_s \left( x_k^\prime , x_k \right)
\psi_{\bit{\scriptstyle u}} (x_1, \dots, x_N)
\, .
\end{align}
Equation \re{HexagonDef} is obtained from the square transition by boosting the top wave function to a different conformal frame $x_n^\prime \to
x_n^{\prime\prime} = x_n^\prime /(1 - x_n^\prime)$, according to the results of Appendix \ref{DecagonParametrization},
such that
\begin{align}\label{boosted-function}
\psi^{\prime}_{\bit{\scriptstyle v}} \left( x_1^\prime, \dots, x_N^\prime \right)
=
\prod\nolimits_{n = 1}^N
\left(
\frac{\partial x_n^{\prime\prime} }{\partial x_n^\prime}
\right)^{1 - s}
\psi_{\bit{\scriptstyle v}} \left( x_1^{\prime\prime}, \dots, x_N^{\prime\prime} \right)
\, .
\end{align}
One can simplify the expression for $H_s (\bit{u} | \bit{v})$ by performing the integrals with respect to unprimed variables $x_n$ as was
discussed in the preceding section, see Eq.~(\ref{first-x-int}),
\begin{align}
H_s (\bit{u} | \bit{v})&=\mathcal{N}_{\bit{\scriptstyle{u}}}\prod_{n=1}^N \mu^{-1}_s(v_n)
\int_0^1 \mathcal{D}^N x^\prime\,
\psi^{\prime\ast}_{\bit{\scriptstyle v}} \left( x_1^\prime, \dots, x_N^\prime \right)
\Psi_{\bit{\scriptstyle u}} (x_1^\prime, \dots, x_N^\prime)\,.
\end{align}
Making a change of variables $x_n=x^\prime_n/(1-x^\prime_n)$ and taking into account the definitions~(\ref{boosted-function})
and~(\ref{psiLambdaRelation}), we can cast $H_s (\bit{u} | \bit{v})$ into the form
\begin{align}
H_s (\bit{u} | \bit{v})=\mathcal{N}_{\bit{\scriptstyle{u}}} \, \mathcal{N}_{\bit{\scriptstyle{v}}}^\ast
 \prod_{n=1}^N \mu^{-1}_s(v_n) \, (\Psi_{\bit{\scriptstyle v}}| \mathbb{T}^s(g^+)\Psi_{\bit{\scriptstyle u}})\,,
\end{align}
where $g^+ = \left( {+1 \ \ 0 \atop -1\ +1} \right)$ and the transformation $\mathbb{T}^s(g)$ is defined in Eq.~(\ref{Tg-definition}).
In explicit form, the transformed wave function reads
\begin{align}
[\mathbb{T}^s(g^+)\Psi_{\bit{\scriptstyle u}}](x_1,\ldots,x_N)=\prod_{n=1}^N (1+x_n)^{- 2s}
\Psi_{\bit{\scriptstyle u}}\left( \frac{x_1}{1+x_1},\ldots, \frac{x_N}{1+x_N} \right)\,.
\end{align}
Using the relation (\ref{s-product-relation}) adopted to the present case
\begin{align}
(\Psi_{\bit{\scriptstyle v}}| \mathbb{T}^s(g^+)\Psi_{\bit{\scriptstyle u}}) = X(\bit{v})\,
\bra{\Psi_{\bit{\scriptstyle v}}} \mathbb{T}^s(g^+) \ket{\Psi_{\bit{\scriptstyle u}}}
\, ,
\end{align}
we can derive the hexagon transition in terms of the scalar product in the upper half-plane
\begin{align}
H_s (\bit{u} | \bit{v})=\mathcal{N}_{\bit{\scriptstyle{u}}} \, \mathcal{N}_{\bit{\scriptstyle{v}}}^\ast \,
{\rm e}^{-\pi \sum_n v_n} \, \bra{\Psi_{\bit{\scriptstyle v}}} \mathbb{T}^s(g^+) \ket{\Psi_{\bit{\scriptstyle u}}}
\,.
\end{align}
Thus the hexagon transition is determined by the matrix element of the group transformation~$\mathbb{T}^s(g^+)$.

\begin{figure}[t]
\begin{center}
\mbox{
\begin{picture}(0,270)(230,0)
\put(0,0){\insertfig{16}{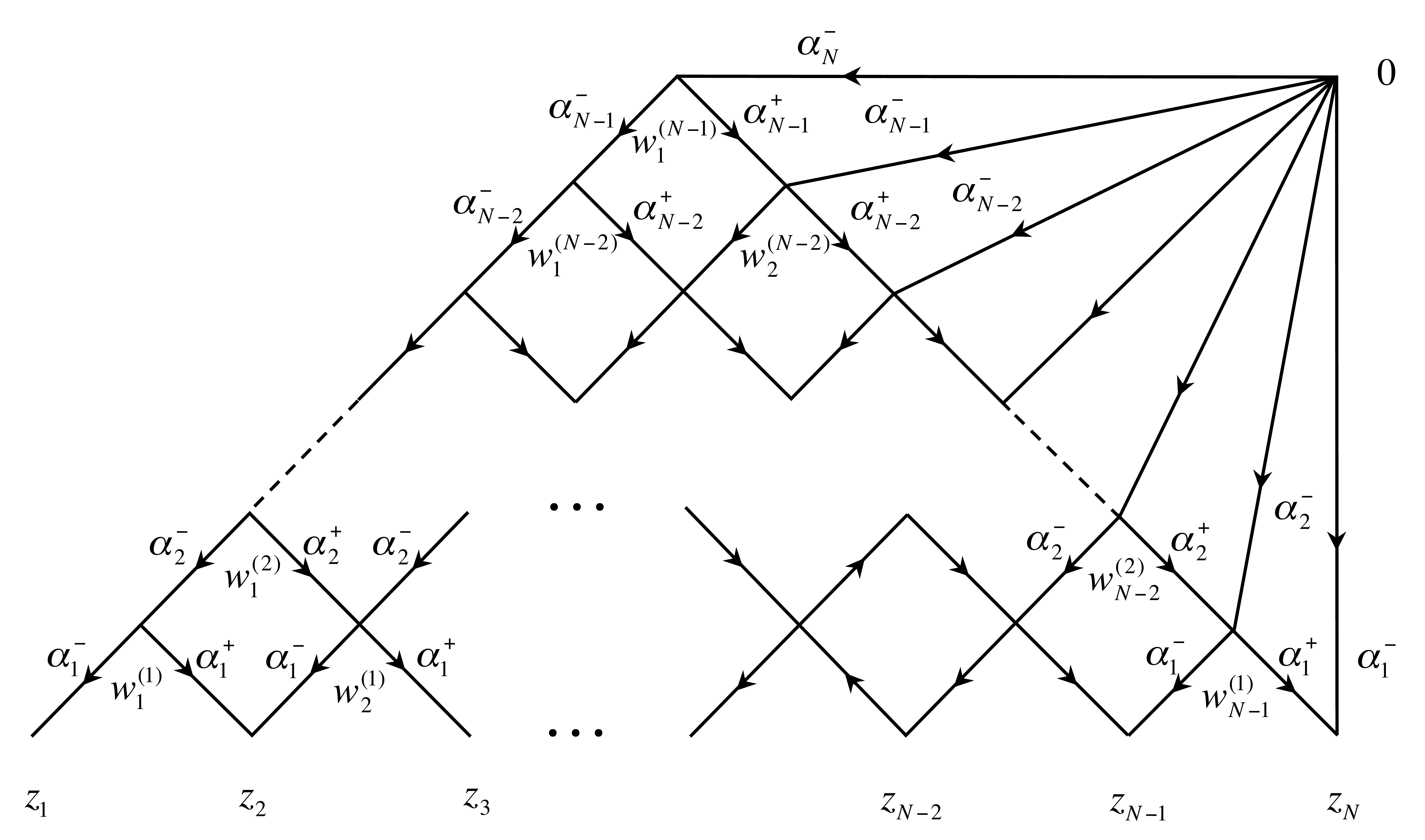}}
\end{picture}
}
\end{center}
\caption{ \label{ANeigenfunction} Diagrammatic representation for $N$-particle function $\widetilde{\Psi}_{\bit{\scriptstyle u}}(z_1,\ldots,z_N)$
introduced in Eq.\ \re{InverseEigenfunction}.}
\end{figure}

\begin{figure}[h]
\begin{center}
\mbox{
\begin{picture}(0,160)(225,0)
\put(0,0){\insertfig{16}{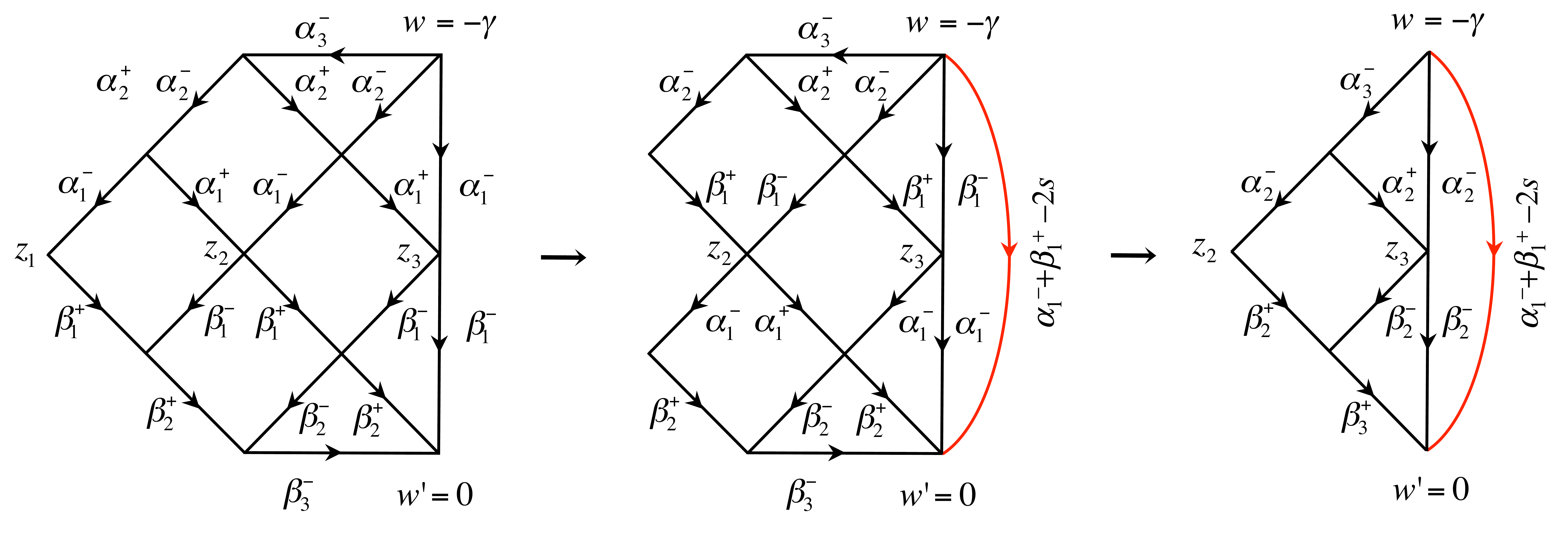}}
\end{picture}
}
\end{center}
\caption{ \label{HexagonRed} The reduction of  $T_\gamma(\bit{u},\bit{v})$ from $N=3$ to $N=2$.}
\end{figure}

In what follows we consider a more general matrix element
\begin{align}
T_\gamma(\bit{u},\bit{v})= \bra{\Psi_{\bit{\scriptstyle v}}} \mathbb{T}^s(g^+_\gamma) \ket{\Psi_{\bit{\scriptstyle u}}}\,,
\end{align}
where $g^+_\gamma=\left( {+1\ \ 0 \atop -\gamma \ + 1} \right)$. In order to calculate it, it is convenient to perform the inversion first,
\begin{align}
\bra{\Psi_{\bit{\scriptstyle v}}} \mathbb{T}^s(g^+_\gamma) \ket{\Psi_{\bit{\scriptstyle u}}}
=
\bra{\mathbb{T}^s(g_I)\Psi_{\bit{\scriptstyle v}}} \mathbb{T}^s(g_I g^+_\gamma g_I^{-1}) \ket{\mathbb{T}^s(g_I)\Psi_{\bit{\scriptstyle u}}}
\, ,
\end{align}
where $g_I= \left( { 0 \  \ +1\atop -1 \ \ 0} \right)$. We immediately find that $g_I g^+_\gamma g_I^{-1} = g^- _\gamma =
\left( {1\ \gamma \atop 0 \ 1} \right)$ and, therefore,
$\mathbb{T}^s( g^-_\gamma )$ generates a translation
\begin{align}\label{shift}
[\mathbb{T}^s( g^-_\gamma )\Psi] (z_1,\ldots,z_N)=\Psi(z_1-\gamma,\ldots,z_N-\gamma)\,.
\end{align}
Under inversion, the eigenfunctions $\Psi_{\bit{\scriptstyle u}}$ transforms as follows
\begin{align}
\label{InverseEigenfunction}
[\mathbb{T}^s(g_I)\Psi_{\bit{\scriptstyle u}}] (z_1,\ldots,z_N)
=
{\rm e}^{-i\pi\sum_n(s+iu_n)}
\widetilde{\Psi}_{\bit{\scriptstyle u}}(z_1,\ldots,z_N)\,.
\end{align}
The diagrammatic representation\footnote{The function $\widetilde{\Psi}_{\bit{\scriptstyle u}}$ is the eigenfunction of the element $A_N$
of the monodromy matrix $\mathbb{T}_N$ since $A_N$ and $D_N$ are related to each other by inversion.} for
$\widetilde{\Psi}_{\bit{\scriptstyle u}}(z_1,\ldots,z_N)$ is shown in Fig.\ \ref{ANeigenfunction}. Its explicit analytic expression can easily be
read off according to the Feynman rules of Appendix \ref{FeynmanAppendix}. Taking into account (\ref{shift}) we find that up to an overall
prefactor, the diagram for the matrix element $T_\gamma(\bit{u},\bit{v})$ (see the left-most graph in Fig.\ \ref{HexagonRed}) is given by the
mirror reflected diagram in Fig.\ \ref{NormLevel1} for the scalar product of two wave functions, where however $w=-\gamma$ is moved away from zero.

The analysis goes along the same lines as the calculation for the norm of the eigenfunctions though the reduction starts from left to right, in the direction
opposite to steps taken in Section \ref{BNorthogonality} since the initial graph defining the current matrix element is a mirror reflection of the one alluded
to above, as shown in Fig.\ \ref{HexagonRed}. A crucial difference from the norm computation arises after the first level of reduction (cf.\ the right-most
graph in Fig.\ \ref{NormLevel1} and the middle graph in Fig.\ \ref{HexagonRed}) and stems from the fact that the (red) line connecting the points $w$ and
$w'$ produces $(-\gamma + i 0)^{i(u_n-v_n)}$ rather than the delta function $\delta(u_n-v_n)$. Taking this into account and collecting all factors due to chain
integrations, we finally obtain
\begin{align}
T_\gamma(\bit{u},\bit{v})=\gamma^{i\sum_n(u_n-v_n)}\, {\rm e}^{\pi\sum_n v_n}\,
\prod_{n,k=1}^N\frac{\Gamma (2s) \Gamma(iv_n-iu_k)}{\Gamma(s+iv_n)\Gamma(s-iu_k)}\,.
\end{align}
Sending $\gamma\to 1$, we find a factorized expression for the $N$-particle hexagon transition
\begin{align}
\label{Hfactorization}
H_s (\bit{u} | \bit{v})
=
\frac{\prod_{n,k=1}^N H_s (u_n|v_k)}{\prod_{n>k}^N H_s (u_n|u_k) \prod_{n<k}^N H_s (v_n|v_k)}
\, ,
\end{align}
where the leading order one-particle hexagon is a generalization of the corresponding expression for the hole excitation,
see Eq.~(\ref{hex-hole-half}),
\begin{align}
H_s (u|v) = \frac{\Gamma (2s) \Gamma (i v - i u)}{\Gamma (s + i v) \Gamma (s - i u)}\,.
\end{align}
This proves the conjecture\footnote{To avoid confusion, we remind that our $u_n$ and $v_n$ have an extra minus sign compared to conventions
in Ref.\  \cite{BasSevVie13}.} for multiparticle transitions put forward in Ref.\ \cite{BasSevVie13}.

\section{Conclusions}

To conclude, in this paper we systematically studied the open spin chain emerging in the analysis of the OPE for the null polygonal Wilson loop. Using
the factorized $R$-matrices we constructed the commuting Hamiltonians and demonstrated that they are identical to the one emerging from the calculation
of the renormalization of the $\Pi$-shaped Wilson contour with elementary field
insertions. We also demonstrated how they emerge from Baxter operators. The main goal of this analysis was to find the eigenfunction of the
Hamiltonian and we constructed them via an iterative procedure for any number of sites. We cast them in multi-dimensional integral form
that is particularly useful for the proof of their orthogonality as well as in explicit applications to multiparticle transitions. Making use of the Baxter
operators, we cast the eigenstates in the SoV form as well. Finally, we demonstrated that our finding yields the factorized form of the multiparticle
hexagon transitions suggested earlier.

There are multiple ways in which we can generalize our results. The consideration can be extended to inhomogeneous case when the spin labels
of representation at different sites are not identical. Further, it should be relatively straightforward to extend our analysis to supersymmetric case
making use of the factorization property of supersymmetric $R$-matrices in terms of intertwiners following the formalism of Ref.\ \cite{BelDerKorMan06}.

\section*{Acknowledgments}

We would like to thank Benjamin Basso for fruitful discussions. The work of A.B.\ was supported by the U.S.\ National Science Foundation
under the grant No.\ PHY-1068286, A.M.\ was supported by the German Research Foundation (DFG), grant BR2021/5-2 and S.D.\ by RFBR grants
12-02-91052, 13-01-12405 and 14-01-00341.

\appendix

\section{Parametrization of polygons}

In Section \ref{OPEpolygons} we are dealing with conformally invariant remainder functions. Working with Minkowski four-dimensional
coordinates $x^\mu$ (with $\mu = 1, \dots, 4$) singles out a particular Poincar\'e subgroup of the $SO(4,2)$ conformal group. One can bypass
this by working in a six-dimensional space $\mathbb{R}^{4,2}$ with projective coordinates $X^M$ (with $M=1,\dots,6$) such that
$X^M \simeq \alpha X^M$ (for a nonzero $\alpha$) and where the Minkowski space-time is realized on a hypercone $X_M X^M = 0$. The
advantage of this space is that the action of the conformal group is realized linearly on $X^M$. This six-dimensional formulation is
particularly useful to pass to the momentum twistor formalism that we will do in the next subsection. The subsequent two subsection are
dedicated to particular parametrizations of Wilson loop polygons that are suitable for their OPE analysis.

\subsection{Coordinates $\leftrightarrow$ twistors}
\label{MomentumTwistorApp}

Here we will present an embedding of the Minkowski space-time in a six-dimensional projective space and introduce momentum twistors
by passing to its spinor representation. This provides with us with a dictionary and an easy conversion of components between different
parametrizations.

Let us start with a six-dimensional Euclidean space where the sigma matrices are defined in terms 't Hooft symbols as \cite{BelDerKorMan03}
\begin{align}
\Sigma^I_{AB} = (\eta^I_{AB}, i \bar\eta^I_{AB})
\, ,
\end{align}
with $\eta$ and $\bar\eta$ being the self-dual and anti-self-dual tensors $\ft{1}{2}\varepsilon^{ABCD} \eta^I_{AB} = \eta^I_{CD}$,
$\ft{1}{2}\varepsilon^{ABCD} \bar\eta^I_{AB} = - \bar\eta^I_{CD}$, respectively. In the matrix form, these read
\begin{align}
&
\Sigma^1_{AB}
=
\left(
\begin{array}{cc}
0                  & \sigma_1 \\
- \sigma_1 & 0
\end{array}
\right)
\, , \quad
\Sigma^2_{AB}
=
\left(
\begin{array}{cc}
0                & - \sigma_3 \\
\sigma_3  & 0
\end{array}
\right)
\, , \quad
\Sigma^3_{AB}
=
\left(
\begin{array}{cc}
i \sigma_2  & 0 \\
0                  & i \sigma_2
\end{array}
\right)
\, , \nonumber\\
&
\Sigma^4_{AB}
=
\left(
\begin{array}{cc}
0                & \sigma_2 \\
\sigma_2  & 0
\end{array}
\right)
\, , \quad\ \ \,
\Sigma^5_{AB}
=
\left(
\begin{array}{cc}
0      & - i \II \\
i \II  & 0
\end{array}
\right)
\, , \quad \,
\Sigma^6_{AB}
=
\left(
\begin{array}{cc}
- \sigma_2  & 0 \\
0                   & \sigma_2
\end{array}
\right)
\, , \nonumber
\end{align}
where the two-by-two blocks are determined by the conventional Pauli matrices. We pass to the $R^{2,4}$ signature of the $SO(4,2)$
conformal group in the following fashion. Define
\begin{align}
\Gamma^0 =  - i \Sigma^4 \, , \
\Gamma^1 = \Sigma^2 \, , \
\Gamma^2 = \Sigma^5 \, , \
\Gamma^3 = \Sigma^1 \, , \
\Gamma^4 = \Sigma^3  \, , \
\Gamma^5 = - i \Sigma^6
\, ,
\end{align}
such that the invariant interval (squared) is given by the determinant
\begin{align}
\det \sum\nolimits_{I = 0}^5 \Gamma ^M X_M = \left( - X_0^2 + \sum\nolimits_{i=1}^3 X_i^2 + X_5^2 - X_6^2 \right)^2
\end{align}
of the matrix
\begin{align}
\label{6DX}
X \equiv
\Gamma^M X_M
=
\left(
\begin{array}{cc}
i \sigma_2 (X_5 + X_6)  & - i \sigma_2 ( \sigma^0 X_0 - \sum_{i=1}^3 \sigma^i X_i) \\ [3mm]
i \sigma_2 ( \sigma^0 X_0 - \sum_{i=1}^3 \sigma^i X_i) &  i \sigma_2 (X_5 - X_6)
\end{array}
\right)
\, .
\end{align}
Pulling $X_5 + X_6$ from $\Gamma^M X_M$ we can identify $x_\mu = X_\mu/(X_5 + X_6)$ with the Poincare coordinates and
$X_6 - X_5 = x^2$ for the light-like six-dimensional vector $X_M X^M = 0$ that provides an embedding of Minkowski space.

The above matrix representation is particularly useful to relate it to momentum twistors. The latter are conventionally defined
as \cite{Hod09}
\begin{align}
\label{PenroseTwistor}
Z_i^A = (\lambda_i^\alpha, \mu_{i, \dot\alpha})
\, ,
\end{align}
such that a line in twistor space corresponds to a point in (dual) space-time \re{6DX}
\begin{align}
X_i = Z_{i - 1} \wedge Z_i
\, ,
\end{align}
or explicitly
\begin{align}
\label{Xi}
X_i =
\vev{i-1, i}
\left(
\begin{array}{cc}
- \varepsilon_{\alpha\beta} & - (x_i)^\alpha{}_{\dot\beta}
\\
(x_i)_{\dot\alpha}{}^\beta & x_i^2 \varepsilon^{\dot\alpha \dot\beta}
\end{array}
\right)
\, ,
\end{align}
where we use a standard notation for the two-dimensional Levi-Civita tensor normalized as $\varepsilon_{12} = - \varepsilon_{\dot{1} \dot{2}} = 1$
and the four-dimensional coordinate is expressed in terms of components of the momentum twistors as
\begin{align}
x_i = \frac{\mu_{i-1}\lambda_i - \mu_i \lambda_{i-1}}{\vev{i-1 i}}
\, .
\end{align}
Notice that in the above construction the infinity bitwistor is defined as
\begin{align}
I^{AB}
=
(i \Sigma^6 -  \Sigma^3)^{AB}
=
\left(
\begin{array}{cc}
- \varepsilon_{\alpha\beta} & 0
\\
0 & 0
\end{array}
\right)
\, ,
\end{align}
such that
\begin{align}
Z_{i-1} I Z_i = \lambda_{i-1}^\alpha \lambda_{i \alpha} \equiv \vev{i-1, i}
\, .
\end{align}

\subsection{Reference square}
\label{RefSquare}

\begin{figure}[t]
\begin{center}
\mbox{
\begin{picture}(0,180)(110,0)
\put(0,0){\insertfig{8}{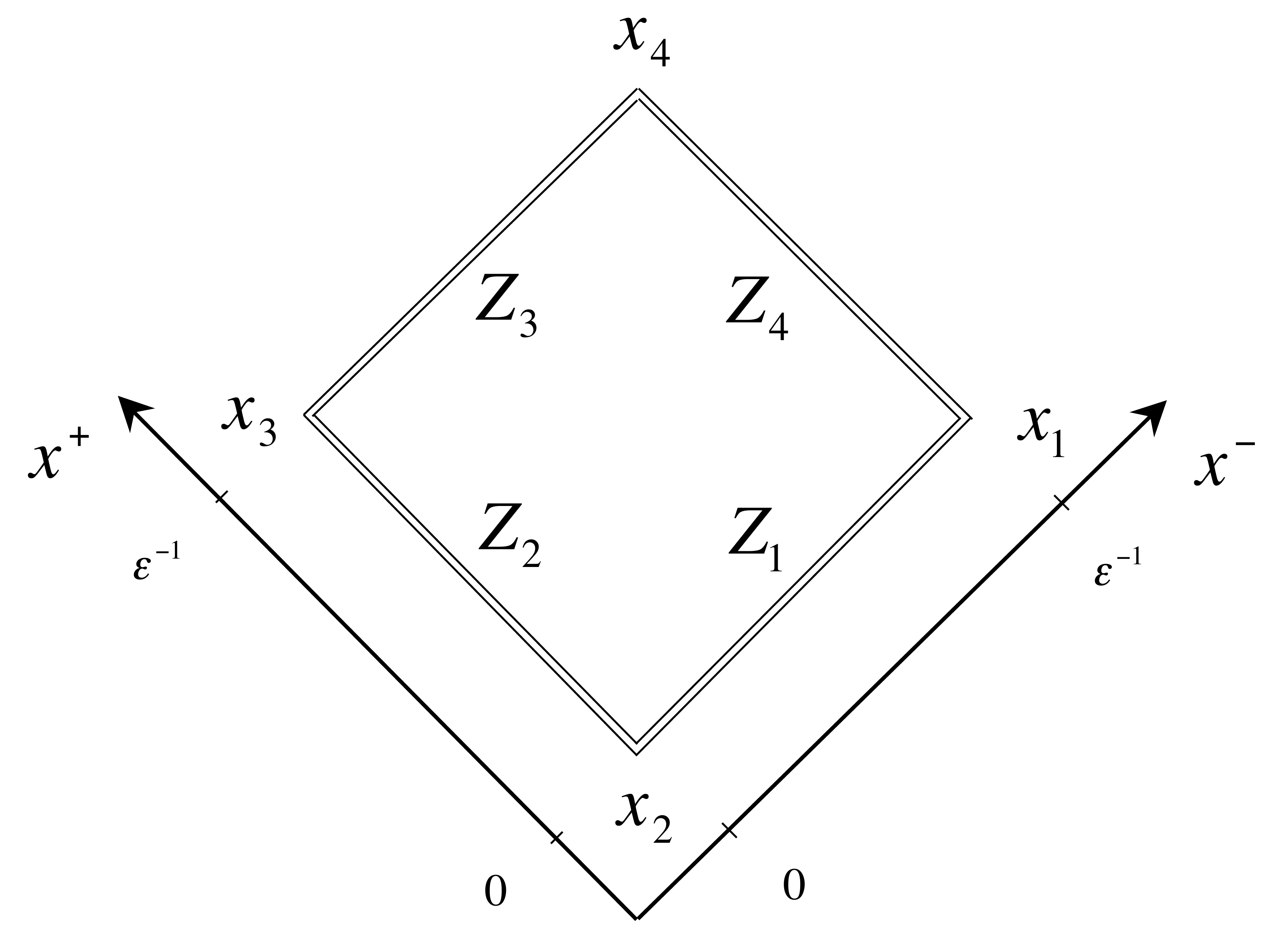}}
\end{picture}
}
\end{center}
\caption{ \label{Square} Reference square.}
\end{figure}

The use of the restricted kinematics for scattering amplitudes implies that the bosonic part of the Wilson loop contour resides on a two-dimensional
plane with two-dimensional vectors parametrized by light-cone coordinates $x = (x^-, x^+)$. Conformal parametrization of higher point polygons relies on
the definition of a reference square that we choose to have the following coordinates of its vertices
\begin{align}
\label{RefSquareX}
x_1 = (1/\varepsilon, 0)
\, , \qquad
x_2 = (0, 0)
\, , \qquad
x_3 = (0, 1/\varepsilon)
\, , \qquad
x_4 = (1/\varepsilon, 1/\varepsilon)
\, ,
\end{align}
where $\varepsilon \to 0$. We would like to take full advantage of linear realization of conformal transformations in the twistor formulation, so that
we use the embedding coordinates
\begin{align}
\label{EmbeddingXfromZ}
X_i^{AB} = \left( Z_{i-1}^A Z_i^B - Z_i^A Z_{i-1}^B \right)/\vev{i-1 i}
\, ,
\end{align}
where the top-right corner is a two-by-two matrix of the Poincare coordinates in question
\begin{align}
\notag
\left(
\begin{array}{cc}
0     & - x^- \\
x^+ & 0
\end{array}
\right)
\, ,
\end{align}
in accord with Eq.\ \re{6DX}. Then, the momentum twistors corresponding to the coordinates \re{RefSquareX} are
\begin{align}
Z_1^T = (1, 0, 0, 0)
\, , \qquad
Z_2^T = (0, 1, 0, 0)
\, , \qquad
Z_3^T = (1, 0, 1/\varepsilon, 0)
\, , \qquad
Z_4^T = (0, 1, 0, 1/\varepsilon)
\, .
\end{align}
Under a conformal transformation
\begin{align}
\label{SforRefSquare1}
S = \left(
\begin{array}{cccc}
{\rm e}^\sigma & & & \\
& {\rm e}^\tau & & \\
& & {\rm e}^{- \sigma} & \\
& & & {\rm e}^{- \tau}
\end{array}
\right)
\, ,
\end{align}
the above twistors transform as
\begin{align}
\label{ConformalTransformation}
Z^\prime = S Z
\, ,
\end{align}
and read
\begin{align}
\notag
Z_1^{\prime T} = ({\rm e}^\sigma , 0, 0, 0)
\, , \qquad
Z_2^{\prime T} = (0, {\rm e}^\tau , 0, 0)
\, , \qquad
Z_3^{\prime T} = ({\rm e}^\sigma, 0, {\rm e}^{- \sigma}/\varepsilon, 0)
\, , \qquad
Z_4^{\prime T} = (0, {\rm e}^\tau, 0, {\rm e}^{- \tau}/\varepsilon)
\, .
\end{align}
Since the twistor space is projective, we can rescale them individually and observe that in the limit $\varepsilon \to 0$,
they are invariant under the conformal $S$-transformation \re{ConformalTransformation}.

If we choose not to stick to the twistor $\leftrightarrow$ coordinate equivalence and thus loose the direct link to
Poincar\'e coordinates, one can use projective invariance to pull out $\varepsilon^{-1}$ from the $Z_3$ and $Z_4$ and
set $\varepsilon = 0$ from the get-go. This gives
\begin{align}
\label{epsilonTozeroTwistors}
Z_1^T = (1, 0, 0, 0)
\, , \qquad
Z_2^T = (0, 1, 0, 0)
\, , \qquad
Z_3^T = (0, 0, 1, 0)
\, , \qquad
Z_4^T = (0, 0, 0, 1)
\, .
\end{align}
Though we are not able to get the absolute $x^\pm_n$ coordinates from these twistors because the denominator in Eq.\ \re{EmbeddingXfromZ}
taken literally is singular, we nevertheless can extract the distances from the square of the interval
\begin{align}
x_{nk}^2 = \frac{\vev{n-1,n, k-1,k}}{\vev{k-1 k}\vev{n-1 n}}  = x_{nk}^- x_{nk}^+
\, ,
\end{align}
where the last equality is valid for two-dimensional kinematics only. In the parametrization used below, the $x^-$ coordinates will be a function of
$\tau$, while $x^+$ of $\sigma$ parameters of the conformal transformation, so the identification will be straightforward (up to an overall
constant factor that cancels in the product of $x^+$ and $x^-$). Notice that with the twistor defined as in \re{epsilonTozeroTwistors}, the invariance of
the reference square under conformal transformation is obvious since under \re{ConformalTransformation} they read
\begin{align}
\notag
Z_1^{\prime T} = ({\rm e}^\sigma, 0, 0, 0)
\, , \qquad
Z_2^{\prime T} = (0, {\rm e}^\tau, 0, 0)
\, , \qquad
Z_3^{\prime T} = (0, 0, {\rm e}^{- \sigma}, 0)
\, , \qquad
Z_4^{\prime T} = (0, 0, 0, {\rm e}^{- \tau})
\, .
\end{align}
and are trivially equivalent to the original ones (\ref{epsilonTozeroTwistors}).

\subsection{Octagon and decagon}
\label{DecagonParametrization}

\begin{figure}[t]
\begin{center}
\mbox{
\begin{picture}(0,180)(240,0)
\put(0,0){\insertfig{17}{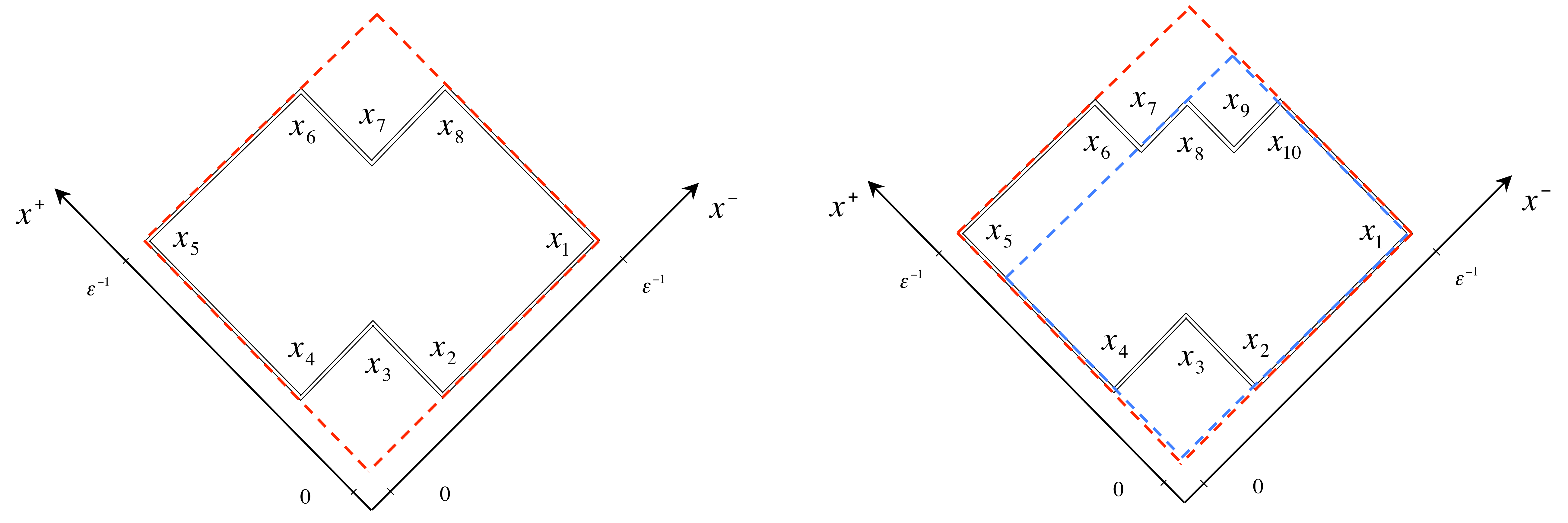}}
\end{picture}
}
\end{center}
\caption{ \label{Decagon} Octagon and decagon with reference squares.}
\end{figure}

Let us now address the tessellation of the octagon and decagon used in Section \ref{OPEpolygons}.
Starting with the octagon, we choose the twistors defining its contour (see Fig.\ \ref{Decagon}) as
\begin{align}
\label{OctagonTwist}
&
Z_1 = (1, 0, 0, 0)
\, , \qquad
Z_2 = (0, 1, 0, -1)
\, , \qquad
Z_3 = (1, 0, -1, 0)
\, , \qquad
Z_4 = (0, 1, 0, 0)
\, , \\
&
Z_5 = (1, 0, 0, \varepsilon^{-1})
\, , \quad\,
Z_6 = (0, 1, 0, 1)
\, , \qquad\ \
Z_7 = (1, 0, 1, 0)
\, , \qquad\ \ \,
Z_{8} = (0, 1, 0, \varepsilon^{-1})
\, , \nonumber
\end{align}
with the reference square formed by $Z_1$, $Z_4$, $Z_5$ and $Z_8$. It is shown by the red dashed contour in Fig.\ \ref{Decagon} and coincides with
the one discussed in the previous section. It is invariant under the transformation (\ref{SforRefSquare1}) as was shown above. By acting with the
$S$-transformation in the top-right corner cusp, we can get all inequivalent octagons. To get the same parametrization, we can act instead with an inverse of
this $S$-transformation on the bottom-left cusp of the polygon,
\begin{align}
Z^\prime_i = S^{-1} (\tau, \sigma) Z_i
\qquad\mbox{for}\qquad i = 2, 3
\, ,
\end{align}
with the transformed momentum twistors being
\begin{align}
&
Z_1 = (1, 0, 0, 0)
\, , \qquad
Z_2 = (0, {\rm e}^{\tau}, 0, -{\rm e}^{-\tau})
\, , \qquad
Z_3 = ({\rm e}^{\sigma}, 0, -{\rm e}^{-\sigma}, 0)
\, , \qquad
Z_4 = (0, 1, 0, 0)
\, , \nonumber\\
&
Z_5 = (1, 0, 0, \varepsilon^{-1})
\, , \quad
Z_6 = (0, 1, 0, 1)
\, , \qquad\qquad \,
Z_7 = (1, 0, 1, 0)
\, , \qquad\qquad \
Z_{8} = (0, 1, 0, \varepsilon^{-1})
\, .
\end{align}
The coordinates of the cusps corresponding to the above twistors can be read of from Eq.\ \re{EmbeddingXfromZ},
\begin{align}
\label{OctagonCoord}
&
x_1 = (\varepsilon^{-1}, 0)
\, , \qquad
x_2 = (-\sigma^{-2 \tau}, 0)
\, , \qquad
x_3 = (-\sigma^{-2 \tau}, - \sigma^{-2 \sigma})
\, , \qquad
x_4 = (0, - \sigma^{-2 \sigma})
\, , \\
&
x_5 = (0, \varepsilon^{-1})
\, , \qquad
x_6 = (1, \varepsilon^{-1})
\, , \qquad\quad \
x_7 = (1, 1)
\, , \qquad\qquad\qquad\ \
x_{8} = (\varepsilon^{-1}, 1)
\, . \nonumber
\end{align}
These are given in Euclidean space. To convert them to the Lorentzian signature (which is shown in the figure), one has to drop all minus signs.

For the decagon, we choose the twistors as
\begin{align}
&
Z_1 = (1, 0, 0, 0)
\, , \qquad
Z_2 = (0, 1, 0, -1)
\, , \qquad
Z_3 = (1, 0, -1, 0)
\, , \qquad
Z_4 = (0, 1, 0, 0)
\, , \nonumber\\
&
Z_5 = (1, 0, 0, \varepsilon^{-1})
\, , \quad
Z_6 = (0, 1, 0, 1)
\, , \qquad\ \ \,
Z_7 = (1, 0, 1, 0)
\, , \qquad\ \ \,
Z_8 = (0, 1, 0, 1)
\, , \\
&\qquad\qquad\qquad\qquad\quad\,
Z_9 = (2, 0, 1, 0)
\, , \qquad\ \ \,
Z_{10} = (0, 1, 0, \varepsilon^{-1})
\, . \nonumber
\end{align}
In this case we have two reference squares, the red and the blue one. The red one is invariant under the transformation (\ref{SforRefSquare1}),
while the blue one is invariant under the following matrix multiplication
\begin{align}
\label{SforRefSquare2}
S^\prime (\tau, \sigma) = \left(
\begin{array}{cccc}
{\rm e}^\sigma & & & \\
& {\rm e}^\tau & & \\
 {\rm e}^{- \sigma} -  {\rm e}^{\sigma} & & {\rm e}^{- \sigma} & \\
& & & {\rm e}^{- \tau}
\end{array}
\right)
\, .
\end{align}
Therefore, we can parametrize all inequivalent decagons by acting with $S$ on the cusps in the top-right corner of the decagon, or which is equivalent to it,
the inverse $S^{-1}$ of the bottom-left cusp $x_3$, also we act with $S^\prime$ on the cusp $x_9$, i.e.,
\begin{align}
&
Z_i^\prime = S^{-1} (\tau_1, \sigma_1) Z_i \, , \qquad\mbox{for}\qquad i = 2, 3
\, , \nonumber\\
&
Z_j^\prime = S^\prime (\tau_2, \sigma_2) Z_i \, , \qquad\ \ \mbox{for}\qquad i = 8, 9
\, .
\end{align}
Such that the corner coordinates are
\begin{align}
\label{DecagonCoord}
&
x_1 = (\varepsilon^{-1}, 0)
\, , \qquad
x_2 = (-\sigma^{-2 \tau_1}, 0)
\, , \qquad\qquad
x_3 = (-\sigma^{-2 \tau_1}, - \sigma^{-2 \sigma_1})
\, , \qquad
x_4 = (0, - \sigma^{-2 \sigma_1})
\, , \nonumber\\
&
x_5 = (0, \varepsilon^{-1})
\, , \qquad
x_6 = (1, \varepsilon^{-1})
\, , \qquad\qquad\quad \ \
x_7 = (1, 1)
\, , \qquad\qquad\qquad\quad\,
x_8 = (\sigma^{2 \tau_1}, 1)
\, , \\
&\qquad\qquad\qquad\qquad
x_9 = (\sigma^{2 \tau_1}, (1 + {\rm e}^{2 \sigma_2})^{-1})
\, , \quad
x_{10} = (\varepsilon^{-1}, (1 + {\rm e}^{2 \sigma_2})^{-1})
\, . \nonumber
\end{align}
The parametrizations \re{OctagonCoord} and \re{DecagonCoord} are used in the main text to cast the remainder functions in a form suitable for OPE
analysis for the octagon \re{W8ConfFrame} and decagon \re{W10ConfFrame}, respectively.

%

\section{Feynman graphs}
\label{FeynmanAppendix}

\begin{figure}[t]
\begin{center}
\mbox{
\begin{picture}(0,50)(240,0)
\put(0,0){\insertfig{17}{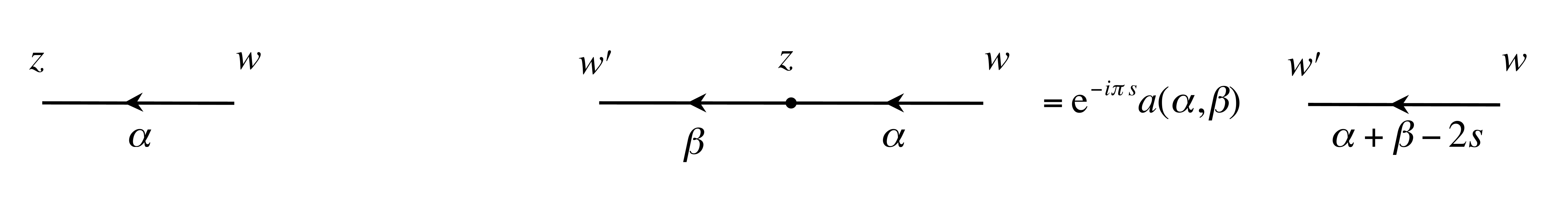}}
\end{picture}
}
\end{center}
\caption{ \label{PropagatorChain} Feynman rules for the propagator \re{Schwinger} on the left and the chain rule \re{ChainRule} on the right.}
\end{figure}

The basic ingredient of the Feynman rules is the propagator, illustrated graphically in Fig.\ \ref{PropagatorChain}. All calculation involving it are easily
performed making use of the Schwinger representation
\begin{align}
\label{Schwinger}
(z - \bar{w})^{- \alpha} = \frac{{\rm e}^{- i \pi \alpha/2}}{\Gamma (\alpha)}
\int_0^\infty d p \, p^{\alpha - 1} {\rm e}^{i p (z - \bar{w})}
\, ,
\end{align}
valid in upper half-plane where $z$ and $w$ have positive imaginary part. Together with the integral representation for the
Dirac delta-function
\begin{align}
\label{Delta}
\int Dz \, {\rm e}^{i \tau z - i \tau^\prime \bar{z}} = \Gamma (2s) \tau^{1- 2s} \delta (\tau - \tau^\prime)
\, ,
\end{align}
and Fourier representation (and its inverse) for  a holomorphic function $\Psi (z)$ in the upper complex half-plane,
\begin{align}\label{PP}
\Psi (z)
=
\int_0^\infty d p \, {\rm e}^{i p z} \, \widetilde\Psi (p)
\, , \qquad
\widetilde\Psi (p)
=
\frac{p^{2s - 1}}{\Gamma (2s)} \int Dz \, {\rm e}^{- i p \bar{z}} \, \Psi (z)
\, ,
\end{align}
make up for all results required for performing explicit computations efficiently.

By virtue of these results, we immediately find the so-called reproducing kernel $\mathcal{K}$ that obeys the following property
\begin{align}
\label{RepKernel}
\int Dw \, \mathcal{K} (z, \bar{w}) \, \Psi (w) = \Psi (z)
\, , \qquad
\mathcal{K} (z, \bar{w}) = {\rm e}^{i \pi s} (z - \bar{w})^{- 2s}
\, .
\end{align}

Let us define the operator of the fractional derivative  $(-i\partial)^a$ as an operator of multiplication by $p^a$ in the momentum
representation. We can easily obtain with the help of Eqs.~(\ref{PP}) the following useful representation
\begin{align}
\label{def2derivative}
\partial^a = (- z)^{- a} \frac{\Gamma (a - z \partial)}{\Gamma (- z \partial)}
\, .
\end{align}
Expanding \re{def2derivative} in the vicinity of $a = 0$, we find the relation
\begin{align}\label{magic}
\ln \partial + \ln (- z) = \psi (- z \partial)
\end{align}
used in the main text.

By means of the Schwinger parametrization and Eq.\ \re{RepKernel}, one can find an expression for the integral in the
upper half of the complex plane in terms of the contour integral along the real axis
\begin{align}
\label{FromZtoLine}
\int D w \frac{{\rm e}^{i \pi s} \Psi (w)}{(z_1 - \bar{w})^\alpha (z_2 - \bar{w})^\beta}
=
\frac{\Gamma (\alpha + \beta)}{\Gamma (\alpha) \Gamma (\beta)}
\int_0^1 d \tau \, \tau^{\alpha - 1} \bar\tau^{\beta - 1} \Psi (\tau z_1 + \bar\tau z_2)
\, ,
\end{align}
which valid for $\alpha + \beta = 2s$.

\begin{figure}[t]
\begin{center}
\mbox{
\begin{picture}(0,100)(240,0)
\put(0,0){\insertfig{17}{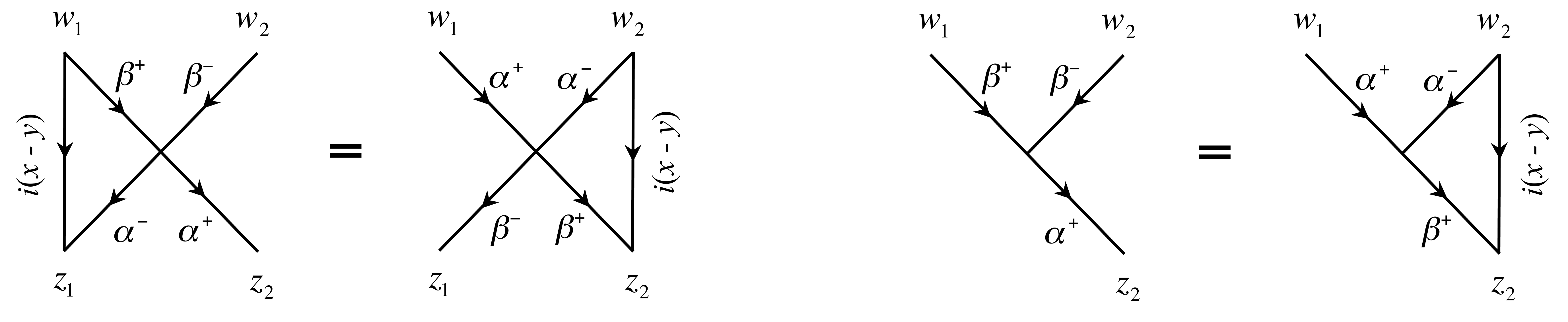}}
\end{picture}
}
\end{center}
\caption{ \label{Permutation} Diagrammatic representation of the permutation identity \re{PermutationIdentity} (on the left) and its form in the
limit when $z_1 \to \infty$ (on the right).}
\end{figure}

Two properties are particularly instrumental in manipulations performed in the main body of the paper. They are:
\begin{itemize}
\item the chain rule
\begin{align}
\label{ChainRule}
\int Dz \, (w^\prime - \bar{z})^{- \beta} (z - \bar{w})^{- \alpha}  = {\rm e}^{- i \pi s}
a (\alpha, \beta) (w^\prime - \bar{w})^{2s-\alpha - \beta }
\, ,
\end{align}
with
\begin{align}
\label{a-factor}
a (\alpha, \beta) = \frac{\Gamma (\alpha + \beta - 2s) \Gamma (2s)}{\Gamma (\alpha) \Gamma (\beta)}
\, ,
\end{align}
shown in Fig.\ \ref{PropagatorChain}, and
\item the permuation identity, illustrated in Fig.\ \ref{Permutation},
\begin{align}
\label{PermutationIdentity}
(z_1 - \bar{w}_1)^{i (y - x)} X (\bit{z}; x | \bit{w}; y)
=
(z_2 - \bar{w}_2)^{i (y - x)} X (\bit{z}; y | \bit{w}; x)
\, ,
\end{align}
for the cross integral
\begin{align}
X (\bit{z}, x | \bit{w}, y)
=
\int D v \, (z_1 - \bar{v})^{- \alpha^-} (z_2 - \bar{v})^{- \alpha^+} (v - \bar{w}_1)^{- \beta^+} (v - \bar{w}_2)^{- \beta^-}
\, ,
\end{align}
where $\bit{z} = (z_1, z_2)$ and $\bit{w} = (w_1, w_2)$ and $\alpha^\pm = s \pm i x$ and $\beta^\pm = s \pm i y$.
\end{itemize}
While the former can be easily obtained by first applying the Schwinger parametrization for the propagators \re{Schwinger} and subsequent use
of the integral representation for the delta function \re{Delta}, the latter requires a few extra steps. Namely, first using Eq.\ \re{FromZtoLine}
with the $\Psi$-function given by the product of the remaining two propagators and joining the latter via the Feynman parametrization
with a parameter $\sigma$, we can conveniently change these two integration variables to $\xi$ and $\rho$, such that $\tau = 1/(1 + \xi)$
and $\sigma = 1/(1 + \rho)$ with both of them varying in the semi-infinite interval $[0, \infty)$. This allows us to rescale them as follows $\xi \to \xi
(\bar{w}_2 - z_2)/(\bar{w}_2  -z_1)$, $\rho \to \rho (\bar{w}_2 - z_2)/(\bar{w}_1 - z_2)$ and evaluate the resulting $\xi$ integral in
terms of the Euler Beta function $B (\alpha^-, \alpha^+)$. The final $\rho$ integral depends on a single cross ratio $R = (\bar{w}_2 - \bar{w}_1)
(z_2 - z_1)/[(\bar{w}_2 - z_2) (\bar{w}_1 - z_1)]$ and yields the hypergeometric function ${_2F_1}$,
\begin{align}
X (\bit{z}, x | \bit{w}, y)
=
{\rm e}^{- i \pi s}
(\bar{w}_1 - z_2)^{\alpha^- - \beta^+} (\bar{w}_1 - z_1)^{- \alpha^-} (\bar{w}_2 - z_2)^{ - \beta^-}
{_2F_1}
\left.\left(
{\beta^-, \alpha^- \atop 2 s}
\right| R \right)
\, .
\end{align}
This representation immediately yields \re{PermutationIdentity}. Another identity used in the main text emerges when one sends the point $z_1$
to infinity. It is shown graphically in the right panel in Fig.\ \ref{Permutation}.

\section{Intertwining relations}
\label{Intappendix}

In this Appendix we provide proofs of intertwining relation (\ref{WNintertwiner}) and eigenvalue relation (\ref{XU}).

\subsection{Intertwining relation~(\ref{WNintertwiner})}

Let us perform a similarity transformation $V_N^{-\lambda} \,\mathbb{T}^{(s)}(u)\, V_N^{\lambda}$ on the monodromy matrix \re{Monodromy}
with $V_N = z_{21}z_{32}\cdots z_{N,N-1}$. In order to proceed, we will use the following factorized expression for the Lax operator
\begin{align}
\mathbb{L}_n(u,s) = \mathbb{M}_n^{(s)}\,
\left(%
\begin{array}{cc}
  1 & -i\partial_n \\
  0 & 1 \\
\end{array}%
\right)\,
\mathbb{N}_n^{(s)}
\end{align}
where the two-by-two matrices are
\begin{align}
\mathbb{M}_n^{(s)}
=
\left(
\begin{array}{cc}
1 & 0 \\
z_n & u-is\\
\end{array}
\right)
\, , \qquad
\mathbb{N}_n^{(s)} = \left(
\begin{array}{cc}
u+is-i & 0 \\
-z_n & 1 \\
\end{array}
\right) \, .
\end{align}
A key ingredient of the calculation can be clearly illustrated on a simplest two-site example. Namely, the product of two Lax operators admits the
following factorized representation
\begin{align}
\mathbb{L}_1(u,s)\,\mathbb{L}_2(u,s)
= \mathbb{M}_1^{(s)}\,
\left(
\begin{array}{cc}
1 & -i\partial_{1} \\
0 & 1 \\
\end{array}
\right) \mathbb{N}_1^{(s)} \mathbb{M}_2^{(s)}
\left(
\begin{array}{cc}
1 & -i\partial_{2} \\
0 & 1 \\
\end{array}
\right)
\mathbb{N}_2^{(s)}\,.
\end{align}
Since the left- and right-most lower-triangular matrices $\mathbb{M}_1^{(s)}$ and $\mathbb{N}_2^{(s)}$ are invariant with respect to the introduced
similarity transformation, we can focus on the middle matrix block,
\begin{align}
\mathbb{N}_1^{(s)}\,\mathbb{M}_2^{(s)} =
\left(
\begin{array}{cc}
u_+-i & 0 \\
z_{21} & u_- \\
\end{array}
\right)\, .
\end{align}
Namely, we find
\begin{align}
&
z_{21}^{-\lambda}\,\mathbb{L}_1(u,s)\,\mathbb{L}_2(u,s)\,
z_{21}^{\lambda}=
\notag\\[2mm]
&\qquad=
\mathbb{M}_1^{(s)}\,\left(
\begin{array}{cc}
1 & -i\partial_{1} \\
0 & 1 \\
\end{array}
\right)\left(
\begin{array}{cc}
u_++i\lambda-i & \lambda(2s-1+\lambda)/z_{21} \\
z_{21} & u_--i\lambda \\
\end{array}
\right)
\left(
\begin{array}{cc}
1 & -i\partial_{2} \\
0 & 1 \\
\end{array}
\right)\,\mathbb{N}_2^{(s)}
\, . 
\end{align}
by moving $z_{21}$ from right to left. Therefore, by setting $\lambda = 1-2s$ we realize that the similarity transformation yields the change of the spin
variable $s \rightarrow 1-s$ in the middle matrix block
$$
z_{21}^{2s-1}
\mathbb{L}_1(u,s) \mathbb{L}_2(u,s)
z_{21}^{1-2s}
=
\mathbb{M}_1^{(s)}
\left(
\begin{array}{cc}
1 & -i\partial_{1} \\
0 & 1 \\
\end{array}
\right)
\mathbb{N}_1^{(1-s)}
\mathbb{M}_2^{(1-s)}
\left(
\begin{array}{cc}
1 & -i\partial_{2} \\
0 & 1 \\
\end{array}
\right) \mathbb{N}_2^{(s)}\,\,.
$$
This calculation clearly demonstrates what happens for three sites and more,
\begin{multline}
z_{32}^{2s-1} \dots z_{N,N-1}^{2s-1}
\mathbb{L}_1 (u,s) \dots \mathbb{L}_N (u,s)\,
z_{21}^{1-2s} \dots z_{N, N-1}^{1-2s}=
\\[2mm]
=
\mathbb{M}_1^{(s)}
\left(
\begin{array}{cc}
1 & - i \partial_1 \\
0 & 1 \\
\end{array}
\right)
\mathbb{N}_1^{(1-s)}\,
\dots
\mathbb{M}_N^{(1-s)}
\left(
\begin{array}{cc}
1 & - i \partial_N \\
0 & 1 \\
\end{array}
\right)
\mathbb{N}_N^{(s)}
\, ,
\end{multline}
namely, while the left- and right-most lower-triangular matrices remain unchanged, for every internal matrix we have the
replace $s \rightarrow 1-s$. Thus, we have
\begin{align}
V_N^{2s-1} \mathbb{T}^{(s)} (u) V_N^{1-2s}
=
\frac{1}{u_-(u_+-i)}
\left(
\begin{array}{cc}
u_+ - i & 0 \\
(2s-1)z_1 & u_- \\
\end{array}%
\right)
\mathbb{T}^{(1-s)}(u)
\left(
\begin{array}{cc}
u_+ - i & 0 \\
(2s-1)z_n & u_- \\
\end{array}
\right) \, ,
\end{align}
where left- and rightmost lower-triangular matrices in this equation are $\mathbb{M}_1^{(s)} [\mathbb{M}_1^{(1-s)}]^{-1}$
and $[ \mathbb{N}_N^{(1-s)} ]^{-1}\,\mathbb{N}_N^{(s)}$, respectively. This is main result of this appendix. It immediately
generates the similarity transformation for the matrix elements of the monodromy matrix \re{Monodromy}. Projecting out the operator
$B^{(s)}_N (u) = \bra{\uparrow} \mathbb{T}^{(s)}_N (u) \ket{\downarrow}$, we get
\begin{align}
V_N^{2s-1} B^{(s)}_N (u) V_N^{1-2s} = B_N^{(1-s)}(u)
\, .
\end{align}
However, for the operator $D_N(u)$ we need to perform an additional similarity transformation on the first site, so that we obtain
instead
\begin{align}
z_1^{2s-1} V_N^{2s-1} D_N^{(s)} (u) V_N^{1-2s}\,z_1^{1-2s} = D_N^{(1-s)}(u)
\,.
\end{align}

\subsection{Formula~(\ref{XU})}

To start with, let us point out that the eigenfunction $\Psi_{\bit{\scriptstyle u}}  (x_1, \dots, x_N)$ defined in Eq.\ \re{ExactWaveFunctLine}
can be written in the nested form corresponding to the diagram shown in Fig.\ \ref{LinePyramid},
\begin{align}
\label{PsiPyramidLambda}
\Psi_{u} (x_1,\ldots, x_N) =
{\rm e}^{- i \pi s N (N-1)/2}
\Lambda_N(u_1)\Lambda_{N - 1}(u_2)\dots \Lambda_{2}(u_{N-1})\, z_N^{- i u_N - s}
\, .
\end{align}
Here the left-most $\Lambda_N$ is an operator with the integral kernel
acting on $N-1$ coordinates of a test function $\Psi (x_2, \dots, x_N)$ as follows
\begin{align}
\label{LambdaN}
\left[\Lambda_N(u)\Psi\right] (x_2, & \ldots, x_N)
=
x_1^{- s - i u} \prod_{k=2}^{N} \int_0^1 d \tau_k \tau_k^{s - i u - 1} \bar\tau_k^{s + i u - 1}
\Psi(\bar\tau_2 x_2 + \tau_2 x_1, \ldots, \bar\tau_{N} x_N + \tau_N x_{N-1} )
\nonumber\\
&= \frac{x_1^{s - iu-1}}{W_N (x_1, \dots, x_N)}
\prod_{n=2}^{N} \int_{x_{n-1}}^{x_n} d x_n^\prime \, (x_n^\prime - x_{n-1})^{s + i u - 1} (x_k - x_k^\prime)^{s - i u - 1}
\Psi (x_2^\prime ,\ldots, x_N^\prime)
\, ,
\end{align}
where the $W_N$ factor is defined in Eq.\ \re{WN}. The other operators $\Lambda_{n < N}$ in Eq.\ \re{PsiPyramidLambda} act only on the last $n-1$
variables, i.e., from $x_{N - n + 2}$ to $x_N$.

We can prove the following identity
\begin{align}
\label{REC}
&
\int_0^\infty \mathcal{D}^{N} x \, W_N (x_1, \dots, x_N) \prod_{n=1}^{N} G_s (x^\prime_n, x_n) \left[\Lambda_N(u)\,\Psi\right](x_2,\ldots, x_N)
\\
&\qquad
=
\mu_s (u)\,\Lambda_N (u)
\int_0^\infty \mathcal{D}^{N -1} x \, W_{N-1} (x_2, \dots, x_N) \prod_{n=2}^{N} G_s (x^\prime_n, x_n) \Psi (x_2,\ldots, x_N)
\, , \nonumber
\end{align}
where the propagator $G_s$ is defined in Eq.\ \re{PropagatorGs}, the one-particle GKP measure  $\mu_s$ is given in Eq.\ \re{musn} and
the integration measure $\mathcal{D}^N x$ is introduced in Eq.\ \re{RealMeasure} with $\mathcal{D}^N x = dx_1 \mathcal{D}^{N-1} x$. This
equation is the main building block in constructing a recursion procedure to prove Eq.\ \re{XU}. The action of the $\Lambda_N$ operator on the
product of propagators can be calculated explicitly yielding
\begin{align}
\Lambda_N(u):\,\,  \prod_{n = 2}^N G_s (x_n^\prime, x_n) \mapsto
(x_1^\prime)^{- s - iu} \prod_{n=2}^N \mu_s (u) (x_{n-1}^\prime + x_n)^{- s + iu}(x_n^\prime + x_n)^{- s - iu}
\, ,
\end{align}
so that the right-hand side of Eq.\ \re{REC} can be rewritten as
\begin{align}
\mu^{N-1}_s (u) (x_1^\prime)^{- s - iu}
\int_0^\infty
\frac{
\mathcal{D}^{N-1} x \, W_{N-1} (x_2, \dots, x_N)
}{
\prod_{n=2}^{N}(x_{n-1}^\prime + x_n)^{s - iu}(x_n^\prime + x_n)^{s + iu}
}
\Psi (x_2, \ldots, x_N)\,.
\end{align}
In the left-hand side of \re{REC} we use the definition of the $\Lambda_N$-operator given in the second line in Eq.\ \re{LambdaN} and obtain
\begin{align}
\label{lhsREC}
\int_0^\infty
\mathcal{D}^{N} x \, x_1^{s - iu - 1} \prod_{n = 1}^{N} G_s (x^\prime_n, x_n)
\,
\prod_{k = 2}^N \int_{x_{k-1}}^{x_k} d y_k\, (y_k-x_{k-1})^{s + i u - 1} (x_k-y_k)^{s - i u - 1} \Psi(y_2, \ldots, y_N)\, .
\end{align}
Here the integration variables are ordered in the following fashion
\begin{align}
x_N\geq y_{N}\geq x_{N-1}\geq \ldots
x_k\geq y_{k}\geq x_{k-1}\geq \ldots x_2\geq y_{2}\geq x_{1}\geq 0
\, .
\end{align}
This constraint can written in terms of step function in two equivalent ways
\begin{align}
&
\theta (x_N > x_{N-1} > \dots > x_1)
\prod_{n = 2}^{N} \theta (x_n > y_n > x_{n - 1})
\nonumber\\
&=
\theta (y_N > y_{N-1} > \dots > y_2)
\,\theta (x_{N} > y_{N})
\,\theta (y_2 > x_{1})
\prod_{n = 2}^{N-1} \theta (y_{n+1} > x_n > y_n)
\, .
\end{align}
Making use of the second representation, it is possible to interchange the order $x$ and $y$ integrals in \re{lhsREC} and get
\begin{align}
\int_0^\infty \mathcal{D}^{N-1} y \,\Psi (y_2,\ldots,y_N)\,
\prod_{n=1}^N \mathcal{J}_n
\, ,
\end{align}
where all $\mathcal{J}_n$ integrals entering this formula  (with $1 \leq n \leq N$) can be calculated explicitly yielding
\begin{align}
&\notag
\mathcal{J}_1 \equiv
\int_{0}^{y_2} dx_1 x_1^{s - iu-1} (y_2 - x_1)^{s + i u - 1} G_s (x^\prime_1, x_1)
=
\mu_s(u) (x^\prime_1)^{- s - iu} (x^\prime_1 + y_2)^{ - s + iu}  y_2^{2s-1}
\, , \\
& \notag
\mathcal{J}_n
\equiv
\int_{y_n}^{y_{n+1}} d x_n \,
(y_{n+1} - x_n)^{s + i u - 1} (x_n - y_n)^{s - i u - 1} G_s (x^\prime_n, x_n)
=
\frac{\mu_s(u)\, y_{n,n-1}^{2s-1}}{(x^\prime_n + y_n)^{s + iu}(x^\prime_n + y_{n + 1})^{s - iu}}
\, , \\
& \notag
\mathcal{J}_N
\equiv
\int_{y_N}^\infty d x_N
\,
(x_N - y_N)^{s - i u - 1} G_s (x^\prime_N, x_N) = \mu_s(u) (x^\prime_N+y_N)^{- s - iu}
\, ,
\end{align}
with $1 < n < N$. Collecting everything together we find that the left-hand side the expression
\begin{align}
\mu^{N-1}_s (u) (x_1^\prime)^{- s - iu}
\int_0^\infty
\frac{\mathcal{D}^{N-1} y \, W_{N-1} (y_2, \dots, y_N)}{\prod_{n=2}^{N}(x^\prime_{n-1} + y_n)^{s - iu}(x^\prime_n + y_n)^{s + iu}}
\Psi (y_2,\ldots, y_N)
\, .
\end{align}
This coincides with right-hand sides of \re{REC} and thus proves the validity of this identity. Thus, starting from the relation
\begin{align}
\int_{0}^\infty d x_N \, x_N^{2s - 1} G_s (x_N^\prime, x_N) x_N^{- i u_N -s} = \mu_s(u_N) (x_N^\prime)^{- i u_N - s}
\, ,
\end{align}
and using recurrence~(\ref{REC}) we obtain the relation~(\ref{XU}) in the following form
\begin{align}
\int_0^\infty \mathcal{D}^N x \, W_N (x_1, \dots, x_N) \prod_{n=1}^N G_s (x^\prime_n, x_n) \Psi_{\bit{\scriptstyle u}} (x_1, \ldots, x_N)
=
\prod_{n=1}^N \mu_s (u_n) \Psi_{\bit{\scriptstyle u}} (x^\prime_1, \ldots, x^\prime_N)
\, .
\end{align}
Changing the variables as $x^\prime_n \to - x^\prime_n$ and using the relation \re{XmX} along with $\mathcal{K}_{w_1, \dots, w_N} (x_1, \dots, x_N)
= {\rm e}^{i \pi s N} \prod_{n=1}^N G_s (- w_n, x_n)$, it proves the relation \re{XU} in question.

\section{Separation of Variables}
\label{SoVappendix}

In this Appendix we provide a complementary construction of the wave functions that has its roots in the SoV formalism by Sklyanin \cite{Skl85}.
It is based on the construction of eigenfunctions of the element $B_N$ of the monodromy matrix and its subsequent use in transformation to a
factorizable representation.

\subsection{Diagonalization of $B_N$}

The 12-element of the $N$-site monodromy matrix $B_N$ encodes a set of $N$ commuting operators $\{\widehat{p}, \widehat{b}_1, \dots,
\widehat{b}_{N-1}\}$,
\begin{align}
B_N (u) = \widehat{p} \prod_{n=1}^{N-1} (u - \widehat{b}_n)
\, ,
\end{align}
with explicitly factorized overall momentum operator $\widehat{p} = \sum_{n = 1}^N i S^0_n$, whose commutation relations follow from the
Yang-Baxter equation with a rational $R$-matrix,
\begin{align}
[\widehat{p}, \widehat{b}_n] = [\widehat{b}_n, \widehat{b}_k] = 0
\, .
\end{align}
To find the eigenfunctions of the operators $\{\widehat{p}, \widehat{b}_1, \dots, \widehat{b}_{N-1}\}$ with eigenvalues
$\{ p, x_1, \dots, x_{N - 1} \}$,
\begin{align}
B_N (u) U_{p, \bit{\scriptstyle x}} (z_1, \dots, z_N) =  p \prod_{n=1}^{N-1} (u - x_n) U_{p, \bit{\scriptstyle x}} (z_1, \dots, z_N)
\end{align}
as in the case of the operator $D_N$, we construct a recursion relation. The starting point of the analysis is the defining equation \re{SLLminus},
where we restore the correct spectral parameters in the Lax operator $\mathbb{L}_1$ making use of the left matrix multiplication
(cf.\ Eq.\ \re{LeftFactUpper})
\begin{align}
\mathbb{L}_1 (u_+, v_-) =  \overline{\mathbb{G}}_1 (v_-|u_-) \mathbb{L}_1 (u_+, u_-)
\, , \qquad
 \overline{\mathbb{G}}_1 (v_+|u_+)
 \equiv
 \left(
\begin{array}{cc}
1 & 0 \\
\frac{u_- - v_-}{u_-} z_1  & \frac{v_-}{u_-}
\end{array}
\right)
\, .
\end{align}
Then we find
\begin{align}
&
\mathcal{S}_N^- (u_+-v_-, u_--v_-)
\mathbb{T}_{N-1}  (u_+, u_-)
\mathbb{L}_N (u_+, v_-)
\nonumber\\
&
=
\overline{\mathbb{G}}_1 (v_-|u_-)
\mathbb{T}_N  (u_+, u_-) \mathcal{S}_N^- (u_+-v_-, u_--v_-)
\, ,
\end{align}
with the $N-1$-site monotromy matrix $\mathbb{T}_{N-1}  (u_+, u_-) = \mathbb{L}_1 (u_+, u_-) \dots \mathbb{L}_{N-1} (u_+, u_-)$.
Projecting out the 12-component of this matrix equation, we find an operator identity relating $N$- and $N-1$-site operators
\begin{align}
&
\mathcal{S}_N^- (u_+-v_-, u_--v_-)
\left( - A_{N-1} (u) i \partial_N + B_{N-1} (u_1) (v_- - i z_N \partial_N) \right)
\nonumber\\
&
=
B_N (u_1) \mathcal{S}_N^- (u_+-v_-, u_--v_-)
\, .
\end{align}
Acting on a function with appropriate properties that allows us to get rid of the first term in the left-hand side, it will define a recursion relation for $B_N$.
Namely, acting on a function independent of the $z_N$ coordinate does the job of eliminating $A_{N-1}$ out from the equation. Then, setting $v_- = u - x_1$,
we get the first level of the recurrence relation
\begin{align}
\label{RecursionForBN}
&
B_N (u) \mathcal{S}_N^- (x_1+is, x_1-is) \Psi (z_1, \dots z_{N - 1})
\nonumber\\
&
=
(u - x_1) \mathcal{S}_N^- (x_1+is, x_1-is) B_{N-1} (u) \Psi (z_1, \dots z_{N - 1})
\, .
\end{align}
Repeating this $N-2$ times, we finally come to the equation
\begin{align}
B_N (u) \mathcal{S}_N^- (x_1+is, x_1-is) \dots
\mathcal{S}_2^- (x_{N-1}+is, x_{N-1}-is)  \Psi (z_1)
=
\prod_{n=1}^{N-1} (u - x_n) B_1 (u)  \Psi (z_1)
\, ,
\end{align}
for the last coordinate $z_1$, where the differential operator acting on it admits a very simple form, i.e., $B_1 (u) = i S^-_1$. The eigenfunction of the $B_1$
operator with the eigenvalue $p$ is $\Psi (z_1) = \exp (i p z_1)$. Therefore, the eigenfunction of $B_N$ is
\begin{align}
U_{p, \bit{\scriptstyle x}} (z_1, \dots, z_N)
&
=
{\rm e}^{-i \pi s N (N-1)/2}
\\
&\times
\mathcal{S}_N^- (x_1 + i s, x_1 - i s) \dots \mathcal{S}_2^- (x_{N-1} + i s, x_{N-1} - is)  {\rm e}^{i p z_1}
\, , \nonumber
\end{align}
where we used the invariance of intertwiners under the shift of of their arguments. This expression can be cast in the integral form and, again, admits the
structure of a pyramid
\begin{align}
U_{p, \bit{\scriptstyle x}} (z_1, \dots, z_N)
&
= \int D w_{1}^{\scriptscriptstyle (1)} \dots D w_{N-1}^{\scriptscriptstyle (1)} \, Y_{x_1} \left( z_1, z_2 | \bar{w}_1^{\scriptscriptstyle (1)} \right)
\dots
Y_{x_1} \left( z_{N-1}, z_N | \bar{w}_{N-1}^{\scriptscriptstyle (1)} \right)
\nonumber\\
&\times \int D w_{1}^{\scriptscriptstyle (2)} \dots D w_{N-2}^{\scriptscriptstyle (2)} \,
\,
Y_{x_2} \left( w_1^{\scriptscriptstyle (1)}, w_2^{\scriptscriptstyle (1)} | \bar{w}_1^{\scriptscriptstyle (2)} \right)
\dots
Y_{x_2} \left( w_{N-2}^{\scriptscriptstyle (1)}, w_{N-1}^{\scriptscriptstyle (1)} | \bar{w}_{N-2}^{\scriptscriptstyle (2)} \right)
\nonumber\\
&\qquad\quad\vdots
\nonumber\\
&\times \int D w_1^{\scriptscriptstyle (N-1)}
\,
Y_{x_{N-1}} \left( w_1^{\scriptscriptstyle (N-2)}, w_2^{\scriptscriptstyle (N-2)} | \bar{w}_1^{\scriptscriptstyle (N-1)} \right)
\exp \left( i p  w_1^{\scriptscriptstyle (N-1)} \right)
\, , \nonumber
\end{align}
in terms of the $Y$-function \re{Yfunction}. This eigenfunction of $B_N$ coincides with the one found in Ref.\ \cite{DerKorMan02} dedicated
to SoV of the $sl(2, \mathbb{R})$ invariant spin chain: in both cases here and there, one diagonalizes the same off-diagonal element of
the monodromy matrix thus yielding identical results. Therefore, we can borrow the result of that work to our present needs. As can be verified
in a straightforward fashion, this eigenfunction obeys the following multidimensional Baxter equation
\begin{align}
\label{MultiBaxterU}
D_N (x_n) U_{p, \bit{\scriptstyle x}} (z_1, \dots, z_N)
=
(x_n - i s)^N U_{p, \bit{\scriptstyle x} - i  \bit{\scriptstyle e}_n} (z_1, \dots, z_N)
\, ,
\end{align}
with the unit vector $\bit{e}_n$ possessing only one nonvanishing component $(\bit{e}_n)_k = \delta_{nk}$. As we will see below, this will
result in similar equations obeyed by the wave function in the SoV representation.

\begin{figure}[t]
\begin{center}
\mbox{
\begin{picture}(0,270)(230,0)
\put(0,0){\insertfig{16}{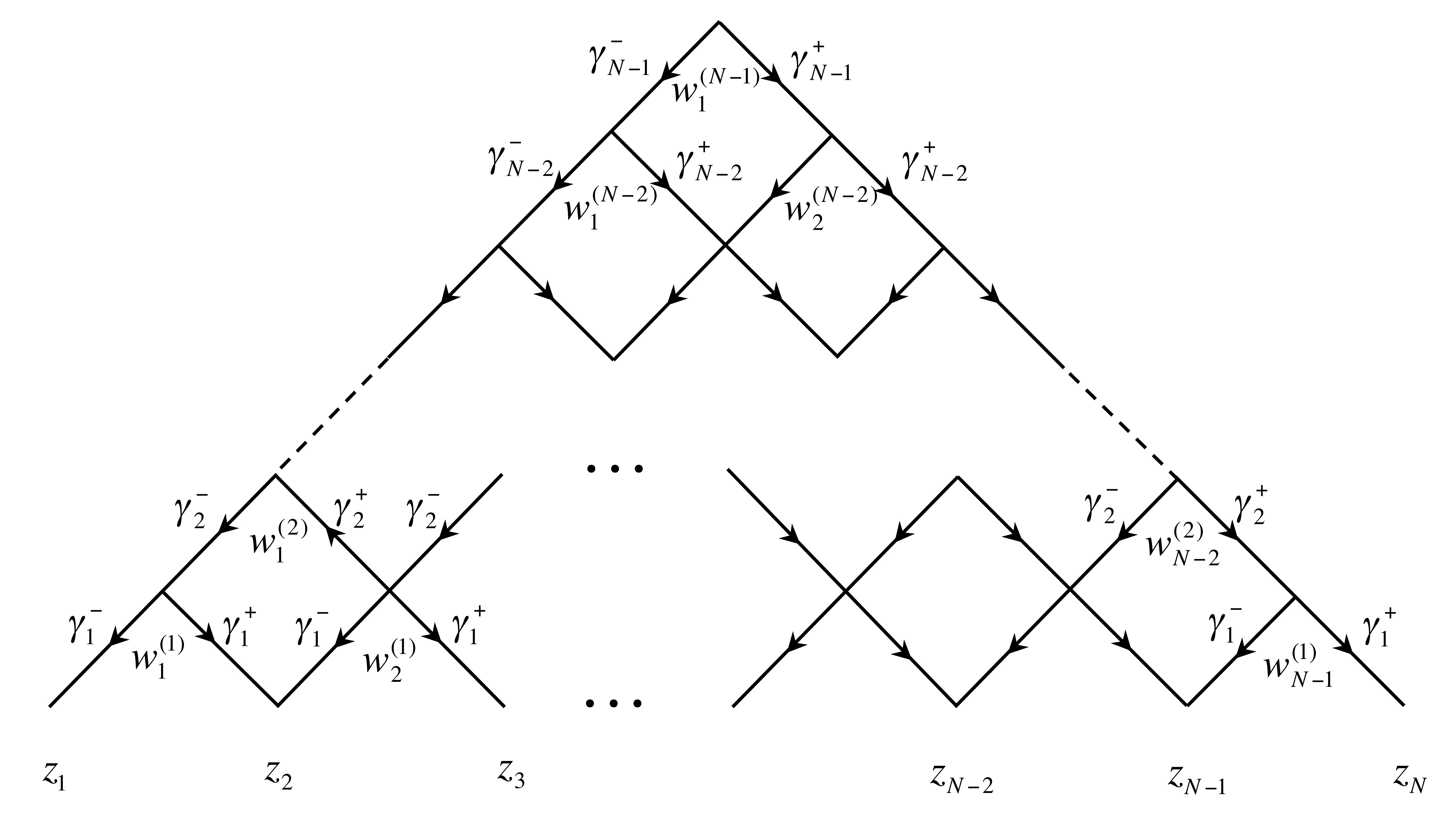}}
\end{picture}
}
\end{center}
\caption{ \label{BeigenstateU} Diagrammatic representation for $B_N$ eigenstate. Here $\gamma_n^\pm \equiv s \pm i x_n$
stands for the power of the propagator and other conventions are the same as in Fig.\ \ref{NparticleWF}.}
\end{figure}

\subsection{Transformation to SoV}

The completeness condition for the wave functions $U_{p, \bit{\scriptstyle x}}$  reads
\begin{align}
\int_0^\infty dp \int_{- \infty}^{\infty} d^{N-1} \bit{x} \, \mathcal{M} (\bit{x}) \,
U^\ast_{p, \bit{\scriptstyle x}} (w_1, \dots , w_N) U_{p, \bit{\scriptstyle x}} (z_1, \dots , z_N)
=
\prod_{n=1}^{N} \mathcal{K} (z_n, \bar{w}_n)
\, ,
\end{align}
where $\mathcal{K}$ is the reproducing kernel \re{RepKernel} and the SoV measure $\mathcal{M} (\bit{x})$ was found to be
\cite{DerKorMan02}
\begin{align}
\mathcal{M} (\bit{x})
=
c_N \prod_{n=1}^{N-1} [\Gamma (s + i x_n) \Gamma (s - i x_n)]^N \prod_{j<k}^{N-1}
[\Gamma (- i x_j + i x_k) \Gamma (- i x_k + i x_j)]^{-1}
\, ,
\end{align}
with the overall factor $c_N = [2^{N-1} \pi^{-(N-1)(N-4)/2} \Gamma (N) \Gamma^{N^2} (2s) ]^{-1}$. Thus, the transformation of the eigenfunction
$\Psi_{\bit{\scriptstyle u}}$ to the SoV basis $\Phi (\bit{x}, p)$, that depends on the SoV coordinates $\bit{x}$ and $p$, can be achieved by
means of the integral representation
\begin{align}
\Psi_{\bit{\scriptstyle u}} (z_1, \dots, z_N)
=
\int_0^\infty dp \int_{- \infty}^{\infty} d^{N-1} \bit{x} \, \mathcal{M} (\bit{x}) \, U_{p, \bit{\scriptstyle x}} (z_1, \dots, z_N) \Phi_{\bit{\scriptstyle u}} (p, \bit{x})
\, .
\end{align}
Making use of the orthogonality condition \cite{DerKorMan02}
\begin{align}
\label{Uorthogonality}
\int D z_1 \dots D z_{N} \, U^\ast_{p^\prime, \bit{\scriptstyle x}^\prime} (z_1, \dots , z_N) U_{p, \bit{\scriptstyle x}} (z_1, \dots , z_N)
=
\mathcal{M}^{-1} (\bit{x}) \delta (p - p^\prime) \delta^{N-1} (\bit{x} - \bit{x}^\prime)
\, ,
\end{align}
we immediately find for the SoV wave function
\begin{align}
\label{SoVwavefunction}
\Phi_{\bit{\scriptstyle u}} (p, \bit{x}) = \int D z_1 \dots D z_N \,
U^\ast_{p, \bit{\scriptstyle x}} (z_1, \dots , z_N) \Psi_{\bit{\scriptstyle u}} (z_1, \dots , z_N)
\, .
\end{align}

\begin{figure}[t]
\begin{center}
\mbox{
\begin{picture}(0,330)(230,0)
\put(0,0){\insertfig{16}{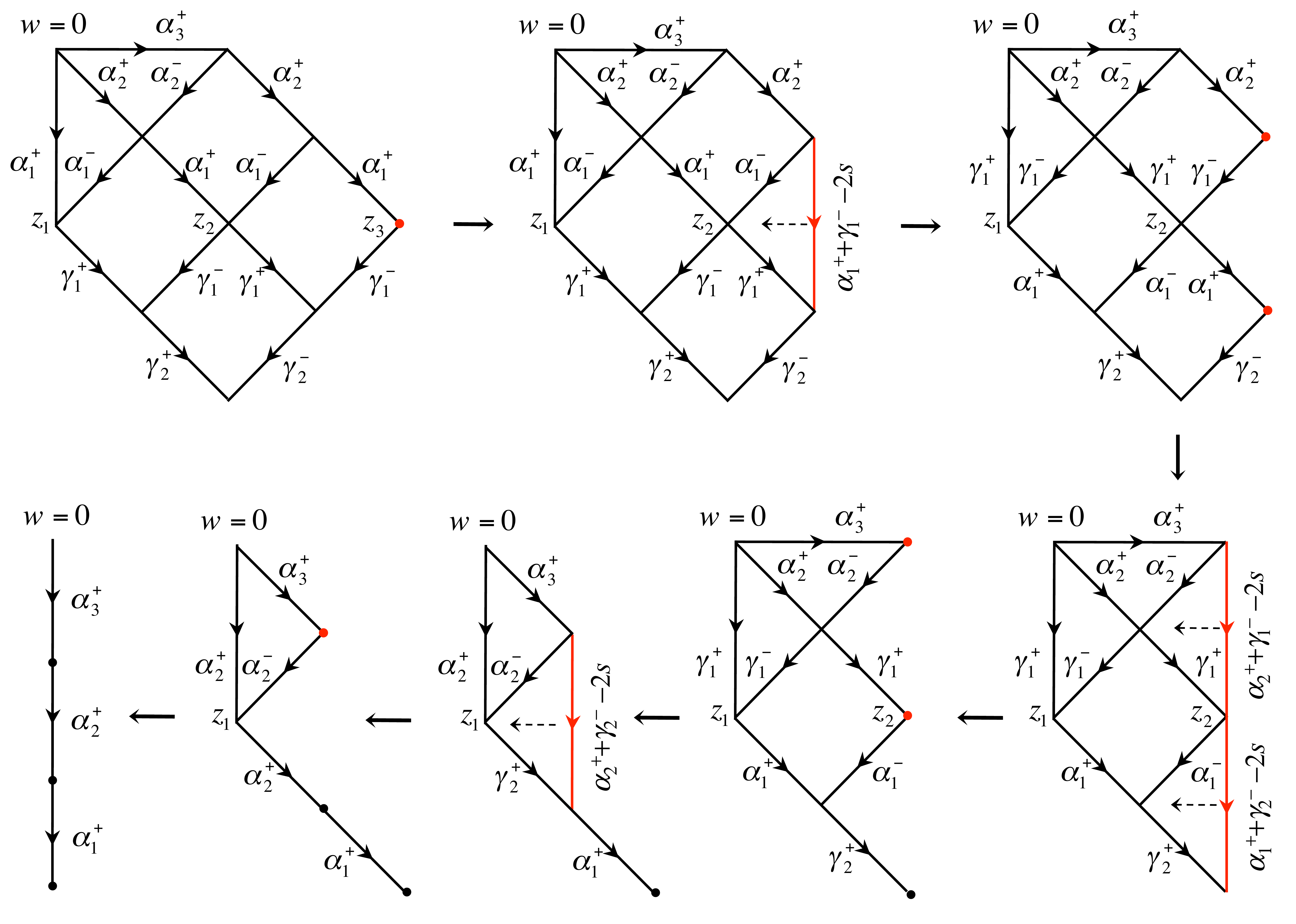}}
\end{picture}
}
\end{center}
\caption{ \label{cphi} Scalar product of $B_N$ and $D_N$ eigenfunctions for $N=3$.}
\end{figure}

As it follows from Eq.\ \re{MultiBaxterU}, the wave function $\Phi_{\bit{\scriptstyle u}} (p, \bit{x})$ satisfies the multidimensional
Baxter equation
\begin{align}
\prod_{k=1}^N(x_n + u_k) \Phi_{\bit{\scriptstyle u}} (p, \bit{x)}
=
(x_n + i s)^N \Phi_{\bit{\scriptstyle u} + i \bit{\scriptstyle e}_n} (p, \bit{x)}
\, ,
\end{align}
and factorizes into the product of $N-1$ Baxter functions (eigenvalues of the Baxter operator),
\begin{align}
\label{SoVwavefunction2}
\Phi_{\bit{\scriptstyle u}} (p, \bit{x}) = c_{\bit{\scriptstyle u}} (p) Q^-_{\bit{\scriptstyle u}} (x_1) \dots Q ^-_{\bit{\scriptstyle u}} (x_{N-1})
\, .
\end{align}
Since both $B_N$ and $D_N$ eigenfunctions are available in an explicit form, we can simply find $c_{\bit{\scriptstyle u}} (p)$ making use
of their integral representation, Eqs.\ \re{Deigenfunction} and \re{SoVwavefunction}, respectively. The calculation uses diagrammatic rules
and follows the steps of Section \ref{BNorthogonality} with minor modification. Namely, one finds (see Fig.\ \ref{cphi} for details)
\begin{align}
\vev{U_{p, \bit{\scriptstyle x}} | \Psi_{\bit{\scriptstyle u}} }
=
c_{\bit{\scriptstyle u}} (p)
\prod_{k=1}^{N - 1} \prod_{n = 1}^N a ( s - i x_k, s + i u_n)
\, ,
\end{align}
where we immediately recognize in the product of $a$-factors \re{a-factor} the product of eigenvalues of the $\mathbb{Q}^-$ operator
\re{QpmOperators}, while the overall coefficient is determined by the nested integral
\begin{align}
c_{\bit{\scriptstyle u}} (p)
= \int \prod_{n=1}^{N} D w_n \, (w_n - \bar{w}_{n+1})^{-s - i u_n} \exp (- i p \bar{w}_1)
\, ,
\end{align}
with $w_{N + 1} = 0$. Its straightforward calculation yields
\begin{align}
c_{\bit{\scriptstyle u}} (p)
=
{\rm e}^{- i (\pi/2) \sum_{n=1}^N (s + i u_n)}
&
\frac{\Gamma^N (2s) p^{2 (N-2) s - \sum_{n=1}^N (s + i u_n)}}{\prod_{n=1}^{N} \Gamma(s + iu_n)}
\, ,
\end{align}
and completes the construction of the normalized eigenfunction in the SoV representation.


\end{document}